\documentclass{ws-p8-50x6-00}
\usepackage{epsf,rotate,cite_unsrt}

\begin{document}

\newcommand{\nwc}{\newcommand}
%
%
\nwc{\cl}  {\clubsuit}
\nwc{\di}  {\diamondsuit}
\nwc{\sps}  {\spadesuit}

\nwc{\fs}  {\footnotesize\em}
\nwc{\ns}  {\normalsize}

\nwc{\hyp} {\hyphenation}
\nwc{\ds}  {\displaystyle}
\nwc{\non} {\nonumber}
\nwc{\lra} {\longrightarrow}
\nwc{\Llra}{\Longleftrightarrow}
\nwc{\lmt} {\longmapsto}
\nwc{\prl} {\partial}
\nwc{\iy}  {\infty}
\nwc{\ol}  {\overline}
\nwc{\ul}  {\underline}
\nwc{\pr}  {\prime}
\nwc{\nnn} {\nonumber \vspace{.2cm} \\ }
\nwc{\Sc}  {{\cal S}}
\nwc{\Lc}  {{\cal L}}
\nwc{\Rc}  {{\cal R}}
\nwc{\Dc}  {{\cal D}}
\nwc{\Oc}  {{\cal O}}
\nwc{\Cc}  {{\cal C}}
\nwc{\Pc}  {{\cal P}}
\nwc{\Mc}  {{\cal M}}
\nwc{\Ec}  {{\cal E}}
\nwc{\Fc}  {{\cal F}}
\nwc{\Hc}  {{\cal H}}
\nwc{\Kc}  {{\cal K}}
\nwc{\Xc}  {{\cal X}}
\nwc{\Gc}  {{\cal G}}
\nwc{\Zc}  {{\cal Z}}
\nwc{\Nc}  {{\cal N}}
\nwc{\fca} {{\cal f}}
\nwc{\xc}  {{\cal x}}
\nwc{\Ac}  {{\cal A}}
\nwc{\Bc}  {{\cal B}}
\nwc{\Uc}  {{\cal U}}
\nwc{\Vc}  {{\cal V}}
%
%
\nwc{\vth} {\vartheta}
\nwc{\eps}{\epsilon}
\nwc{\si} {\sigma}
\nwc{\Gm} {\Gamma}
\nwc{\gm} {\gamma}
\nwc{\bt} {\beta}
\nwc{\La} {\Lambda}
\nwc{\la} {\lambda}
\nwc{\om} {\omega}
\nwc{\Om} {\Omega}
\nwc{\dt} {\delta}
\nwc{\Si} {\Sigma}
\nwc{\Dt} {\Delta}
\nwc{\vph}{\varphi}
\nwc{\zt} {\zeta}
%
%
\def\tr{\mathop{\rm tr}}
\def\Tr{\mathop{\rm Tr}}
\def\Det{\mathop{\rm Det}}
\def\Im{\mathop{\rm Im}}
\def\Re{\mathop{\rm Re}}
\def\secder#1#2#3{{\partial^2 #1\over\partial #2 \partial #3}}
\def\bra#1{\left\langle #1\right|}
\def\ket#1{\left| #1\right\rangle}
\def\VEV#1{\left\langle #1\right\rangle}
\def\gdot#1{\rlap{$#1$}/}
\def\abs#1{\left| #1\right|}
\def\pr#1{#1^\prime}
\def\ltap{\raisebox{-.4ex}{\rlap{$\sim$}} \raisebox{.4ex}{$<$}}
\def\gtap{\raisebox{-.4ex}{\rlap{$\sim$}} \raisebox{.4ex}{$>$}}
\nwc{\Id}  {{\bf 1}}
\nwc{\sgn}  {{\rm sgn}}
\nwc{\diag} {{\rm diag}}
\nwc{\inv}  {{\rm inv}}
\nwc{\mod}  {{\rm mod}}
\nwc{\hal} {\frac{1}{2}}
\nwc{\tpi}  {2\pi i}
\def\contract{\makebox[1.2em][c]{
        \mbox{\rule{.6em}{.01truein}\rule{.01truein}{.6em}}}}
\def\slash#1{#1\!\!\!/\!\,\,}

\def\APP#1{Acta Phys.~Pol.~{\bf #1}}
\def\AP#1{Ann. Phys.~{\bf #1}}
\def\CMP#1{Comm. Math. Phys.~{\bf #1}}
\def\CNPP#1{Comm. Nucl. Part. Phys.~{\bf #1}}
\def\EPJ#1{Eur. Phys. J.~{\bf #1}}
\def\HPA#1{Helv. Phys. Acta~{\bf #1}}
\def\IJMP#1{Int. J. Mod. Phys.~{\bf #1}}
\def\JP#1{J. Phys.~{\bf #1}}
\def\MPL#1{Mod. Phys. Lett.~{\bf #1}}
\def\NP#1{Nucl. Phys.~{\bf #1}}
\def\NPPS#1{Nucl. Phys. Proc. Suppl.~{\bf #1}}
\def\NC#1{Nuovo Cim.~{\bf #1}}
\def\PH#1{Physica {\bf #1}}
\def\PL#1{Phys. Lett.~{\bf #1}}
\def\PR#1{Phys. Rev.~{\bf #1}}
\def\PRP#1{Phys. Rep.~{\bf #1}}
\def\PRL#1{Phys. Rev. Lett.~{\bf #1}}
\def\PNAS#1{Proc. Nat. Acad. Sc.~{\bf #1}}
\def\PTP#1{Progr. Theor. Phys.~{\bf #1}}
\def\RMP#1{Rev. Mod. Phys.~{\bf #1}}
\def\RNC#1{Riv. Nuo. Cim.~{\bf #1}}
\def\ZP#1{Z. Phys.~{\bf #1}}

\def \Msol {M_\odot}
\def\eV {\,{\rm  eV}}     
\def\KeV {\,{\rm  KeV}}     
\def\MeV {\,{\rm  MeV}}
\def\GeV {\,{\rm  GeV}}     
\def\TeV {\,{\rm  TeV}}     
\def\fm {\,{\rm  fm}}     

\def \lta {\mathrel{\vcenter
     {\hbox{$<$}\nointerlineskip\hbox{$\sim$}}}}
\def \gta {\mathrel{\vcenter
     {\hbox{$>$}\nointerlineskip\hbox{$\sim$}}}} 
\newsavebox{\nnin} \sbox{\nnin}{$\hspace{1mm}\in\kern -.8em /
                   \hspace{1mm}$}
\newcommand{\nin}{\usebox{\nnin}}
 
\newcommand{\sub}{\subset}
\newsavebox{\nnsub} \sbox{\nnsub}{$\hspace{1mm}\sub\kern -.9em /
            \hspace{1mm}$}
\newcommand{\nsub}{\usebox{\nnsub}}
%
%
\def\KK{{\rm I\kern -.2em  K}}
\def\NN{{\rm I\kern -.16em N}}
\def\RR{{\rm I\kern -.2em  R}}
\def\ZZ{Z \kern -.43em Z}
\def\QQ{{\rm \kern .25em
             \vrule height1.4ex depth-.12ex width.06em\kern-.31em Q}}
\def\CC{{\rm \kern .25em
             \vrule height1.4ex depth-.12ex width.06em\kern-.31em C}}
\def\ZZZ{Z\kern -0.31em Z}
 
\nwc{\iddq}  {\int\frac{d^d q}{(2\pi)^d}}

\nwc{\cst}     {SU_L(3)\times SU_R(3)}
\nwc{\csN}     {SU_L(N)\times SU_R(N)}
\nwc{\rmcl}    {{\rm cl}}
\nwc{{\rmeff}}   {{\rm eff}}

\title{Flow Equations for Phase Transitions in Statistical Physics and QCD}

\author{D.-U.~Jungnickel\thanks{E-mail:
    D.Jungnickel@thphys.uni-heidelberg.de} and C.~Wetterich\thanks{E-mail:
    C.Wetterich@thphys.uni-heidelberg.de}}

\address{Institut f\"ur Theoretische Physik, Universit\"at Heidelberg,
  Philosophenweg 16\\ 69120 Heidelberg, Germany}


\maketitle

\abstracts{We review the formalism of the effective average action in quantum
  field theory which corresponds to a coarse grained free energy in
  statistical mechanics. The associated exact renormalization group equation
  and possible nonperturbative approximations for its solution are discussed.
  We describe in detail $O(N)$--symmetric scalar theories in two and three
  dimensions. These ideas are also applied to QCD where one observes the
  consecutive emergence of mesonic bound states and spontaneous chiral
  symmetry breaking as the coarse graining scale is lowered. We finally
  present a study of the chiral phase transition in two flavor QCD. A
  precision estimate of the universal critical equation of state for the
  three--dimensional $O(4)$ Heisenberg model is presented.  We explicitly
  connect the $O(4)$ universal behavior near the critical temperature and zero
  quark mass with the physics at zero temperature and a realistic pion mass.
  For realistic quark masses the pion correlation length near $T_c$ turns out
  to be smaller than its zero temperature value.  }

\section{Effective average action}
\label{EffectiveAverageAction}

Quantum chromodynamics (QCD) describes qualitatively different physics
at different length scales. At short distances the relevant degrees of
freedom are quarks and gluons which can be treated perturbatively. At
long distances we observe hadrons, and an essential part of the
dynamics can be encoded in the masses and interactions of mesons. Any
attempt to deal with this situation analytically and to predict the
meson properties from the short distance physics (as functions of the
strong gauge coupling $\alpha_s$ and the current quark masses $m_q$)
has to bridge the gap between two qualitatively different effective
descriptions. Two basic problems have to be mastered for an
extrapolation from short distance QCD to mesonic length scales:
\begin{itemize}
\item The effective couplings change with scale. This does not only
  concern the running gauge coupling, but also the coefficients of
  non--renormalizable operators as, for example, four quark operators.
  Typically, these non--renormalizable terms become important in the
  momentum range where $\alpha_s$ is strong and deviate substantially
  from their perturbative values. Consider the four--point function
  which results after integrating out the gluons. For heavy quarks it
  contains the information about the shape of the heavy quark
  potential whereas for light quarks the complicated spectrum of light
  mesons and chiral symmetry breaking are encoded in it. At distance
  scales around $1{\rm fm}$ one expects that the effective action
  resembles very little the form of the classical QCD action which is
  relevant at short distances.
\item Not only the couplings, but even the relevant variables or
  degrees of freedom are different for long distance and short
  distance QCD. It seems forbiddingly difficult to describe the
  low--energy scattering of two mesons in a language of quarks and
  gluons only. An appropriate analytical field theoretical method
  should be capable of introducing field variables for composite
  objects such as mesons.
\end{itemize}
A conceptually very appealing idea for our task is the block--spin
action.$^{\citen{Kad66-1,Wil71-1}}$ It realizes that physics with a given
characteristic length scale $l$ is conveniently described by a functional
integral with an ultraviolet (UV) cutoff $\Lambda$ for the momenta. Here
$\Lambda$ should be larger than $l^{-1}$ but not necessarily by a large
factor. The Wilsonian effective action $S_\Lambda^{\rm W}$ replaces then the
classical action in the functional integral. It is obtained by integrating out
the fluctuations with momenta $q^2\gta\Lambda^2$. An exact renormalization
group equation$^{\citen{Wil71-1,WH73-1,Wei76-1,Pol84-1,Has86-1}}$ describes
how $S_\Lambda^{\rm W}$ changes with the UV cutoff $\Lambda$.

We will use here the somewhat different but related concept of the effective
average action$^{\citen{Wet91-1}}$ $\Gamma_k$ which, in the language of
statistical physics, is a coarse grained free energy with coarse graining
scale $k$. The effective average action is based on the quantum field
theoretical concept of the effective action$^{\citen{Sch51-1}}$ $\Gamma$ which
is obtained by integrating out all quantum fluctuations.  The effective action
contains all information about masses, couplings, form factors and so on,
since it is the generating functional of the $1PI$ Green functions. The field
equations derived from $\Gamma$ are exact including all quantum effects. For a
field theoretical description of thermal equilibrium this concept is easily
generalized to a temperature dependent effective action which includes now
also the thermal fluctuations. In statistical physics $\Gamma$ describes the
free energy as a functional of some convenient (space dependent) order
parameter, for instance the magnetization. In particular, the behavior of
$\Gamma$ for a constant order parameter (the effective potential) specifies
the equation of state. The effective average action $\Gamma_k$ is a simple
generalization of the effective action, with the distinction that only quantum
fluctuations with momenta $q^2\gta k^2$ are included. This can be achieved by
introducing an explicit infrared cutoff $\sim k$ in the functional integral
defining the partition function (or the generating functional for the
$n$--point functions).  Typically, this IR--cutoff is quadratic in the fields
and modifies the inverse propagator, for example by adding a mass--like term
$\sim k^2$. The effective average action can then be defined in complete
analogy to the effective action (via a Legendre transformation of the
logarithm of the partition function). The mass--like term in the propagator
suppresses the contributions from the small momentum modes with $q^2\lta k^2$
and $\Gamma_k$ accounts effectively only for the fluctuations with $q^2\gta
k^2$.

Following the behavior of $\Gamma_k$ for different $k$ is like looking at
the world through a microscope with variable resolution: For large $k$ one
has a very precise resolution $\sim k^{-1}$ but one also studies
effectively only a small volume $\sim k^d$. Taking in QCD the coarse
graining scale $k$ much larger than the confinement scale guarantees that
the complicated nonperturbative physics does not play a role yet.  In this
case, $\Gamma_k$ will look similar to the classical action, typically with
a running gauge coupling evaluated at the scale $k$.  (This does not hold
for Green functions with much larger momenta $p^2\gg k^2$ where the
relevant IR cutoff is $p$, and the effective coupling is $\alpha_s(p)$.)
For lower $k$ the resolution is smeared out and the detailed information of
the short distance physics can be lost.  (Again, this does not concern
Green functions at high momenta.)  On the other hand, the ``observable
volume'' is increased and long distance aspects such as collective
phenomena become visible. In a theory with a physical UV cutoff $\Lambda$
we may associate $\Gamma_\Lambda$ with the classical action $S$ since no
fluctuations are effectively included. By definition, the effective average
action equals the effective action for $k=0$, $\Gamma_{0}=\Gamma$, since
the infrared cutoff is absent. Thus $\Gamma_k$ interpolates between the
classical action $S$ and the effective action $\Gamma$ as $k$ is lowered
from $\Lambda$ to zero.  The ability to follow the evolution
$k\rightarrow0$ is equivalent to the ability to solve the quantum field
theory.

For a formal description we will consider in the first two sections a
model with real scalar fields $\chi^a$ (the index $a$ labeling
internal degrees of freedom) in $d$ Euclidean dimensions with
classical action $S[\chi]$. We define the generating functional for
the connected Green functions by
\begin{equation}
  W_k[J]=\ln\int\Dc\chi\exp\left\{-S_k[\chi]+\int d^d x
  J_a(x)\chi^a(x)\right\}
\end{equation} 
where we have added to the classical action an IR cutoff $\Delta_kS$
\begin{equation}
  S_k[\chi] = S[\chi]+\Dt S_k[\chi]
\end{equation}
which is quadratic in the fields and conveniently formulated in
momentum space
\begin{equation}
  \Dt S_k[\chi] = \hal\int\frac{d^d q}{(2\pi)^d}
  R_k(q^2)\chi_a(-q)\chi^a(q)\; .  
\end{equation}
Here $J_a$ are the usual scalar sources introduced to define
generating functionals and $R_k(q^2)$ denotes an appropriately chosen
(see below) IR cutoff function. We require that $R_k(q^2)$ vanishes
rapidly for $q^2\gg k^2$ whereas for $q^2\ll k^2$ it behaves as
$R_k(q^2)\simeq k^2$. This means that all Fourier components of
$\chi^a$ with momenta smaller than the IR cutoff $k$ acquire an
effective mass $m_\rmeff\simeq k$ and therefore decouple while the
high momentum components of $\chi^a$ should not be affected by $R_k$.
The classical fields
\begin{equation}
  \Phi^a\equiv\VEV{\chi^a}=\frac{\dt W_k[J]}{\dt J_a} 
\end{equation} 
now depend on $k$. In terms of $W_k$ the effective average action is
defined via a Legendre transformation
\begin{equation}
  \label{AAA60}
  \Gm_k[\Phi]=-W_k[J]+\int d^dx J_a(x)\Phi^a(x)- 
  \Delta S_k[\Phi] \; .
\end{equation}
In order to define a reasonable coarse grained free energy we have subtracted
in~(\ref{AAA60}) the infrared cutoff piece. This guarantees that the only
difference between $\Gamma_k$ and $\Gamma$ is the effective IR cutoff in the
fluctuations. Furthermore, this has the consequence that $\Gamma_k$ does not
need to be convex whereas a pure Legendre transform is always convex by
definition. (The coarse grained free energy becomes convex$^{\citen{RW90-1}}$
only for $k\rightarrow0$.) This is very important for the description of phase
transitions, in particular first order ones.  One notes
\begin{equation}\begin{array}{lcl}
 \ds{\lim_{k\rightarrow0}R_k(q^2)=0} &\Rightarrow& 
 \ds{\lim_{k\rightarrow0}\Gm_k[\Phi]=\Gm[\Phi]}\nnn
 \ds{\lim_{k\rightarrow\La}R_k(q^2)=\infty} &\Rightarrow&
 \ds{\lim_{k\rightarrow\La}\Gm_k[\Phi]=S[\Phi]}
 \label{ConditionsForRk}
\end{array}\end{equation}
for a convenient choice of $R_k$ like
\begin{equation}
 R_k(q^2)=Z_{\Phi,k} q^2
 \frac{1+e^{-q^2/k^2}-e^{-q^2/\La^2}}
 {e^{-q^2/\La^2}-e^{-q^2/k^2}}\; .
 \label{Rk}
\end{equation}
Here $Z_{\Phi,k}$ denotes the wave function renormalization to be
specified below and we will often use for $R_k$ the limit
$\Lambda\rightarrow\infty$
\begin{equation}
  \label{AAA61}
  R_k(q^2)=\frac{Z_{\Phi,k} q^2}
  {e^{q^2/k^2}-1}\; .
\end{equation}
We note that the property $\Gamma_\Lambda=S$ is not essential since the short
distance laws may be parameterized by $\Gamma_\Lambda$ as well as by $S$. In
addition, for momentum scales much smaller than $\Lambda$ universality implies
that the precise form of $\Gamma_\Lambda$ is irrelevant, up to the values of a
few relevant renormalized couplings.

A few properties of the effective average action are worth mentioning:
\begin{enumerate}
\item All symmetries of the model which are respected by the IR cutoff
  $\Delta_k S$ are automatically symmetries of $\Gamma_k$. In
  particular, this concerns translation and rotation invariance and
  one is not plagued by many of the problems encountered by a
  formulation of the block--spin action on a lattice.
\item In consequence, $\Gamma_k$ can be expanded in terms of
  invariants with respect to these symmetries with couplings depending
  on $k$. For the example of a scalar theory one may use a derivative
  expansion ($\rho=\Phi^a\Phi_a/2$)
  \begin{equation}
    \label{AAA63}
    \Gamma_k=\int d^d x\left\{
    U_k(\rho)+\frac{1}{2}Z_{\Phi,k}(\rho)
    \prl^\mu\Phi_a\prl_\mu\Phi^a+\ldots\right\}
  \end{equation}
  and expand further in powers of $\rho$
  \begin{eqnarray}
    \label{AAA64}
    \ds{U_k(\rho)} &=& \ds{
    \frac{1}{2}\ol{\la}_k\left(
    \rho-\rho_0(k)\right)^2+
    \frac{1}{6}\ol{\gamma}_k\left(
    \rho-\rho_0(k)\right)^3+\ldots}\nnn
    \ds{Z_{\Phi,k}(\rho)} &=& \ds{
    Z_{\Phi,k}(\rho_0)+Z_{\Phi,k}^\prime(\rho_0)
    \left(\rho-\rho_0\right)+\ldots}\; 
  \end{eqnarray}
  where $\rho_0$ denotes the ($k$--dependent) minimum of the effective average
  potential $U_k(\rho)$.  We see that $\Gamma_k$ describes infinitely many
  running couplings.  ($Z_{\Phi,k}$ in~(\ref{Rk}) can be identified with
  $Z_{\Phi,k}(\rho_0)$.)
\item There is no problem incorporating chiral fermions since a chirally
  invariant cutoff $R_k$ can be formulated.$^{\citen{Wet90-1,CKM97-1}}$
\item Gauge theories can be formulated along similar
  lines$^{\citen{RW93-1,Bec96-1,BAM94-1,EHW94-1,Wet95-2,EHW96-1,Ell98-1,
      Ell98-2,LP98-1,Sim98-1}}$ even though $\Delta_k S$ may not be gauge
  invariant.\footnote{For a manifestly gauge invariant formulation in terms of
    Wilson loops see$^{\citen{Mor98-1}}$.} In this case the usual Ward
  identities receive corrections for which one can derive closed
  expressions.$^{\citen{EHW94-1}}$ These corrections vanish for
  $k\rightarrow0$.
\item The high momentum modes are very effectively integrated out because of
  the exponential decay of $R_k$ for $q^2\gg k^2$.  Nevertheless, it is
  sometimes technically easier to use a cutoff without this fast decay
  property (e.g.~$R_k\sim k^2$ or $R_k\sim k^4/q^2$). In the latter case one
  has to be careful with possible remnants of an incomplete integration of the
  short distance modes.  Also our cutoff does not introduce any
  non--analytical behavior as would be the case for a sharp
  cutoff.$^{\citen{Wet91-1}}$\footnote{For results using a sharp cutoff
    function see$^{\citen{Alf94-1}}$.}
\item Despite a similar spirit and many analogies there remains also a
  conceptual difference to the Wilsonian effective action $S_\Lambda^{\rm W}$.
  The Wilsonian effective action describes a set of different actions
  (parameterized by $\Lambda$) for one and the same theory --- the $n$--point
  functions are independent of $\Lambda$ and have to be computed from
  $S_\Lambda^{\rm W}$ by further functional integration. In contrast,
  $\Gamma_k$ describes the effective action for different theories --- for any
  value of $k$ the effective average action is related to the generating
  functional of a theory with a different action $S_k=S+\Delta_k S$. The
  $n$--point functions depend on $k$. The Wilsonian effective action does not
  generate the $1PI$ Green functions.$^{\citen{KKS92-1}}$
\item Because of the incorporation of an infrared cutoff, $\Gamma_k$ is
  closely related to an effective action for averages of
  fields,$^{\citen{Wet91-1}}$ where the average is taken over a volume $\sim
  k^d$.
\end{enumerate}

\section{Exact renormalization group equation}
\label{AnExactRGE}

The dependence of $\Gamma_k$ on the coarse graining scale $k$ is governed by
an exact renormalization group equation (ERGE)$^{\citen{Wet93-1}}$
\begin{equation}
  \prl_t\Gm_k[\Phi] =  \hal\Tr\left\{\left[
  \Gm_k^{(2)}[\Phi]+R_k\right]^{-1}\prl_t R_k\right\} \; .
  \label{ERGE}
\end{equation}
Here $t=\ln(k/\La)$ with some arbitrary momentum scale $\La$, and the
trace includes a momentum integration as well as a summation over
internal indices, $\Tr=\int\frac{d^d q}{(2\pi)^d}\sum_a$. The second
functional derivative $\Gm_k^{(2)}$ denotes the {\em exact} inverse
propagator
\begin{equation}
 \label{AAA69}
 \left[\Gm_k^{(2)}\right]_{ab}(q,q^\prime)=
 \frac{\dt^2\Gm_k}{\dt\Phi^a(-q)\dt\Phi^b(q^\prime)}\; .
\end{equation}
The flow equation~(\ref{ERGE}) can be derived from~(\ref{AAA60}) in a
straightforward way using
\begin{eqnarray}
  \label{AAA70}
  \ds{\prl_t \left.\Gamma_k\right|_\Phi} &=& \ds{
  -\prl_t \left.W_k\right|_J-
  \prl_t\Delta_kS[\Phi]}\nnn
  &=& \ds{
  \frac{1}{2}\Tr\left\{\prl_t R_k\left(
  \VEV{\Phi\Phi}-\VEV{\Phi}\VEV{\Phi}\right)
  \right\}}\nnn
  &=& \ds{
  \frac{1}{2}\Tr\left\{
  \prl_t R_k W^{(2)}_k\right\} }
\end{eqnarray}
and
\begin{eqnarray}
  \label{AAA71}
  \ds{W_{k,ab}^{(2)}(q,q^\prime)} &=& \ds{
  \frac{\delta^2 W_k}
  {\delta J^a(-q)\delta J^b(q^\prime)} }\nnn
  \ds{\frac{\delta^2 W_k}
  {\delta J_a(-q)\delta J_b(q^\prime)}
  \frac{\delta^2\left(\Gamma_k+\Delta_k S\right)}
  {\delta\Phi_b(-q^{\prime})\delta\Phi_c
  (q^{\prime\prime})}  } &=& \ds{
  \delta_{ac}\delta_{q q^{\prime\prime}} }\; .
\end{eqnarray}
It has the form of a renormalization group improved one--loop
expression.$^{\citen{Wet91-1}}$ Indeed, the one--loop formula for $\Gamma_k$
reads
\begin{equation}
  \label{AAA72}
  \Gamma_k[\Phi]=S[\Phi]+
  \frac{1}{2}\Tr\ln\left(
  S^{(2)}[\Phi]+R_k\right)
\end{equation}
with $S^{(2)}$ the second functional derivative of the {\em classical} action,
similar to~(\ref{AAA69}).  (Remember that $S^{(2)}$ is the field dependent
classical inverse propagator. Its first and second derivative with respect to
the fields describe the classical three-- and four--point vertices,
respectively.) Taking a $t$--derivative of~(\ref{AAA72}) gives a one--loop
flow equation very similar to~(\ref{ERGE}) with $\Gamma_k^{(2)}$ replaced by
$S^{(2)}$. It may seem surprising, but it is nevertheless true, that the
renormalization group improvement $S^{(2)}\ra\Gamma_k^{(2)}$ promotes the
one--loop flow equation to an exact nonperturbative flow equation which
includes the effects from all loops as well as all contributions which are
non--analytical in the couplings like instantons, etc.! For practical
computations it is actually often quite convenient to write the flow
equation~(\ref{ERGE}) as a formal derivative of a renormalization group
improved one--loop expression
\begin{equation}
  \label{AAA73}
  \prl_t\Gamma_k=
  \frac{1}{2}\Tr\tilde{\prl}_t\ln\left(
  \Gamma_k^{(2)}+R_k\right)
\end{equation}
with $\tilde{\prl}_t$ acting only on $R_k$ and not on $\Gamma_k^{(2)}$, i.e.
$\tilde{\prl}_t=\left(\prl R_k/\prl t\right)\left(\prl/\prl R_k\right)$. Flow
equations for $n$--point functions follow from appropriate functional
derivatives of~(\ref{ERGE}) or~(\ref{AAA73}) with respect to the fields. For
their derivation it is sufficient to evaluate the corresponding one--loop
expressions (with the vertices and propagators derived from $\Gamma_k$) and
then to take a formal $\tilde{\prl}_t$--derivative.  (If the one--loop
expression is finite or properly regularized the $\tilde{\prl}_t$--derivative
can be taken after the evaluation of the trace.) This permits the use of
(one--loop) Feynman diagrams and standard perturbative techniques in many
circumstances. Most importantly, it establishes a very direct connection
between the solution of flow--equations and perturbation theory. If one uses
on the right hand side of~(\ref{ERGE}) a truncation for which the propagator
and vertices appearing in $\Gamma_k^{(2)}$ are replaced by the ones derived
from the classical action, but with running $k$--dependent couplings, and then
expands the result to lowest non--trivial order in the coupling constants one
recovers standard renormalization group improved one--loop perturbation
theory. The formal solution of the flow equation can also be employed for the
development of a systematically resummed perturbation
theory.$^{\citen{Wet96-1}}$

For a choice of the cutoff function similar to~(\ref{AAA61}) the momentum
integral contained in the trace on the right hand side of the flow equation is
both infrared and ultraviolet finite. Infrared finiteness arises through the
presence of the infrared regulator $\sim R_k$. We note that all eigenvalues of
the matrix $\Gamma_k^{(2)}+R_k$ must be positive semi--definite. The proof
follows from the observation that the functional $\Gamma_k+\Delta_k S$ is
convex since it is obtained from $W_k$ by a Legendre transform. On the other
hand, ultraviolet finiteness is related to the fast decay of $\prl_t R_k$ for
$q^2\gg k^2$. This expresses the fact that only a narrow range of fluctuations
with $q^2\simeq k^2$ contributes effectively if the infrared cutoff $k$ is
lowered by a small amount. If for some other choice of $R_k$ the right hand
side of the flow equation would not remain UV finite this would indicate that
the high momentum modes have not yet been integrated out completely in the
computation of $\Gamma_k$.  Since the flow equation is manifestly finite this
can be used to define a regularization scheme. The ``ERGE--scheme'' is
specified by the flow equation, the choice of $R_k$ and the ``initial
condition'' $\Gamma_\Lambda$. This is particularly important for gauge
theories where other regularizations in four dimensions and in the presence of
chiral fermions are difficult to construct. For gauge theories
$\Gamma_\Lambda$ has to obey appropriately modified Ward identities. In the
context of perturbation theory a first proposal how to regularize gauge
theories by use of flow equations can be found in.$^{\citen{Bec96-1,BAM94-1}}$
We note that in contrast to previous versions of exact renormalization group
equations there is no need in the present formulation to construct an
ultraviolet momentum cutoff --- a task known to be very difficult in
non--Abelian gauge theories.

Despite the conceptual differences between the Wilsonian effective action
$S_\Lambda^{\rm W}$ and the effective average action $\Gamma_k$ the exact flow
equations describing the $\Lambda$--dependence of $S_\Lambda^{\rm W}$ and the
$k$--dependence of $\Gamma_k$ are simply related. Polchinski's continuum
version of the Wilsonian flow equation$^{\citen{Pol84-1}}$ can be transformed
into~(\ref{ERGE}) by means of a Legendre transform and a suitable variable
redefinition.$^{\citen{BAM93-1}}$

Even though intuitively simple, the replacement of the (RG--improved)
classical propagator by the full propagator turns the solution of the flow
equation~(\ref{ERGE}) into a difficult mathematical problem: The evolution
equation is a functional differential equation. Once $\Gamma_k$ is expanded in
terms of invariants (e.g.~Eqs.(\ref{AAA63}), (\ref{AAA64})) this is equivalent
to a coupled system of non--linear partial differential equations for
infinitely many couplings. General methods for the solution of functional
differential equations are not developed very far. They are mainly restricted
to iterative procedures that can be applied once some small expansion
parameter is identified.  This covers usual perturbation theory in the case of
a small coupling, the $1/N$--expansion or expansions in the dimensionality
$4-d$ or $2-d$. It may also be extended to less familiar expansions like a
derivative expansion which is related in critical three dimensional scalar
theories to a small anomalous dimension. In the absence of a clearly
identified small parameter one nevertheless needs to truncate the most general
form of $\Gamma_k$ in order to reduce the infinite system of coupled
differential equations to a (numerically) manageable size. This truncation is
crucial. It is at this level that approximations have to be made and, as for
all nonperturbative analytical methods, they are often not easy to control.
The challenge for nonperturbative systems like low momentum QCD is to find
flow equations which (a) incorporate all the relevant dynamics such that
neglected effects make only small changes, and (b) remain of manageable size.
The difficulty with the first task is a reliable estimate of the error. For
the second task the main limitation is a practical restriction for numerical
solutions of differential equations to functions depending only on a small
number of variables.  The existence of an exact functional differential flow
equation is a very useful starting point and guide for this task. At this
point the precise form of the exact flow equation is quite important.
Furthermore, it can be used for systematic expansions through enlargement of
the truncation and for an error estimate in this way.  Nevertheless, this is
not all. Usually, physical insight into a model is necessary to device a
useful nonperturbative truncation!

So far, two complementary approaches to nonperturbative truncations
have been explored: an expansion of the effective Lagrangian in powers
of derivatives ($\rho\equiv\frac{1}{2}\Phi_a\Phi^a$)
\begin{equation}
\label{DerExp}
  \Gamma_k[\Phi]=\int d^d x\left\{
  U_k(\rho)+\frac{1}{2}Z_{\Phi,k}(\rho)
  \prl_\mu\Phi^a\prl^\mu\Phi_a+
  \frac{1}{4}Y_{\Phi,k}(\rho)\prl_\mu\rho
  \prl^\mu\rho+
  \Oc(\prl^4)\right\}
\end{equation}
or one in powers of the fields
\begin{equation}
\label{FieldExp}
  \Gamma_k[\Phi]=
  \sum_{n=0}^\infty\frac{1}{n!}\int
  \left(\prod_{j=0}^n d^d x_j
  \left[\Phi(x_j)-\Phi_0\right]\right)
  \Gamma_k^{(n)}(x_1,\ldots,x_n)\; .
\end{equation}
If one chooses$^{\citen{Wet91-1}}$\footnote{See also~$^{\citen{AMSST98-1}}$
  for the importance of expanding around $\Phi=\Phi_0$ instead of $\Phi=0$.}
$\Phi_0$ as the $k$--dependent VEV of $\Phi$, the series~(\ref{FieldExp})
starts effectively at $n=2$. The flow equations for the $1PI$ $n$--point
functions $\Gamma_k^{(n)}$ are obtained by functional differentiation
of~(\ref{ERGE}).  Such flow equations have been discussed earlier from a
somewhat different viewpoint.$^{\citen{Wei76-1}}$ They can also be interpreted
as a differential form of Schwinger--Dyson equations.$^{\citen{DS49-1}}$

The formation of mesonic bound states, which typically appear as poles in the
(Minkowskian) four--quark Green function, is most efficiently described by
expansions like~(\ref{FieldExp}).  This is also the form needed to compute the
nonperturbative momentum dependence of the gluon propagator and the heavy
quark potential.$^{\citen{Wet95-2,EHW96-1,BBW97-1}}$ On the other hand, a
parameterization of $\Gamma_k$ as in~(\ref{DerExp}) seems particularly suited
for the study of phase transitions. The evolution equation for the average
potential $U_k$ follows by evaluating~(\ref{DerExp}) for constant $\Phi$. In
the limit where the $\Phi$--dependence of $Z_{\Phi,k}$ is neglected and
$Y_{\Phi,k}=0$ one finds$^{\citen{Wet91-1}}$ for the $O(N)$--symmetric scalar
model
\begin{eqnarray}
  \label{AAA80}
  \ds{\prl_t U_k(\rho)} &=& \ds{
        \frac{1}{2}
        \int\frac{d^d q}{(2\pi)^d}
        \frac{\prl R_k}{\prl t}\Bigg(
        \frac{N-1}{Z_{\Phi,k}q^2+R_k+U_k^\prime}
        }\nnn
  &+& \ds{
        \frac{1}{Z_{\Phi,k}q^2+R_k(q)+U_k^\prime+
        2\rho U_k^{\prime\prime}}\Bigg)
        }
\end{eqnarray}
with $U_k^\prime\equiv\frac{\prl U_k}{\prl\rho}$, etc. One observes the
appearance of $\rho$--dependent mass terms in the effective propagators of the
right hand side of~(\ref{AAA80}). Once $\eta_\Phi\equiv-\prl_t\ln Z_{\Phi,k}$
is determined$^{\citen{Wet91-1}}$ in terms of the couplings parameterizing
$U_k$ this is a partial differential equation for a function $U_k$ depending
on two variables $k$ and $\rho$ which can be solved
numerically.$^{\citen{ABB95-1,BTW96-1,Tet96-1,BW96-1}}$ (The Wilson--Fisher
fixed point relevant for a second order phase transition ($d=3$) corresponds
to a scaling solution$^{\citen{TW94-1,Mor94-1}}$ where $\prl_t U_k=0$.) A
suitable truncation of a flow equation of the type~(\ref{DerExp}) will play a
central role in the description of chiral symmetry breaking below.

It should be mentioned at this point that the weakest point in the ERGE
approach seems to be a reliable estimate of the truncation error in a
nonperturbative context. This problem is common to all known analytical
approaches to nonperturbative phenomena and appears often even within
systematic (perturbative) expansions. One may hope that the existence of an
exact flow equation could also be of some help for error estimates. An obvious
possibility to test a given truncation is its enlargement to include more
variables --- for example, going one step higher in a derivative expansion.
This is similar to computing higher orders in perturbation theory and limited
by practical considerations. As an alternative, one may employ different
truncations of comparable size --- for instance, by using different
definitions of retained couplings. A comparison of the results can give a
reasonable picture of the uncertainty if the used set of truncations is wide
enough. In this context we should also note the dependence of the results on
the choice of the cutoff function $R_k(q)$.  Of course, for $k\ra0$ the
physics should not depend on a particular choice of $R_k$ and, in fact, it
does not for full solutions of~(\ref{ERGE}). Different choices of $R_k$ just
correspond to different trajectories in the space of effective average actions
along which the unique IR limit $\Gamma[\Phi]$ is reached. Once approximations
are used to solve the ERGE~(\ref{ERGE}), however, not only the trajectory but
also its end point will depend on the precise definition of the function
$R_k$. This is very similar to the renormalization scheme dependence usually
encountered in perturbative computations of Green functions. One may use this
scheme dependence as a tool to study the robustness of a given approximation
scheme.

Before applying a new nonperturbative method to a complicated theory like QCD
it should be tested for simpler models. A good criterion for the capability of
the ERGE to deal with nonperturbative phenomena concerns the critical behavior
in three dimensional scalar theories.  In a first step the well known results
of other methods for the critical exponents have been reproduced within a few
percent accuracy.$^{\citen{TW94-1}}$ The ability of the method to produce new
results has been demonstrated by the computation of the critical equation of
state for Ising and Heisenberg models$^{\citen{BTW96-1}}$ which has been
verified by lattice simulations.$^{\citen{Tsy94-1}}$ This has been extended to
first order transitions in matrix models$^{\citen{BW96-1}}$ or for the Abelian
Higgs model relevant for superconductors.$^{\citen{BLL95-1,Tet96-1}}$
Analytical investigations of the high temperature phase transitions in $d=4$
scalar theories ($O(N)$--models) have correctly described the second order
nature of the transition,$^{\citen{TW93-1}}$ in contrast to earlier attempts
within high temperature perturbation theory.

For an extension of the flow equations to Abelian and non--Abelian gauge
theories we refer the reader
to.$^{\citen{RW93-1,Bec96-1,BAM94-1,EHW94-1,Wet95-2,EHW96-1,BBW97-1}}$ The
other necessary piece for a description of low--energy QCD, namely the
transition from fundamental (quark and gluon) degrees of freedom to composite
(meson) fields within the framework of the ERGE can be found
in.$^{\citen{EW94-1}}$ We will describe the most important aspects of this
formalism for mesons below.

\section{Scalar $O(N)$ models in two and three dimensions}
\label{ScalarONModels}

The universal critical behavior of many systems of statistical mechanics is
described by the field theory for scalars with $O(N)$ symmetry. This covers
the gas--liquid and many chemical transitions described by Ising models with a
discrete symmetry $Z_2\equiv O(1)$, superfluids with continuous abelian
symmetry $O(2)$, Heisenberg models for magnets with $N=3$, etc. In $2<d\le4$
dimensions all these models have a continuous second order phase transition.
In two dimensions one observes a second order transition for the Ising model,
a Kosterlitz--Thouless phase transition$^{\citen{KT73-1}}$ for $N=2$ and no
phase transition for non--abelian symmetries $N\ge3$. It is
known$^{\citen{MW66-1}}$ that a continuous symmetry cannot be broken in two
dimensions in the sense that the expectation value of the unrenormalized
scalar field vanishes in the limit of vanishing sources,
$\Phi_a=\VEV{\chi_a(x)}=0$. We will see that all this diversity can already be
described using an extremely simple truncation of the flow equation for
$O(N)$--symmetric linear $\Phi^4$--model.

We start from the flow equation~(\ref{AAA80}) for arbitrary $d$ and
introduce the anomalous dimension
\begin{equation}
  \eta_\Phi=-\frac{\prl}{\prl t}\ln Z_{\Phi,k}
\end{equation}
and the threshold function
\begin{equation}
  l_0^d(w;\eta_\Phi)=\frac{1}{4}
  v_d^{-1}k^{-d}Z_{\Phi,k}^{-1}
  \int\frac{d^d q}{(2\pi)^d}
  \frac{\prl_t R_k}
  {q^2+Z_{\Phi,k}^{-1}R_k(q)+k^2 w}
\end{equation}
where
\begin{equation}
  v_d^{-1}=2^{d+1}\pi^{d/2}
  \Gamma(d/2)\; .
\end{equation}
The threshold function decays fast for $w\gg1$ and therefore expresses the
decoupling of heavy particles with mass $\ol{M}_i^2\gg Z_{\Phi,k}k^2$ from the
renormalization group flow. Its precise shape depends on the choice of $R_k$.
It contains a term linear in $\eta_\Phi$ from $Z_{\Phi,k}^{-1}\prl_t
R_k=\prl_t(R_k/Z_{\Phi,k})-\eta_\Phi(R_k/Z_{\Phi,k})$, cf.~eq.~(\ref{AAA61}).
In terms of this threshold function one has
\begin{equation}
  \label{eq:BBB001}
  \prl_t U_k=2v_d k^d\Bigg\{
  l_0^d(\frac{U_k^\prime+2\rho U_k^{\prime\prime}}{Z_{\Phi,k}k^2};
  \eta_\Phi)+
  (N-1)l_0^d(\frac{U_k^\prime}{Z_{\Phi,k}k^2};\eta_\Phi)
  \Bigg\}\; .
\end{equation}
For given $\eta_\Phi(k)$ (see below) this is a partial differential equation
for a function of two variables $U_k(\rho)$. It may be solved numerically for
arbitrary $d$ and $N$. For given ``initial conditions'' specifying
$U_\La(\rho)$ for some microscopic length scale $\La^{-1}$ the free energy
results from the solution for $k\ra0$. In particular, the expectation value
$\VEV{\chi}$ corresponds to the minimum of $U_k$ for $k\ra0$. Denoting the
running location of the potential minimum by $\rho_0(k)$, spontaneous symmetry
breaking occurs for $\rho_0(0)>0$, whereas $\rho_0(0)=0$ corresponds to the
symmetric or disordered phase. The dependence on temperature $T$ is reflected
by the $T$--dependence of the mass term or the location of the minimum at the
scale $\La$. Varying $\rho_0(\La)$ one may detect the different phases. In
particular, a second order phase transition corresponds to a critical value
$\rho_0(\La)=\rho_{0*}$ such that for $\rho_0(\La)>\rho_{0*}$ spontaneous
symmetry breaking occurs, whereas for $\rho_0(\La)<\rho_{0*}$ one ends in the
symmetric phase with $\rho_0(0)=0$ and $U_0^\prime(0)>0$. Near the critical
temperature $T_c$ one may linearize $\rho_0(\La)=\rho_{0*}+A(T_c-T)$.

Instead of describing the results of a numerical solution$^{\citen{BTW96-1}}$
of eq.~(\ref{eq:BBB001}) we make here a very simple polynomial approximation
to $U_k$, namely
\begin{equation}
  \label{eq:BBB002}
  U_k(\rho)=\frac{1}{2}\ol{\la}_k
  \left(\rho-\rho_0(k)\right)^2\; .
\end{equation}
All characteristic features of the flow can already be seen in the
approxi\-mation.$^{\citen{Wet91-1}}$ Furthermore, we also approximate
$l_0^d(w;\eta_\Phi)$ by $l_0^d(w)=l_0^d(w;0)$ which is well justified for
small $\eta_\Phi$. The flow of the potential minimum can be inferred from the
identity
\begin{equation}
  \label{eq:BBB003}
  0=\frac{d}{d t}U_k^\prime(\rho_0(k))=
  \prl_t U_k^\prime(\rho_0(k))+
  U_k^{\prime\prime}(\rho_0(k))\prl_t\rho_0(k)\; .
\end{equation}
Here $\prl_t U_k^\prime(\rho)$ is the partial $t$--derivative with $\rho$ held
fixed which is computed by differentiating eq.~(\ref{eq:BBB001}) with respect
to $\rho$. Defining the additional threshold functions by
\begin{eqnarray}
  \label{eq:BBB004}
  \ds{l_1^d(w;\eta_\Phi)} &=& \ds{
    -\frac{\prl}{\prl w}l_0^d(w;\eta_\Phi)
    }\nnn
  \ds{l_{n+1}^d(w;\eta_\Phi)} &=& \ds{
    -\frac{1}{n}\frac{\prl}{\prl w}
    l_n^d(w;\eta_\Phi)\;,\;\;\;
    n\ge1
    }
\end{eqnarray}
and constants $l_0^d=l_n^d(0;0)$ one obtains
\begin{equation}
  \label{eq:BBB005}
  \prl_t\rho_0=2v_d k^{d-2}Z_{\Phi,k}^{-1}
  \Bigg\{3l_1^d(\frac{2\rho_0\ol{\la}}{Z_{\Phi,k}k^2})+
  (N-1)l_1^d\Bigg\}\; .
\end{equation}
We conclude that $\rho_0(k)$ always decreases as the infrared cutoff $k$ is
lowered. For $d=2$ and $N\ge2$ only a $Z$--factor increasing without bounds
can prevent $\rho_0(k)$ from reaching zero at some value $k>0$. We will see
that this happens for $N=2$ in the low temperature phase.

It is convenient to introduce renormalized dimensionless fields and a
dimensionless potential
\begin{equation}
  \label{eq:BBB006}
  \tilde{\rho}=Z_{\Phi,k}k^{2-d}\rho\; ,\;\;\;
  u=k^{-d} U_k
\end{equation}
and corresponding renormalized dimensionless couplings as
\begin{equation}
  \label{eq:BBB007}
  \kappa=Z_{\Phi,k}k^{2-d}\rho_0\;,\;\;\;
  \la=Z_{\Phi,k}^{-2}k^{d-4}\ol{\la}\; .
\end{equation}
This yields the scaling form$^{\citen{TW94-1}}$ of the flow
equation~(\ref{eq:BBB001})
\begin{eqnarray}
  \label{eq:BBB008}
  \ds{\prl_t u_{|_{\tilde{\rho}}}} &=& \ds{
    -d u+(d-2+\eta_\Phi)\tilde{\rho}u^\prime
    }\nnn
  &+& \ds{
    2v_d\Bigg\{
    l_0^d(u^\prime+2\tilde{\rho}u^{\prime\prime};\eta_\Phi)+
    (N-1)l_0^d(u^\prime,\eta_\Phi)\Bigg\}
    }
\end{eqnarray}
where $u^\prime$ denotes $\prl u/\prl\tilde{\rho}$. All explicit scale and
wave function renormalization factors have disappeared from this partial
differential equation\footnote{The critical solution $\prl_t u=0$ is an
  ordinary differential equation for $u(\tilde{\rho})$ which can be solved
  numerically$^{\citen{Mor94-1,Mor94-2}}$.} for $u(\tilde{\rho},t)$.  For our
simple truncation one has
\begin{equation}
  \label{eq:BBB009}
  u=\frac{1}{2}\la(\tilde{\rho}-\kappa)^2
\end{equation}
and the flow equations for $\kappa$ and $\la$ read
\begin{eqnarray}
  \label{eq:BBB010}
  \ds{\prl_t\kappa} &=& \ds{
    \beta_\kappa=(2-d-\eta_\Phi)\kappa+
    2v_d\Bigg\{3l_1^d(2\la\kappa)+
    (N-1)l_1^d\Bigg\}
    }\\[2mm]
  \label{eq:BBB010a}
  \ds{\prl_t\la} &=& \ds{
    \beta_\la=(d-4+2\eta_\Phi)\la+
    2v_d\la^2\Bigg\{9l_2^d(2\la\kappa)+
    (N-1)l_2^d\Bigg\}
    }\; .
\end{eqnarray}
In this truncation the anomalous dimension $\eta_\Phi$ is given
by$^{\citen{Wet91-1}}$
\begin{equation}
  \label{eq:BBB011}
  \eta_\Phi=\frac{16v_d}{d}\la^2\kappa
  m_{2,2}^d(0,2\la\kappa)
\end{equation}
where $m_{2,2}^d$ is another threshold function with the property
\begin{equation}
  \label{eq:BBB012a}
  \eta_\Phi=\frac{1}{4\pi\kappa}\;\;{\rm for}\;\;
  d=2\;,\;\;\la\kappa\gg1\; .
\end{equation}

We want to demonstrate next that the two differential
equations~(\ref{eq:BBB010}), (\ref{eq:BBB010a}) describe all phase transitions
in two or three dimensions correctly. For $d=3$ one finds a fixed point
$(\kappa_*,\la_*)$ where $\beta_\kappa=\beta_\la=0$. From the generic form
$\beta_\la=-\la+\la^2(c_1+c_2(\la\kappa))$ one concludes that $\la$
essentially corresponds to an infrared stable coupling which is attracted
towards its fixed point value $\la_*$ as $k$ is lowered. On the other hand,
$\beta_\kappa=-\kappa+c_3+c_4(\la\kappa)$ shows that $\kappa$ is essentially
an infrared unstable or relevant coupling. Starting for given $\la_\La$ with
$\kappa_\La=\kappa_*(\la_\La)+\dt\kappa_\La$, $\dt\kappa_\La=a(T_c-T)$, $a>0$,
one either ends in the symmetric phase for $\dt\kappa_\La<0$ or spontaneous
symmetry breaking occurs for $\dt\kappa_\La>0$. The fixed point or, more
generally, the scaling solution with $\prl_t u=0$ corresponds precisely to the
critical temperature of a second order phase transition. Critical exponents
can be computed from solutions in the vicinity of the scaling solution. The
index $\nu$ characterizes the divergence of the correlation length for $T\ra
T_c$, i.e., $\xi\sim m_R^{-1}\sim\abs{T-T_c}^{-\nu}$ with
$m_R^2=\lim_{k\ra0}Z_{\Phi,k}^{-1}\left[U_k^\prime(\rho_0)+2\rho_0
  U_k^{\prime\prime}(\rho_0)\right]$. It corresponds to the negative
eigenvalue of the ``stability matrix'' $A_{i
  j}=(\prl\beta_i/\prl\la_j)(\kappa_*,\la_*)$ with $\la_i\equiv(\kappa,\la)$.
(This can be generalized for more than two couplings.) The critical exponent
$\eta$ determines the long distance behavior of the two--point function for
$T=T_c$. It is simply given by the anomalous dimension at the fixed point,
$\eta=\eta_\Phi(\kappa_*,\la_*)$. It is remarkable that already in a very
simple polynomial truncation the critical exponents come out with reasonable
accuracy.$^{\citen{TW94-1}}$ (See also$^{\citen{BHLM95-1}}$ for a discussion
of the $N=1$ case in three dimensions.)

In two dimensions the term linear in $\kappa$ vanishes in $\beta_\kappa$. This
changes the fixed point structure dramatically as can be seen from
\begin{equation}
  \label{eq:BBB012}
  \lim_{\kappa\ra\infty}\beta_\kappa=
  \frac{N-2}{4\pi}
\end{equation}
where $l_1^2=1$ was used. Since $\beta_\kappa$ is always positive for
$\kappa=0$ a fixed point requires that $\beta_\kappa$ becomes negative for
large $\kappa$. This is the case for the Ising
model$^{\citen{Mor98-1,KNP98-1}}$ where $N=1$. On the other hand, for a
non--abelian symmetry with $N\ge3$ no fixed point and therefore no phase
transition occurs. The location of the minimum always reaches zero for some
value $k_s>0$. The only phase corresponds to a linear realization of $O(N)$
with $N$ degenerate masses $m_R\sim k_s$. It is interesting to note that the
limit $\kappa\ra\infty$ describes the non--linear sigma model. The
non--abelian coupling $g$ of the non--linear model is related to $\kappa$ by
$g^2=1/(2\kappa)$ and eq.~(\ref{eq:BBB012}) reproduces the standard one--loop
beta function for $g$
\begin{equation}
  \label{eq:BBB013}
  \frac{\prl g^2}{\prl t}=
  -\frac{N-2}{2\pi}g^4
\end{equation}
which is characterized by asymptotic freedom.$^{\citen{Pol75-1}}$ The
``confinement scale'' where the coupling $g$ becomes strong can be associated
with $k_s$.  The strongly interacting physics of the non--linear model finds a
simple description in terms of the symmetric phase of the linear
$O(N)$--model!$^{\citen{Wet91-1}}$

Particularly interesting is the abelian continuous symmetry for $N=2$. Here
$\beta_\kappa$ vanishes for $\kappa\ra\infty$ and $\kappa$ becomes a marginal
coupling. Using the exact renormalization group approach outlined above one
actually finds$^{\citen{GW95-1}}$ a behavior consistent with a second order
phase transition with $\eta\simeq0.25$ near the critical trajectory. The low
temperature phase ($\kappa_\La>\kappa_*$) is special since it has many
characteristics of the phase with spontaneous symmetry breaking, despite the
fact that $\rho_0(k\ra0)$ must vanish according to the Mermin--Wagner
theorem.$^{\citen{MW66-1}}$ There is a massless Goldstone--type boson
(infinite correlation length) and one massive mode. Furthermore, the exponent
$\eta$ depends on $\kappa_\La$ or the temperature
(cf.~eq.~(\ref{eq:BBB012a})), since $\kappa$ flows only marginally. These are
the characteristic features of a Kosterlitz--Thouless phase
transition.$^{\citen{KT73-1}}$ The puzzle of the Goldstone boson in the low
temperature phase despite the absence of spontaneous symmetry breaking is
solved by the observation that the wave function renormalization never stops
running with k:
\begin{equation}
  \label{eq:BBB014}
  Z_{\Phi,k}=\ol{Z}\left(\frac{k}{\La}\right)^{-\eta}\; .
\end{equation}
Even though the renormalized field $\chi_R=Z_{\Phi,k}^{1/2}\chi$ acquires a
non--zero expectation value $\VEV{\chi_R}=\sqrt{2\kappa}$, for $k\ra0$ the
unrenormalized order parameter vanishes due to the divergence of $Z_{\Phi,k}$,
\begin{equation}
  \label{eq:BBB015}
  \VEV{\chi(k)}=\sqrt{\frac{2\kappa}{\ol{Z}}}
  \left(\frac{k}{\La}\right)^{\frac{1}{4\pi\kappa}}\; .
\end{equation}
Also the inverse Goldstone boson propagator behaves as
$(q^2)^{1-1/(8\pi\kappa)}$ and circumvents Coleman's no--go
theorem$^{\citen{Col73-1}}$ for free massless scalar fields in two dimensions.
It is remarkable that all these features arise from the solution of a simple
one--loop type equation without ever invoking nonperturbative vortex
configurations. This enhances our confidence that the nonperturbative aspects
are indeed already caught by relatively simple truncations of the effective
average action.

Having found all important qualitative features of the phase transition
already in a very rough truncation the way is open for systematic quantitative
improvements. We report here only the results of a recent investigation
regarding the critical behavior in the universality class of the
three--dimensional Ising model in next to leading order of a systematic
derivative expansion.$^{\citen{SW98-1}}$ This truncation is given by
eq.~(\ref{DerExp}) with $Y_{\Phi,k}=0$ and used for a numerical solution of
the flow equation. One expects the flow of the potential and the exponent
$\nu$ to be described quite accurately in this approximation: Only the
momentum dependence of the wave function renormalization is neglected which is
governed by the small anomalous dimension. The uncertainties for $\eta$ are
more substantial. One finds $\nu=0.631$, $\eta=0.047$ to be compared with the
averages of several other methods $\nu=0.630$, $\eta=0.035$ which are
presumably more precise.  Results of a similar quality are found for the
universal amplitude ratios and the Widom scaling form of the critical equation
of state.

\section{Chiral symmetry breaking in QCD}
\label{ChiralSymmetryBreaking}

The strong interaction dynamics of quarks and gluons at short distances or
high energies is successfully described by quantum chromodynamics (QCD). One
of its most striking features is asymptotic freedom$^{\citen{GW73-1}}$ which
makes perturbative calculations reliable in the high energy regime. On the
other hand, at scales around a few hundred $\MeV$ confinement sets in. As a
consequence, the low--energy degrees of freedom in strong interaction physics
are mesons, baryons and glueballs rather than quarks and gluons. When
constructing effective models for these IR degrees of freedom one usually
relies on the symmetries of QCD as a guiding principle, since a direct
derivation of such models from QCD is still missing. The most important
symmetry of QCD, its local color $SU(3)$ invariance, is of not much help here,
since the IR spectrum appears to be color neutral.  When dealing with bound
states involving heavy quarks the so called ``heavy quark symmetry'' may be
invoked to obtain approximate symmetry relations between IR
observables.$^{\citen{Neu94-1}}$ We will rather focus here on the light scalar
and pseudoscalar meson spectrum and therefore consider QCD with only the light
quark flavors $u$, $d$ and $s$. To a good approximation the masses of these
three flavors can be considered as small in comparison with other typical
strong interaction scales.  One may therefore consider the chiral limit of QCD
(vanishing current quark masses) in which the classical QCD Lagrangian does
not couple left-- and right--handed quarks. It therefore exhibits a global
chiral invariance under $U_L(N)\times U_R(N)=SU_L(N)\times SU_R(N)\times
U_V(1)\times U_A(1)$ where $N$ denotes the number of massless quarks ($N=2$ or
$3$) which transform as
\begin{eqnarray}
  \ds{\psi_R\equiv\frac{1-\gm_5}{2}\psi} &\longrightarrow&
  \ds{\Uc_R \psi_R\; ;\;\;\;\Uc_R\in U_R(N)}\nnn
  \ds{\psi_L\equiv\frac{1+\gm_5}{2}\psi} &\longrightarrow& \ds{\Uc_L
  \psi_L\; ;\;\;\;\Uc_L\in U_L(N)}\; .
 \label{ChiralTransoformation}
\end{eqnarray}
Even for vanishing quark masses only the diagonal $SU_V(N)$ vector--like
subgroup can be observed in the hadron spectrum (``eightfold
way''). The symmetry $SU_L(N)\times SU_R(N)$ must therefore be
spontaneously broken to $SU_V(N)$
\begin{equation}
  SU_L(N)\times SU_R(N)\longrightarrow
  SU_{L+R}(N)\equiv SU_V(N)\; .
  \label{CSBPattern}
\end{equation}
Chiral symmetry breaking is one of the most prominent features of strong
interaction dynamics and phenomenologically well
established,$^{\citen{Leu95-1}}$ though a rigorous derivation of this
phenomenon starting from first principles is still missing. In particular, the
chiral symmetry breaking~(\ref{CSBPattern}) predicts for $N=3$ the existence
of eight light parity--odd (pseudo--)Goldstone bosons: $\pi^0$, $\pi^\pm$,
$K^0$, $\ol{K}^0$, $K^\pm$ and $\eta$.  Their comparably small masses are a
consequence of the explicit chiral symmetry breaking due to small but
non--vanishing current quark masses.  The axial Abelian subgroup
$U_A(1)=U_{L-R}(1)$ is broken in the quantum theory by an anomaly of the
axial--vector current. This breaking proceeds without the occurrence of a
Goldstone boson.$^{\citen{Hoo86-1}}$ Finally, the $U_V(1)=U_{L+R}(1)$ subgroup
corresponds to baryon number conservation.

The light pseudoscalar and scalar mesons are thought of as color neutral
quark--antiquark bound states $\Phi^{ab}\sim \ol{\psi}_L^b \psi_R^a$,
$a,b=1,\ldots,N$, which therefore transform under chiral
rotations~(\ref{ChiralTransoformation}) as
\begin{equation} 
  \Phi\longrightarrow
  \Uc_R\Phi\Uc_L^\dagger\; .  
\end{equation} 
Hence, the chiral symmetry breaking pattern~(\ref{CSBPattern}) is realized if
the meson potential develops a VEV
\begin{equation}
  \label{AAA50}
  \VEV{\Phi^{ab}}=\ol{\si}_0\dt^{ab}\;
  ;\;\;\; \ol{\si}_0\neq0\; .  
\end{equation} 
One of the most crucial and yet unsolved problems of strong
interaction dynamics is to derive an effective field theory for the
mesonic degrees of freedom directly from QCD which exhibits this
behavior.

\section{A semi--quantitative picture}
\label{ASemiQuantitativePicture}

Before turning to a quantitative description of chiral symmetry
breaking using flow equations it is perhaps helpful to give a brief
overview of the relevant scales which appear in relation to this
phenomenon and the physical degrees of freedom associated to them.
Some of this will be explained in more detail in the remainder of
these lectures whereas other parts are rather well established
features of strong interaction physics.

At scales above approximately $1.5\GeV$, the relevant degrees of
freedom of strong interactions are quarks and gluons and their
dynamics appears to be well described by perturbative QCD. At somewhat
lower energies this changes dramatically. Quark and gluon bound states
form and confinement sets in. Concentrating on the physics of scalar
and pseudoscalar mesons~\footnote{One may assume that all other bound
  states are integrated out. We will comment on this issue below.}
there are three important momentum scales which appear to be rather
well separated:
\begin{itemize}
\item The compositeness scale $k_\Phi$ at which mesonic
  $\ol{\psi}\psi$ bound states form because of the increasing strength
  of the strong interaction. It will turn out to be somewhere in the
  range $(600-700)\MeV$.
\item The chiral symmetry breaking scale $k_{\chi SB}$ at which the chiral
  condensate $\VEV{\ol{\psi}^b\psi^a}$ or $\VEV{\Phi^{ab}}$ assumes a
  non--vanishing value, therefore breaking chiral symmetry according
  to~(\ref{CSBPattern}).  This scale is found to be around $(400-500)\MeV$.
  For $k$ below $k_{\chi SB}$ the quarks acquire constituent masses
  $M_q\simeq350\MeV$ due to their Yukawa coupling to the chiral
  condensate~(\ref{AAA50}).
\item The confinement scale $\Lambda_{\rm QCD}\simeq200\MeV$ which
  corresponds to the Landau pole in the perturbative evolution of the
  strong coupling constant $\alpha_s$. In our context, this is the
  scale where possible deviations of the effective quark propagator
  from its classical form and multi--quark interactions not included
  in the meson physics may become very important.
\end{itemize}
For scales $k$ in the range $k_{\chi SB}\lta k\lta k_\Phi$ the most
relevant degrees of freedom are mesons and quarks. Typically, the
dynamics in this range is dominated by the strong Yukawa coupling $h$
between quarks and mesons: $h^2/(4\pi)\gg\alpha_s$.  One may therefore
assume that the dominant QCD effects are included in the meson physics
and consider a simple model of quarks and mesons only.  As one evolves
to scales below $k_{\chi SB}$ the Yukawa coupling decreases whereas
$\alpha_s$ increases. Of course, getting closer to $\Lambda_{\rm QCD}$
it is no longer justified to neglect the QCD effects which go beyond
the dynamics of effective meson degrees of freedom. On the other hand,
the final IR value of the Yukawa coupling $h$ is fixed by the typical
values of constituent quark masses $M_q\simeq350\MeV$ to be
$h^2/(4\pi)\simeq4.5$. One may therefore speculate that the domination
of the Yukawa interaction persists down to scales $k\simeq M_q$ at
which the quarks decouple from the evolution of the mesonic degrees of
freedom altogether due to their mass. Of course, details of the
gluonic interactions are expected to be crucial for an understanding
of quark and gluon confinement. Strong interaction effects may
dramatically change the momentum dependence of the quark $n$--point
functions for $k$ around $\Lambda_{\rm QCD}$.  Yet, as long as one is
only interested in the dynamics of the mesons one is led to expect
that these effects are quantitatively no too important. Because of the
effective decoupling of the quarks and therefore the whole colored
sector the details of confinement have only little influence on the
mesonic flow equations for $k\lta\Lambda_{\rm QCD}$. We conclude that
there are good prospects that the meson physics can be described by an
effective action for mesons and quarks for $k<k_\Phi$. The main part
of the work presented here is concerned with this effective quark
meson model.

In order to obtain this effective action at the compositeness scale
$k_\Phi$ from short distance QCD two steps have to be carried out. In
a first step one computes at the scale $k_p\simeq1.5\GeV$ an effective
action involving only quarks. This step integrates out the gluon
degrees of freedom in a ``quenched approximation''. More precisely,
one solves a truncated flow equation for QCD with quark and gluon
degrees of freedom in presence of an effective infrared cutoff
$k_p\simeq1.5\GeV$ in the quark propagators. The exact flow equation
to be used for this purpose is obtained by lowering the infrared
cutoff $R_k$ for the gluons to zero while keeping the one for the
quarks fixed. Subsequently, the gluons are eliminated by solving the
field equations for the gluon fields as functionals of the quarks.
This will result in a non--trivial momentum dependence of the quark
propagator and effective non--local four and higher quark
interactions. Because of the infrared cutoff $k_p$ the resulting
effective action for the quarks resembles closely the one for heavy
quarks (at least for Euclidean momenta). The dominant effect is the
appearance of an effective quark potential (similar to the one for the
charm quark) which describes the effective four--quark interactions.
For the effective quark action at $k_p$ we only retain this
four--quark interaction in addition to the two--point
function, while neglecting $n$--point functions involving six and more
quarks. 

\begin{figure}[htb]
  \unitlength1.0cm
  \begin{picture}(10.,8.)
    \put(-0.7,7.0){\bf $\ds{q^2V(q^2)}$}
    \put(5.2,-0.2){\bf $\sqrt{q^2}/\GeV$}
    \put (4.0,7.0) {\footnotesize\rm 2-loop}
    \put (1.25,2.8) {\footnotesize\rm Richardson}
    \put (1.7,3.05) {\vector(1,1){1.05}}
    \put (8.25,2.15) {\footnotesize\rm Richardson}
    \put (8.2,2.1) {\vector(-1,-1){1.0}}
    \put(0.0,0.0){
      \epsfysize=8.cm
      \epsffile{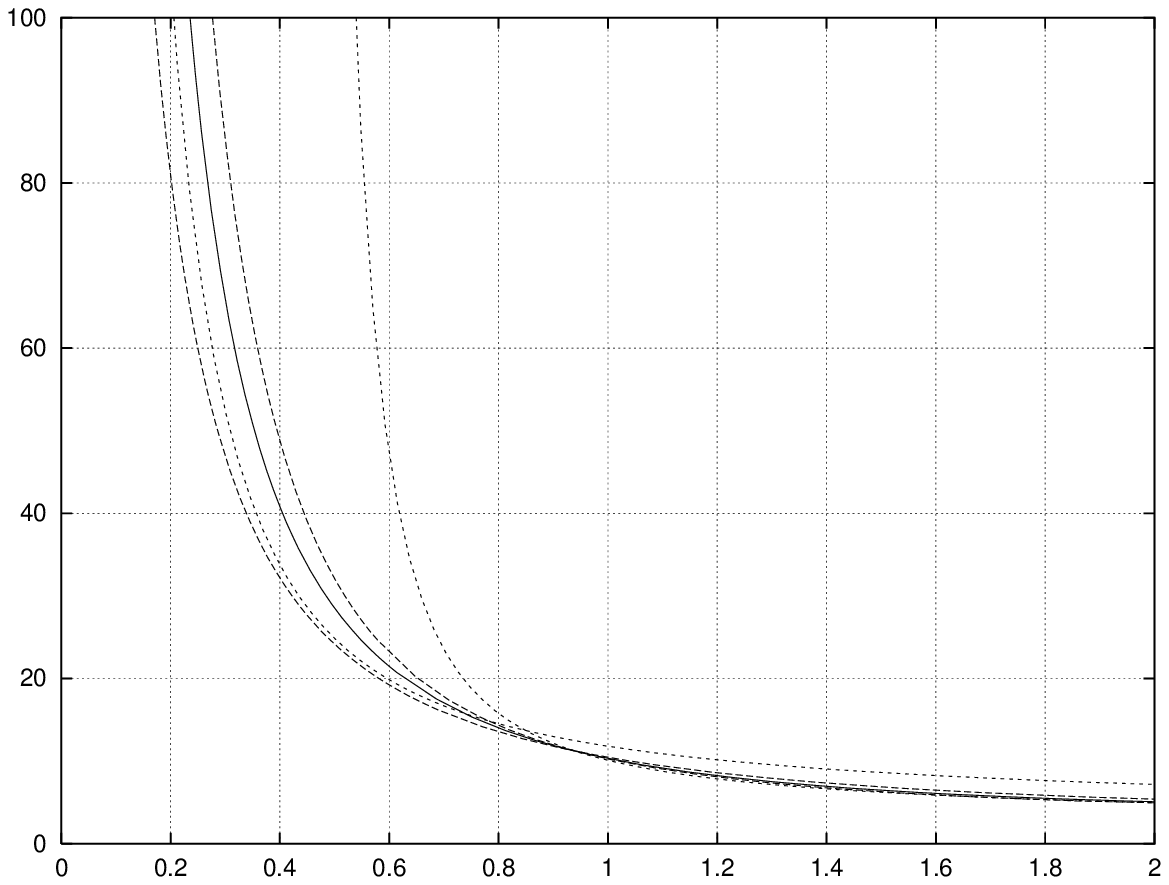}
      }
  \end{picture}
\caption{Heavy
  quark potential for different values of $\alpha_\star$.  The solid
  line corresponds to $\alpha_\star = 1.5$ whereas the dashed lines
  are for $\alpha_\star = 1$ (lower dashed line) and $\alpha_\star =
  2$ (upper dashed line).  Also shown are the two loop potential and
  the Richardson fit as dotted lines.}
\label{HQP} 
\end{figure}

For typical momenta larger than $\Lambda_{\rm QCD}$ a reliable computation of
the effective quark action should be possible by using in the quark--gluon
flow equation a relatively simple truncation. The result$^{\citen{BBW97-1}}$
for the Fourier transform of the potential $V(q^2)$ is shown in
fig.~\ref{HQP}.
(See Ref.$^{\citen{BBW97-1}}$ for the relation between the four--quark
interaction and the heavy quark potential which involves a rescaling in
dependence on $k_p$.)  A measure for the truncation uncertainties is the
parameter $\alpha_*$ which corresponds to the value to which an appropriately
defined running strong coupling $\alpha_s$ evolves for $k\ra0$. We see that
these uncertainties affect principally the low--$q^2$ region. For high values
of $q^2$ the potential is very close to the perturbative two--loop potential
whereas for intermediate $q^2$ relevant for quarkonium spectra it is quite
close to phenomenologically acceptable potentials (e.g.~the Richardson
potential.$^{\citen{Ric79-1}}$) The inverse quark propagator is found in this
computation to remain very well approximated by the simple classical momentum
dependence $\slash{q}$.

In the second step one has to lower the infrared cutoff in the effective
non--local quark model in order to extrapolate from $k_p$ to $k_\Phi$.  This
task can be carried out by means of the exact flow equation for quarks only,
starting at $k_p$ with an initial value $\Gamma_{k_p}[\psi]$ as obtained after
integrating out the gluons. For fermions the trace in~(\ref{ERGE}) has to be
replaced by a supertrace in order to account for the minus sign related to
Grassmann variables.$^{\citen{Wet90-1}}$ A first investigation in this
direction$^{\citen{EW94-1}}$ has used a truncation with a chirally invariant
four quark interaction whose most general momentum dependence was retained
\begin{eqnarray}
 \ds{\Gm_k} &=& \ds{\int\frac{d^4 p}{(2\pi)^4}
 \ol{\psi}_a^i(p)Z_{\psi,k}(p)\left[
 \slash{p}\delta^{ab}+m^{ab}(p)\gamma_5+
 i\tilde{m}^{ab}(p)\right]\psi_{ib}(p)}\nnn 
  &+& \ds{
 \frac{1}{2}\int\left(\prod_{l=1}^4
 \frac{d^4 p_l}{(2\pi)^4}\right)
 \left(2\pi\right)^4\delta(p_1+p_2-p_3-p_4)}\nnn
 &\times& \ds{
 \la_k^{(\psi)}(p_1,p_2,p_3,p_4)
 \left\{
 \left[\ol{\psi}_a^i(-p_1)\psi_i^b(p_2)\right]
 \left[\ol{\psi}_b^j(p_4)\psi_j^a(-p_3)\right]
 \right. }\nnn
 && \ds{\left.\hspace{2.95cm}
 -\left[\ol{\psi}_a^i(-p_1)\gamma_5\psi_i^b(p_2)\right]
 \left[\ol{\psi}_b^j(p_4)\gamma_5\psi_j^a(-p_3)\right]
 \right\} }\; .
 \label{QCDFourFermi}
\end{eqnarray}
Here $i,j$ run from one to $N_c$ which is the number of quark colors.  The
indices $a,b$ label the different light quark flavors and run from $1$ to $N$.
The matrices $m$ and $\tilde{m}$ are hermitian and $m+i\tilde{m}\gamma_5$
forms therefore the most general quark mass matrix. (Our chiral
conventions$^{\citen{Wet90-1}}$ where the hermitian part of the mass matrix is
multiplied by $\gamma_5$ may be somewhat unusual but they are quite convenient
for Euclidean calculations.) The ansatz~(\ref{QCDFourFermi}) does not
correspond to the most general chirally invariant four--quark interaction. It
neglects similar interactions in the $\rho$--meson and pomeron channels which
are also obtained from a Fierz transformation of the heavy quark
potential.$^{\citen{Wet95-2}}$ With $V(q^2)$ the heavy quark potential in a
Fourier representation, the initial value at $k_p=1.5\GeV$ was taken as
($\hat{Z}_{\psi,k}=Z_{\psi,k}(p^2=-k_p^2)$)
\begin{equation}
  \label{FFCBC}
  \la_{k_p}^{(\psi)}(p_1,p_2,p_3,p_4)
  \hat{Z}_{\psi,k_p}^{-2}=
  \frac{1}{2}V((p_1-p_3)^2)=
  \frac{2\pi\alpha_s}{(p_1-p_3)^2}+
  \frac{8\pi\la}{\left((p_1-p_3)^2\right)^2}\; .
\end{equation}
This corresponds to an approximation by a one gluon exchange term
$\sim\alpha_s(k_p)$ and a string tension $\la\simeq0.18\GeV^2$ and is in
reasonable agreement with the form computed recently$^{\citen{BBW97-1}}$ from
the solution of flow equations (see fig\ref{HQP}).  In the simplified
ansatz~(\ref{FFCBC}) the string tension introduces a second scale in addition
to $k_p$ and it becomes clear that the incorporation of gluon fluctuations is
a crucial ingredient for the emergence of mesonic bound states. For a more
precise treatment$^{\citen{BBW97-1}}$ of the four--quark interaction at the
scale $k_\Phi$ this second scale is set by the running of $\alpha_s$ or
$\Lambda_{\rm QCD}$.

The evolution equation for the function $\la_k^{(\psi)}$ for $k<k_p$ can be
derived from the fermionic version of~(\ref{ERGE}) and the
truncation~(\ref{QCDFourFermi}). Since $\la_k^{(\psi)}$ depends on six
independent momentum invariants it is a partial differential equation for a
function depending on seven variables and has to be solved
numerically.$^{\citen{EW94-1}}$ The ``initial value''~(\ref{FFCBC})
corresponds to the $t$--channel exchange of a ``dressed'' colored gluonic
state and it is by far not clear that the evolution of $\la_k^{(\psi)}$ will
lead at lower scales to a momentum dependence representing the exchange of
colorless mesonic bound states. Yet, at the compositeness scale
\begin{equation}
\label{kphi}
  k_\Phi\simeq630\MeV
\end{equation}
one finds an approximate factorization
\begin{equation}
\label{BSFact}
  \la_{k_\Phi}^{(\psi)}(p_1,p_2,p_3,p_4)=
  g(p_1,p_2)\tilde{G}(s)g(p_3,p_4)+\ldots
\end{equation}
which indicates the formation of mesonic bound states.  Here $g(p_1,p_2)$
denotes the amputated Bethe--Salpeter wave function and $\tilde{G}(s)$ is the
mesonic bound state propagator displaying a pole--like structure in the
$s$--channel if it is continued to negative $s=(p_1+p_2)^2$. The dots indicate
the part of $\la_k^{(\psi)}$ which does not factorize and which will be
neglected in the following. In the limit where the momentum dependence of $g$
and $\tilde{G}$ is neglected we recover the four--quark interaction of the
Nambu--Jona-Lasinio model.$^{\citen{NJL61-1,Bij95-1}}$ It is therefore not
surprising that our description of the dynamics for $k<k_\Phi$ will parallel
certain aspects of other investigations of this model, even though we are not
bound to the approximations used typically in such studies (large--$N_c$
expansion, perturbative renormalization group, etc.).

It is clear that for scales $k\lta k_\Phi$ a description of strong interaction
physics in terms of quark fields alone would be rather inefficient. Finding
physically reasonable truncations of the effective average action should be
much easier once composite fields for the mesons are introduced.  The exact
renormalization group equation can indeed be supplemented by an exact
formalism for the introduction of composite field variables or, more
generally, a change of variables.$^{\citen{EW94-1}}$ For our purpose, this
amounts in practice to inserting at the scale $k_\Phi$ the identities
\begin{eqnarray}
  1 &\sim& \ds{
    \int{\cal D}\si_A
    \exp\Bigg\{ -\tr
    \left(\si_A^\dagger -
      K_A^\dagger \tilde{G}-m_A^\dagger-
      {\cal O}^\dagger \tilde{G}\right)
    \frac{1}{2\tilde{G}}
    }\nnn
  && \ds{\hspace{2.6cm}\times
    \left(\si_A -\tilde{G}K_A -m_A-
      \tilde{G}{\cal O} \right)\Bigg\} 
    }\nnn
  \label{identity}
  1 &\sim& \ds{
    \int{\cal D}\si_H
    \exp\Bigg\{-\tr
    \left(\si_H^\dagger -
      K_H^\dagger \tilde{G}-m_H^\dagger-
      {\cal O}^{(5)\dagger} \tilde{G}\right)
    \frac{1}{2\tilde{G}}
    }\nnn
  && \ds{\hspace{2.6cm}\times
    \left(\si_H -\tilde{G}K_H -m_H-
      \tilde{G}{\cal O}^{(5)} \right)\Bigg\} 
    }\nonumber
\end{eqnarray}
into the functional integral which formally defines the quark
effective average action. Here we have used the shorthand notation
$A^\dagger G B\equiv\int\frac{d^d q}{(2\pi)^d}A_a^*(q)G^{ab}(q)
B_b(q)$, and $K_{A,H}$ are sources for the collective fields
$\si_{A,H}$ which correspond in turn to the anti-hermitian and
hermitian parts of the meson field $\Phi$. They are associated to the
fermion bilinear operators ${\cal O} [\psi]$, ${\cal O}^{(5)}[\psi ]$
whose Fourier components read
\begin{eqnarray}
 \ds{{\cal O}_{\;\; b}^a (q)} &=& \ds{ 
 -i\int\frac{d^4 p}{(2\pi)^4} g(-p,p+q)
 \ol{\psi}^a (p)\psi_b (p+q) }\nnn
 \ds{{\cal O}_{\;\;\;\;\;\; b}^{(5)a} (q)} &=& \ds{ 
 -\int\frac{d^4 p}{(2\pi)^4} g(-p,p+q)
 \ol{\psi}^a (p)\gamma_5\psi_b (p+q) }\; .
\end{eqnarray}
The choice of $g(-p,p+q)$ as the bound state wave function renormalization and
of $\tilde{G}(q)$ as its propagator guarantees that the four--quark
interaction contained in~(\ref{identity}) cancels the dominant factorizing
part of the QCD--induced non--local four--quark interaction
Eqs.(\ref{QCDFourFermi}), (\ref{BSFact}). In addition, one may choose
\begin{eqnarray}
 \ds{m_{Hab}^T} &=& \ds{
 m_{ab}(0)g^{-1}(0,0)Z_{\psi,k_\Phi}(0)}\nnn
 \ds{m_{Aab}^T} &=& \ds{
 \tilde{m}_{ab}(0)g^{-1}(0,0)Z_{\psi,k_\Phi}(0)}
\end{eqnarray}
such that the explicit quark mass term cancels out for $q=0$. The
remaining quark bilinear is $\sim m(q)-m(0)Z_{\psi,k_\Phi}(0)g(-q,q)/
[Z_{\psi,k_\Phi}(q)g(0,0)]$. It vanishes for zero momentum and will be
neglected in the following. Without loss of generality we can take $m$
real and diagonal and $\tilde{m}=0$. 

In consequence, we have replaced at the scale $k_\Phi$ the effective quark
action~(\ref{QCDFourFermi}) with~(\ref{BSFact}) by an effective quark meson
action given by
\begin{eqnarray}
 \ds{\hat{\Gamma}_k} &=& \ds{
 \Gamma_k-\frac{1}{2}\int d^4 x\tr
 \left(\Phi^\dagger\jmath+\jmath^\dagger\Phi\right)}\nnn
 \ds{\Gamma_{k}} &=& \displaystyle{ \int d^4 x
 U_k(\Phi,\Phi^\dagger)
  \label{EffActAnsatz}
 }\\[2mm]
 &+& \displaystyle{ 
 \int\frac{d^4 q}{(2\pi)^d}\Bigg\{
 Z_{\Phi,k}(q) q^2 \tr\left[
 \Phi^\dagger (q)\Phi (q)\right] +
 Z_{\psi,k}(q)\ol{\psi}_a(q)
 \gamma^\mu q_\mu \psi^a (q)
 }\nonumber\vspace{.2cm}\\
 &+& \displaystyle{
 \int\frac{d^4 p}{(2\pi)^d}\ol{h} _k (-q,q-p)}\nnn
 &\times& \ds{
 \ol{\psi}^a(q) \left(
 \frac{1+\gamma_5}{2}\Phi _{ab}(p)-
 \frac{1-\gamma_5}{2}\Phi_{ab}^\dagger (-p) \right)
 \psi^b (q-p) \Bigg\}\nonumber \; .}
\end{eqnarray}
At the scale $k_\Phi$ the inverse scalar propagator is related to
$\tilde{G}(q)$ in~(\ref{BSFact}) by
\begin{equation}
 \tilde{G}^{-1}(q^2) = 2\ol{m}^2_{k_\Phi} +
 2Z_{\Phi,k_\Phi}(q)q^2\; .
\end{equation}
This fixes the term in $U_{k_\Phi}$ which is quadratic in $\Phi$ to be
positive, $U_{k_\Phi}=\ol{m}^2_{k_\Phi}\tr\Phi^\dagger\Phi+\ldots$.  The
higher order terms in $U_{k_\Phi}$ cannot be determined in the
approximation~(\ref{QCDFourFermi}) since they correspond to terms involving
six or more quark fields.  The initial value of the Yukawa coupling
corresponds to the ``quark wave function in the meson'' in~(\ref{BSFact}),
i.e.
\begin{equation}
 \ol{h} _{k_\Phi}(-q,q-p) = g(-q,q-p)
\end{equation}
which can be normalized with $\ol{h}_{k_\Phi}(0,0)=g(0,0)=1$.
We observe that the explicit chiral symmetry breaking from
non--vanishing current quark masses appears now in the form of a meson
source term with
\begin{equation}
  \label{AAA22}
  \jmath=2\ol{m}^2_{k_\Phi} Z_{\psi,k_\Phi}(0)
  g^{-1}(0,0)\left(
  m_{ab}+i\tilde{m}_{ab}\right)=
  2Z_{\psi,k_\Phi}\ol{m}^2_{k_\Phi}
  {\rm diag}(m_u,m_d,m_s)\; .
\end{equation}
This induces a non--vanishing $\VEV{\Phi}$ and an effective quark mass $M_q$
through the Yukawa coupling. We note that the current quark mass $m_q$ and the
constituent quark mass $M_q\sim\ol{h}_k\VEV{\Phi}$ are identical at the scale
$k_\Phi$. (By solving the field equation for $\Phi$ as a functional of
$\ol{\psi}$, $\psi$ (with $U_k=\ol{m}^2_k\tr\Phi^\dagger\Phi$) one recovers
from~(\ref{EffActAnsatz}) the effective quark action~(\ref{QCDFourFermi}).
For a generalization beyond the approximation of a four--quark interaction or
a quadratic potential see Ref.$^{\citen{BJW97-1}}$.)  Spontaneous chiral
symmetry breaking can be described in this language by a non--vanishing
$\VEV{\Phi}$ in the limit $\jmath\ra0$. Because of spontaneous chiral symmetry
breaking the constituent quark mass $M_q$ can differ from zero even for
$m_q=0$. It is a nice feature of our formalism that it provides for a unified
description of the concepts of the current and the constituent quark masses.
As long as the effective average potential $U_k$ has a unique minimum at
$\Phi=0$ there is simply no difference between the two. The running of the
current quark mass in the pure quark model should be equivalent in the quark
meson language to the running of $\ol{h}_k\VEV{\Phi}(k)$.  (A verification of
this property would actually provide a good check for the truncation errors.)
Nevertheless, the formalism is now adapted to account for the quark mass
contribution from chiral symmetry breaking since the absolute minimum of $U_k$
may be far from the perturbative one for small $k$. We also note that
$\Gamma_k$ in~(\ref{EffActAnsatz}) is chirally invariant.  The explicit chiral
symmetry breaking in $\hat{\Gamma}_k$ appears only in the form of a linear
source term which is independent of $k$ and does not affect the flow of
$\Gamma_k$.  The flow equation for $\Gamma_k$ therefore respects chiral
symmetry even in the presence of quark masses. This leads to a considerable
simplification.

At the scale $k_\Phi$ the propagator $\tilde{G}$ and the wave function
$g(-q,q-p)$ should be optimized for a most complete elimination of terms
quartic in the quark fields. In the present context we will, however, neglect
the momentum dependence of $Z_{\psi,k}$, $Z_{\Phi,k}$ and $\ol{h}_k$.  The
mass $\ol{m}_{k_\Phi}$ was found in$^{\citen{EW94-1}}$ for the simple
truncation~(\ref{QCDFourFermi}) with $Z_{\psi,k_\Phi}=1$, $m=\tilde{m}=0$ to
be $\ol{m}_{k_\Phi}\simeq120\MeV$.  In view of the possible large truncation
errors we will take this only as an order of magnitude estimate. Below we will
consider the range $\ol{m}_{k_\Phi}=(45-120)\MeV$ for which chiral symmetry
breaking can be obtained in a two flavor model.  Furthermore, we will assume,
as usually done in large--$N_c$ computations within the NJL--model, that
$Z_{\Phi,k_\Phi}\equiv Z_{\Phi,k_\Phi}(q=0)\ll1$.  The quark wave function
renormalization $Z_{\psi,k}\equiv Z_{\psi,k}(q=0)$ is set to one at the scale
$k_\Phi$ for convenience.  For $k<k_\Phi$ we will therefore study an effective
action for quarks and mesons in the truncation
\begin{eqnarray}
  \ds{\Gm_k} &=& \ds{
    \int d^4x\Bigg\{
    Z_{\psi,k}\ol{\psi}_a i\slash{\prl}\psi^a+
    Z_{\Phi,k}\tr\left[\prl_\mu\Phi^\dagger\prl^\mu\Phi\right]+
    U_k(\Phi,\Phi^\dagger)
    }\nnn
  &+& \ds{
    \ol{h}_k\ol{\psi}^a\left(\frac{1+\gm_5}{2}\Phi_{ab}-
    \frac{1-\gm_5}{2}(\Phi^\dagger)_{ab}\right)\psi^b
    \Bigg\} }
  \label{GammaEffective}
\end{eqnarray}
with compositeness conditions
\begin{eqnarray}
 \ds{U_{k_\Phi}(\Phi,\Phi^\dagger)} &=& \ds{
 \ol{m}^2_{k_\Phi}\tr\Phi^\dagger\Phi-
 \frac{1}{2}\ol{\nu}_{k_\Phi}\left(
 \det\Phi+\det\Phi^\dagger\right)}\nnn
 &+& \ds{
 \frac{1}{2}\ol{\la}_{1,k_\Phi}\left(\tr\Phi^\dagger\Phi\right)^2+
 \frac{N-1}{4}\ol{\la}_{2,k_\Phi}\tr
 \left(\Phi^\dagger\Phi-\frac{1}{N}\tr\Phi^\dagger\Phi\right)^2+
 \ldots}\nnn
 \ds{\ol{m}^2_{k_\Phi}} &\equiv& \ds{\frac{1}{2\tilde{G}(0)}\simeq
 (45\MeV)^2-(120\MeV)^2}\nnn
 \ds{\ol{h}_{k_\Phi}} &=&
 Z_{\psi,k_\Phi}=1\nnn
 \ds{Z_{\Phi,k_\Phi}} &\ll& 1\; .
 \label{CompositenessConditions}
\end{eqnarray}
As a consequence, the initial value of the renormalized Yukawa coupling which
is given by
$h_{k_\Phi}=\ol{h}_{k_\Phi}Z_{\psi,k_\Phi}^{-1}Z_{\Phi,k_\Phi}^{-1/2}$ is
large!  Note that we have included in the potential an explicit $U_A(1)$
breaking term~\footnote{The anomaly term in the fermionic effective average
  action has been computed in.$^{\citen{Paw96-1}}$} $\sim\ol{\nu}_k$ which
mimics the effect of the chiral anomaly of QCD to leading order in an
expansion of the effective potential in powers of $\Phi$. Because of the
infrared stability of the evolution of $\Gamma_k$ which will be discussed
below the precise form of the potential, i.e.~the values of the quartic
couplings $\ol{\la}_{i,k_\Phi}$ and so on, will turn out to be unimportant.

We have refrained here for simplicity from considering four quark operators
with vector and pseudo--vector spin structure.  Their inclusion is
straightforward and would lead to vector and pseudo--vector mesons in the
effective action~(\ref{GammaEffective}).  We will concentrate first on two
flavors and consider only the two limiting cases $\ol{\nu}_k=0$ and
$\ol{\nu}_k\ra\infty$.  We also omit first the explicit quark masses and study
the chiral limit $\jmath=0$.  Because of the positive mass term
$\ol{m}^2_{k_\Phi}$ one has at the scale $k_\Phi$ a vanishing expectation
value $\VEV{\Phi}=0$ (for $\jmath=0$). There is no spontaneous chiral symmetry
breaking at the compositeness scale.  This means that the mesonic bound states
at $k_\Phi$ and somewhat below are not directly connected to chiral symmetry
breaking.

The question remains how chiral symmetry is broken. We will try to answer it
by following the evolution of the effective potential $U_k$ from $k_\Phi$ to
lower scales using the exact renormalization group method outlined above with
the compositeness conditions~(\ref{CompositenessConditions}) defining the
initial values.  In this context it is important that the formalism for
composite fields$^{\citen{EW94-1}}$ also induces an infrared cutoff in the
meson propagator. The flow equations are therefore exactly of the
form~(\ref{ERGE}) (except for the supertrace), with quarks and mesons treated
on an equal footing.  In fact, one would expect that the large renormalized
Yukawa coupling will rapidly drive the scalar mass term to negative values as
the IR cutoff $k$ is lowered.$^{\citen{Wet90-1}}$ This will then finally lead
to a potential minimum away from the origin at some scale $k_{\rm \chi
  SB}<k_\Phi$ such that $\VEV{\Phi}\neq0$. The ultimate goal of such a
procedure, besides from establishing the onset of chiral symmetry breaking,
would be to extract phenomenological quantities, like $f_\pi$ or meson masses,
which can be computed in a straightforward manner from $\Gamma_k$ in the IR
limit $k\ra0$.

At first sight, a reliable computation of $\Gamma_{k\ra0}$ seems a
very difficult task. Without a truncation $\Gamma_k$ is described by
an infinite number of parameters (couplings, wave function
renormalizations, etc.) as can be seen if $\Gamma_k$ is expanded in
powers of fields and derivatives. For instance, the pseudoscalar and
scalar meson masses are obtained as the poles of the exact propagator,
$\lim_{k\ra0}\Gamma_k^{(2)}(q)|_{\Phi=\VEV{\Phi}}$, which receives
formally contributions from terms in $\Gamma_k$ with arbitrarily high
powers of derivatives and the expectation value $\si_0$.  Realistic
nonperturbative truncations of $\Gamma_k$ which reduce the problem to
a manageable size are crucial.  We will argue in the following that
there may be a twofold solution to this problem:
\begin{itemize}
\item Due to an IR fixed point structure of the flow equations in the
  symmetric regime, i.e. for $k_{\chi SB}<k<k_\Phi$, the values of
  many parameters of $\Gamma_k$ for $k\ra0$ will be approximately
  independent of their initial values at the compositeness scale
  $k_{\Phi}$. For small enough $Z_{\Phi,k_\Phi}$ only a few relevant
  parameters ($\ol{m}^2_{k_\Phi}$, $\ol{\nu}_{k_\Phi}$) need to be
  computed accurately from QCD. They can alternatively be determined
  from phenomenology.
\item One can show that physical observables like meson masses, decay
  constants, etc., can be expanded in powers of the quark masses within the
  linear meson model.$^{\citen{JW96-1,JW96-3,JW97-1}}$ This is similar to the
  way it is usually done in chiral perturbation theory.$^{\citen{Leu95-1}}$ To
  a given finite order of this expansion only a finite number of terms of a
  simultaneous expansion of $\Gamma_k$ in powers of derivatives and $\Phi$ are
  required if the expansion point is chosen properly.
\end{itemize}
In combination, these two results open the possibility for a perhaps
unexpected degree of predictive power within the linear meson model.

We wish to stress, though, that a perturbative treatment of the model
at hand, e.g., using perturbative RG techniques, cannot be expected to
yield reliable results. The renormalized Yukawa coupling is expected
to be large at the scale $k_\Phi$. Even the IR value of $h$ is still
relatively big
\begin{equation} 
  \label{IRh}
  \lim_{k\ra0}h=\frac{2M_q}{f_\pi}\simeq7.5
\end{equation} 
and $h$ increases with increasing $k$. The dynamics of the linear
meson model is therefore clearly nonperturbative for all scales $k\leq
k_\Phi$.

\section{Flow equations for the linear quark meson model}

We will next turn to the ERGE analysis of the linear meson model which was
introduced in the last section\footnote{For a study of chiral symmetry
  breaking in QED using related exact renormalization group techniques see
  Ref.$^{\citen{AMST97-1}}$.}.  In a first approach we will attack the problem
at hand by truncating $\Gm_k$ in such a way that it contains all
perturbatively relevant and marginal operators, i.e.  those with canonical
dimensions. This has the advantage that the flow of a small number of
couplings permits quantitative insight into the relevant mechanisms, e.g., of
chiral symmetry breaking. More quantitative precision will be obtained once we
generalize our truncation (see below) for the $O(4)$--model by allowing for
the most general effective average potential $U_k$.  The effective potential
$U_k$ is a function of only four $\csN$ invariants for $N=3$:
\begin{eqnarray}
  \label{Invariants}
  \ds{\rho} &=&
  \ds{\tr\Phi^\dagger\Phi}\nnn \ds{\tau_2} &=& \ds{
  \frac{N}{N-1}\tr\left(\Phi^\dagger\Phi- \frac{1}{N}\rho\right)^2}\nnn
  \ds{\tau_3} &=& \ds{ \tr\left(\Phi^\dagger\Phi-
  \frac{1}{N}\rho\right)^3}\nnn \ds{\xi} &=&
  \ds{\det\Phi+\det\Phi^\dagger}\; .  
\end{eqnarray}
The invariant $\tau_3$ is only independent for $N\ge3$. For $N=2$ it can be
eliminated by a suitable combination of $\tau_2$ and $\rho$.  The additional
$U_A(1)$ breaking invariant $\omega=i(\det\Phi-\det\Phi^\dagger)$ is $\Cc\Pc$
violating and may therefore appear only quadratically in $U_k$. It is
straightforward to see that $\omega^2$ is expressible in terms of the
invariants~(\ref{Invariants}).  We may expand $U_k$ as a function of these
invariants around its minimum, i.e. $\rho=\rho_0\equiv N\ol{\si}_0^2$,
$\xi=\xi_0=2\ol{\si}_0^N$ and $\tau_2=\tau_3=0$ where
\begin{equation}
  \label{Sigma_0}
  \ol{\si}_0\equiv\frac{1}{N}\tr\VEV{\Phi}\; .
\end{equation}
The expansion coefficients are the $k$--dependent couplings of the
model. In our first version we only keep couplings of canonical
dimension $d_c\leq4$. This yields in the chirally symmetric regime,
i.e., for $k_{\rm\chi SB}\leq k\leq k_\Phi$ where $\ol{\si}_0=0$
\begin{equation}
  \label{SymmetricPotential}
  U_k=\ol{m}^2_k\rho+\hal\ol{\la}_{1,k}\rho^2+
  \frac{N-1}{4}\ol{\la}_{2,k}\tau_2-
  \frac{1}{2}\ol{\nu}_k\xi
\end{equation}
whereas in the SSB regime for $k\leq k_{\rm\chi SB}$ we have
\begin{equation}
 U_k=\hal\ol{\la}_{1,k}\left[\rho-N\ol{\si}_{0.k}^2\right]^2+
 \frac{N-1}{4}\ol{\la}_{2,k}\tau_2+
 \hal\ol{\nu}_k\left[\ol{\si}_{0,k}^{N-2}\rho-\xi\right]\, .
 \label{EffPotentialSSB}
\end{equation}

Before continuing to compute the nonperturbative beta functions for these
couplings it is worthwhile to pause here and emphasize that naively
(perturbatively) irrelevant operators can by no means always be neglected.
The most prominent example for this is QCD itself. It is the very assumption
of our treatment of chiral symmetry breaking (substantiated by the results
of$^{\citen{EW94-1}}$) that the momentum dependence of the coupling constants
of some six--dimensional quark operators $(\ol{\psi}\psi)^2$ develop poles in
the $s$--channel indicating the formation of mesonic bound states. On the
other hand, it is quite natural to assume that $\Phi^6$ or $\Phi^8$ operators
are not really necessary to understand the properties of the potential in a
neighborhood around its minimum.  Yet, truncating higher dimensional operators
does not imply the assumption that the corresponding coupling constants are
small. In fact, this could only be expected as long as the relevant and
marginal couplings are small as well. What is required, though, is that their
{\em influence} on the evolution of those couplings kept in the truncation,
for instance, the set of equations~(\ref{FlowEquationsSSB}) below, is small.
A comparison with the results of the later sections for the full potential
$U_k$ (i.e., including arbitrarily many couplings in the formal expansion)
will provide a good check for the validity of this assumption.  In this
context it is perhaps also interesting to note that the
truncation~(\ref{GammaEffective}) includes the known one--loop beta functions
of a small coupling expansion as well as the leading order result of the
large--$N_c$ expansion of the $U_L(N)\times U_R(N)$ model.$^{\citen{BHJ94-1}}$
This should provide at least some minimal control over this truncation, even
though we believe that our results are significantly more accurate.

Inserting the truncation Eqs.(\ref{GammaEffective}),
(\ref{SymmetricPotential}), (\ref{EffPotentialSSB}) into~(\ref{ERGE}) reduces
this functional differential equation for infinitely many variables to a
finite set of ordinary differential equations. This yields, in particular, the
beta functions for the couplings $\ol{\la}_{1,k}$, $\ol{\la}_{2,k}$,
$\ol{\nu}_k$ and $\ol{m}^2_k$ or $\ol{\si}_{0,k}$. Details of the calculation
can be found in Ref.$^{\citen{JW95-1}}$. We will refrain here from presenting
the full set of flow equations but rather illustrate the main results with a
few examples. Defining dimensionless renormalized VEV and coupling constants
\begin{eqnarray}
  \ds{\kappa} &=& \ds{Z_{\Phi,k}
    N\ol{\si}_{0,k}^2 k^{-2}}\nnn 
  \ds{h^2} &=& \ds{
    \ol{h}_k^2 Z_{\Phi,k}^{-1}Z_{\psi,k}^{-2}}\nnn 
  \ds{\la_i} &=& \ds{
    \ol{\la}_{i,k} Z_{\Phi,k}^{-2}\; ;\;\;\; i=1,2}\nnn 
  \ds{\nu} &=& \ds{
    \ol{\nu}_k Z_{\Phi,k}^{-\frac{N}{2}}k^{N-4}}
  \label{DimensionlessCouplings}
\end{eqnarray}
one finds, e.g., for the spontaneous symmetry breaking (SSB) regime
and $\ol{\nu}_k=0$ 
\begin{eqnarray}
  \ds{ \frac{d}{d t}\kappa } &=& \ds{
    -(2+\eta_\Phi )\kappa + \frac{1}{16\pi^2} \Bigg\{ N^2l_1^4(0;\eta_\Phi)
    +3l_1^4 (2\la_1 \kappa;\eta_\Phi) }\nnn 
  &+& \ds{ 
    (N^2-1)\left[
    1+\frac{\la_2}{\la_1}\right] l_1^4 (\la_2\kappa;\eta_\Phi)-4N_c
    \frac{h^2}{\la_1} 
    l_{1}^{(F)4} (\frac{1}{N}h^2 \kappa;\eta_\psi) \Bigg\} }\nnn
  \ds{\frac{d}{d t}\la_1 } &=& \ds{ 
    2\eta_\Phi \la_1
    +\frac{1}{16\pi^2} \Bigg\{ N^2\la_1^2 
    l_2^4(0;\eta_\Phi) +9\la_1^2 l_2^4
    (2\la_1\kappa;\eta_\Phi) }\nnn 
  &+& \ds{ 
    (N^2-1)\left[\la_1 +\la_2\right]^2
    l_2^4 (\la_2\kappa;\eta_\Phi)-
    4\frac{N_c}{N}h^4 l_{2}^{(F)4}
    (\frac{1}{N}h^2\kappa;\eta_\psi) 
    \Bigg\} \label{FlowEquationsSSB} }\\[2mm]
  \ds{\frac{d}{d t}\la_2 } &=& \ds{ 
    2\eta_\Phi \la_2
    +\frac{1}{16\pi^2} \Bigg\{ \frac{N^2}{4}\la_2^2 l_2^4(0;\eta_\Phi) +
    \frac{9}{4}(N^2-4)\la_2^2 l_2^4 (\la_2\kappa;\eta_\Phi) }\nnn 
  &-& \ds{ 
    \hal N^2 \la_2^2 l_{1,1}^4(0,\la_2\kappa;\eta_\Phi) + 
    3[\la_2+4\la_1]\la_2
    l_{1,1}^4(2\la_1\kappa,\la_2\kappa;\eta_\Phi) }\nnn 
  &-& \ds{ 8\frac{N_c}{N}h^4
    l_{2}^{(F)4} (\frac{1}{N}h^2\kappa;\eta_\psi)\Bigg\} }\nnn 
  \ds{\frac{d}{d t} h^2} &=& \ds{ 
    \left[ d-4+2\eta_\psi +\eta_\Phi\right] h^2 -
      \frac{4}{N}v_d h^4 \left\{
      N^2 l_{1,1}^{(FB)d} (\frac{1}{N}\kappa h^2,0;
      \eta_\psi,\eta_\Phi) \right. }\nonumber\vspace{.2cm}\\ 
  &-& \ds{ 
    (N^2-1)l_{1,1}^{(FB)d} (\frac{1}{N}\kappa
    h^2,\kappa\la_2; \eta_\psi,\eta_\Phi)}\nnn
  &-& \ds{\left.
    l_{1,1}^{(FB)d} (\frac{1}{N}\kappa h^2,2\kappa\la_1;
    \eta_\psi,\eta_\Phi) \right\}\nonumber } 
\end{eqnarray} 
Here $\eta_\Phi=-\frac{d}{d t}\ln Z_{\Phi,k}$, $\eta_\psi=-\frac{d}{d t}\ln
Z_{\psi,k}$ are the meson and quark anomalous dimensions,
respectively.$^{\citen{JW95-1}}$ The symbols $l_n^d$, $l_{n_1,n_2}^d$ and
$l_n^{(F)d}$ denote bosonic and fermionic mass threshold functions,
respectively, which are defined in Ref.$^{\citen{JW95-1}}$. They describe the
decoupling of massive modes and provide an important nonperturbative
ingredient. For instance, the bosonic threshold functions
\begin{equation}
 \label{AAA85n}
 l_n^d(w;\eta_\Phi)=\frac{n+\delta_{n,0}}{4}v_d^{-1}
 k^{2n-d}\int\frac{d^d q}{(2\pi)^d}
 \frac{1}{Z_{\Phi,k}}\frac{\prl R_k}{\prl t}
 \frac{1}{\left[ P(q^2)+k^2w\right]^{n+1}}
\end{equation}
involve the inverse average propagator
$P(q^2)=q^2+Z_{\Phi,k}^{-1}R_k(q^2)$ where the infrared cutoff is
manifest. These functions decrease $\sim w^{-(n+1)}$ for $w\gg1$.
Since typically $w=M^2/k^2$ with $M$ a mass of the model, the main
effect of the threshold functions is to cut off fluctuations of
particles with masses $M^2\gg k^2$. Once the scale $k$ is changed
below a certain mass threshold, the corresponding particle no longer
contributes to the evolution and decouples smoothly.

Within our truncation the beta functions~(\ref{FlowEquationsSSB}) for the
dimensionless couplings look almost the same as in one--loop perturbation
theory. There are, however, two major new ingredients which are crucial for
our approach: First, there is a new equation for the running of the mass term
in the symmetric regime or for the running of the potential minimum in the
regime with spontaneous symmetry breaking. This equation is related to the
quadratic divergence of the mass term in perturbation theory and does not
appear in the Callan--Symanzik$^{\citen{CS70-1}}$ or
Coleman--Weinberg$^{\citen{CW73-1}}$ treatment of the renormalization group.
Obviously, these equations are the key for a study of the onset of spontaneous
chiral symmetry breaking as $k$ is lowered from $k_\Phi$ to zero.  Second, the
most important nonperturbative ingredient in the flow equations for the
dimensionless Yukawa and scalar couplings is the appearance of effective mass
threshold functions like~(\ref{AAA85n}) which account for the decoupling of
modes with masses larger than $k$.  Their form is different for the symmetric
regime (massless fermions, massive scalars) or the regime with spontaneous
symmetry breaking (massive fermions, massless Goldstone bosons).  Without the
inclusion of the threshold effects the running of the couplings would never
stop and no sensible limit $k\rightarrow0$ could be obtained because of
unphysical infrared divergences. The threshold functions are not arbitrary but
have to be computed carefully. The mass terms appearing in these functions
involve the dimensionless couplings. Expanding the threshold functions in
powers of the mass terms (or the dimensionless couplings) makes their
non--perturbative content immediately visible.  It is these threshold
functions which will make it possible below to use one--loop type formulae for
the necessarily nonperturbative computation of the critical behavior of the
(effectively) three--dimensional $O(4)$--symmetric scalar model.

\section{The chiral anomaly and the  $O(4)$--model}
\label{ChiralAnomaly}

We have seen how the mass threshold functions in the flow equations
describe the decoupling of heavy modes from the evolution of
$\Gamma_k$ as the IR cutoff $k$ is lowered. In the chiral limit with
two massless quark flavors ($N=2$) the pions are the massless
Goldstone bosons.  Below, once we will consider the flow of the full
effective average potential without resorting to a quartic truncation,
we will also consider the more general case of non--vanishing current
quark masses. For the time being, the effect of the physical pion mass
of $m_\pi\simeq140\MeV$, or equivalently of the two small but
non--vanishing current quark masses, can easily be mimicked by
stopping the flow of $\Gamma_k$ at $k=m_\pi$ by hand.  This situation
changes significantly once the strange quark is included. Now the
$\eta$ and the four $K$ mesons appear as additional massless Goldstone
modes in the spectrum. They would artificially drive the running of
$\Gamma_k$ at scales $m_\pi\lta k\lta500\MeV$ where they should
already be decoupled because of their physical masses. It is therefore
advisable to focus on the two flavor case $N=2$ as long as the chiral
limit of vanishing current quark masses is considered.

It is straightforward to obtain an estimate of the (renormalized) coupling
$\nu$ which parameterizes the explicit $U_A(1)$ breaking due to the chiral
anomaly. From~(\ref{EffPotentialSSB}) we find
\begin{equation}
  \label{AAA01}
  m_{\eta^\prime}^2=\frac{N}{2}\nu\si_0^{N-2}\simeq1\GeV
\end{equation}
which translates for $N=2$ for $k\ra0$ into
\begin{equation}
  \label{AAA02}
  \nu(k\ra0)\simeq1\GeV\; .
\end{equation}
This suggests that $\nu\ra\infty$ can be considered as a realistic
limit.  An important simplification occurs for $N=2$ and
$\nu\ra\infty$, related to the fact that for $N=2$ the chiral group
$SU_L(2)\times SU_R(2)$ is (locally) isomorphic to $O(4)$. Thus, the
complex $(\bf 2,\bf 2)$ representation $\Phi$ of $SU_L(2)\times
SU_R(2)$ may be decomposed into two vector representations,
$(\si,\pi^k)$ and $(\eta^\prime,a^k)$ of $O(4)$:
\begin{equation}
 \Phi=\hal\left(\si-i\eta^\prime\right)+
 \hal\left( a^k+i\pi^k\right)\tau_k \; .
\end{equation}
For $\nu\ra\infty$ the masses of the $\eta^\prime$ and the $a^k$ diverge and
these particles decouple. We are then left with the original $O(4)$ symmetric
linear $\si$--model of Gell--Mann and Levy$^{\citen{GML60-1}}$ coupled to
quarks. The flow equations of this model have been derived
previously$^{\citen{Wet90-1,Wet91-1}}$ for the truncation of the effective
action used here. For mere comparison we also consider the opposite limit
$\nu\ra0$.  Here the $\eta^\prime$ meson becomes an additional Goldstone boson
in the chiral limit which suffers from the same problem as the $K$ and the
$\eta$ in the case $N=3$.  Hence, we may compare the results for two different
approximate limits of the effects of the chiral anomaly:
\begin{itemize}
\item the $O(4)$ model corresponding to $N=2$ and $\nu\ra\infty$
\item the $U_L(2)\times U_R(2)$ model corresponding to $N=2$ and
  $\nu=0$.
\end{itemize}
For the reasons given above we expect the first situation to be closer
to reality. In this case we may imagine that the fluctuations of the
kaons, $\eta$, $\eta^\prime$ and the scalar mesons (as well as vector
and pseudovector mesons) have been integrated out in order to obtain
the initial values of $\Gamma_{k_\Phi}$ --- in close analogy to the
integration of the gluons for the effective quark action
$\Gamma_{k_p}[\psi]$ discussed above. We will keep the initial values
of the couplings $\ol{m}^2_{k_\Phi}$, $\ol{\la}_{1,k_\Phi}$ and
$Z_{\Phi,k_\Phi}$ as free parameters. Our results should be
quantitatively accurate to the extent to which the local polynomial
truncation is a good approximation.

\section{Infrared stability}
\label{InfraredStability}

\begin{center}
  \begin{figure}[htb]
    \unitlength1.0cm
    \begin{picture}(10.,7.)
      \put(0.2,5.){\bf $\ds{\frac{m,\si_0}{\MeV}}$}
      \put(6.0,0.2){\bf $k/\MeV$}
      \put(6.5,2.5){\bf $\si_0$}
      \put(9.0,4.){\bf $m$}
      \put(-1.0,-9.5){
        \epsfysize=18.cm
        \epsffile{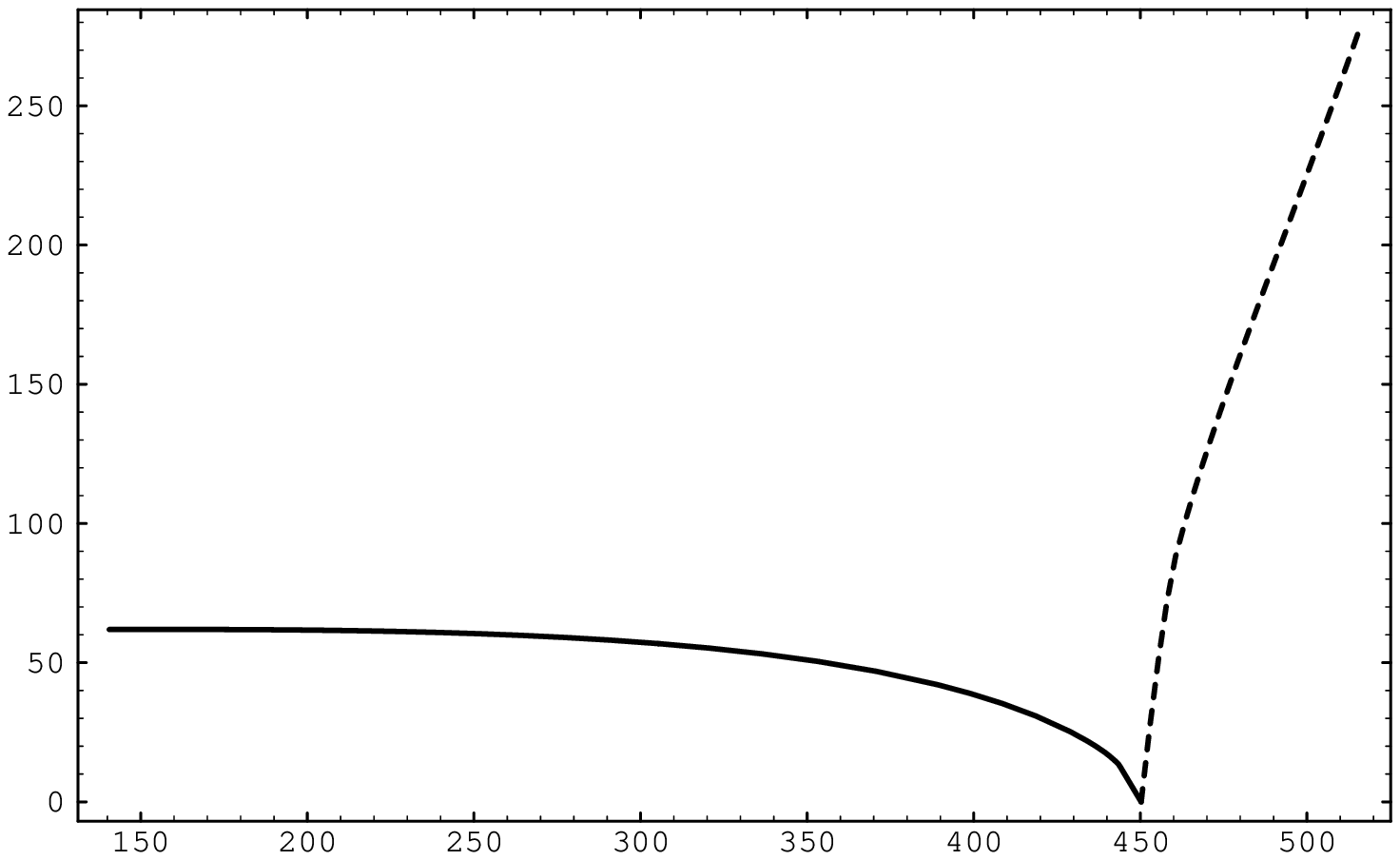}
        }
    \end{picture}
\caption{Evolution of the renormalized mass $m$ in the symmetric
  regime (dashed line) and the vacuum expectation value
  $\si_0=Z_{\Phi,k}^{1/2}\ol{\si}_{0,k}$ of the scalar field in the
  SSB regime (solid line) as functions of $k$ for the $U_L(2)\times
  U_R(2)$ model. Initial values at $k=k_\Phi$ are $\la_{i,I}=0$,
  $h^2_I=300$ and $\ol{m}_{k_\Phi}=63\MeV$.}
\label{Fig1} 
\end{figure}
\end{center}

\ref{FlowEquationsSSB} and the corresponding set of flow equations for the
symmetric regime constitute a coupled system of ordinary differential
equations which can be integrated numerically.  Similar equations can be
computed for the $O(4)$ model where $N=2$ and the coupling $\la_2$ is absent.
We have neglected in a first step the dependence of all threshold functions
appearing in the flow equations on the anomalous dimensions. This dependence
will be taken into account below once we abandon the quartic potential
approximation.  The most important result is that chiral symmetry breaking
indeed occurs for a wide range of initial values of the parameters including
the presumably realistic case of large renormalized Yukawa coupling and a bare
mass $\ol{m}_{k_\Phi}$ of order $100\MeV$. A typical evolution of the
renormalized meson mass $m$ with $k$ is plotted in fig.~\ref{Fig1}.
Driven by the strong Yukawa coupling, $m$ decreases rapidly and goes
through zero at a scale $k_{\chi{\rm SB}}$ not far below $k_\Phi$.
Here the system enters the SSB regime and a non--vanishing
(renormalized) VEV $\si_0$ for the meson field $\Phi$ develops. The
evolution of $\si_0$ with $k$ turns out to be reasonably stable
already before scales $k\simeq m_\pi$ where the evolution is stopped.
We take this result as an indication that our truncation of the
effective action $\Gm_k$ leads at least qualitatively to a
satisfactory description of chiral symmetry breaking. The reason for
the relative stability of the IR behavior of the VEV (and all other
couplings) is that the quarks acquire a constituent mass
$M_q=h\si_0\simeq350\MeV$ in the SSB regime. As a consequence they
decouple once $k$ becomes smaller than $M_q$ and the evolution is then
dominantly driven by the massless Goldstone bosons.  This is also
important in view of potential confinement effects expected to become
important for the quark $n$--point functions for $k$ around $\La_{\rm
  QCD}\simeq200\MeV$.  Since confinement is not explicitly included in
our truncation of $\Gamma_k$, one might be worried that such effects
could spoil our results completely. Yet, as discussed in some more
detail above, only the colored quarks should feel confinement and they
are no longer important for the evolution of the meson couplings for
$k$ around $200\MeV$.  One might therefore hope that a precise
treatment of confinement is not crucial for this approach to chiral
symmetry breaking.

\begin{center}
  \begin{figure}
    \unitlength1.0cm
    \begin{picture}(10.,8.)
      \put(0.5,5.5){\bf $\ds{\frac{f_\pi}{\MeV}}$}
      \put(6.0,0.2){\bf$\ds{\frac{\ol{m}_{k_\Phi}^2}{k_\Phi^2}}$}
      \put(8.0,4.5){\bf $O(4)$}
      \put(4.0,2.5){\bf $U_L(2)\times U_R(2)$}
      \put(-1.0,-9.5){
        \epsfysize=18.cm
        \epsffile{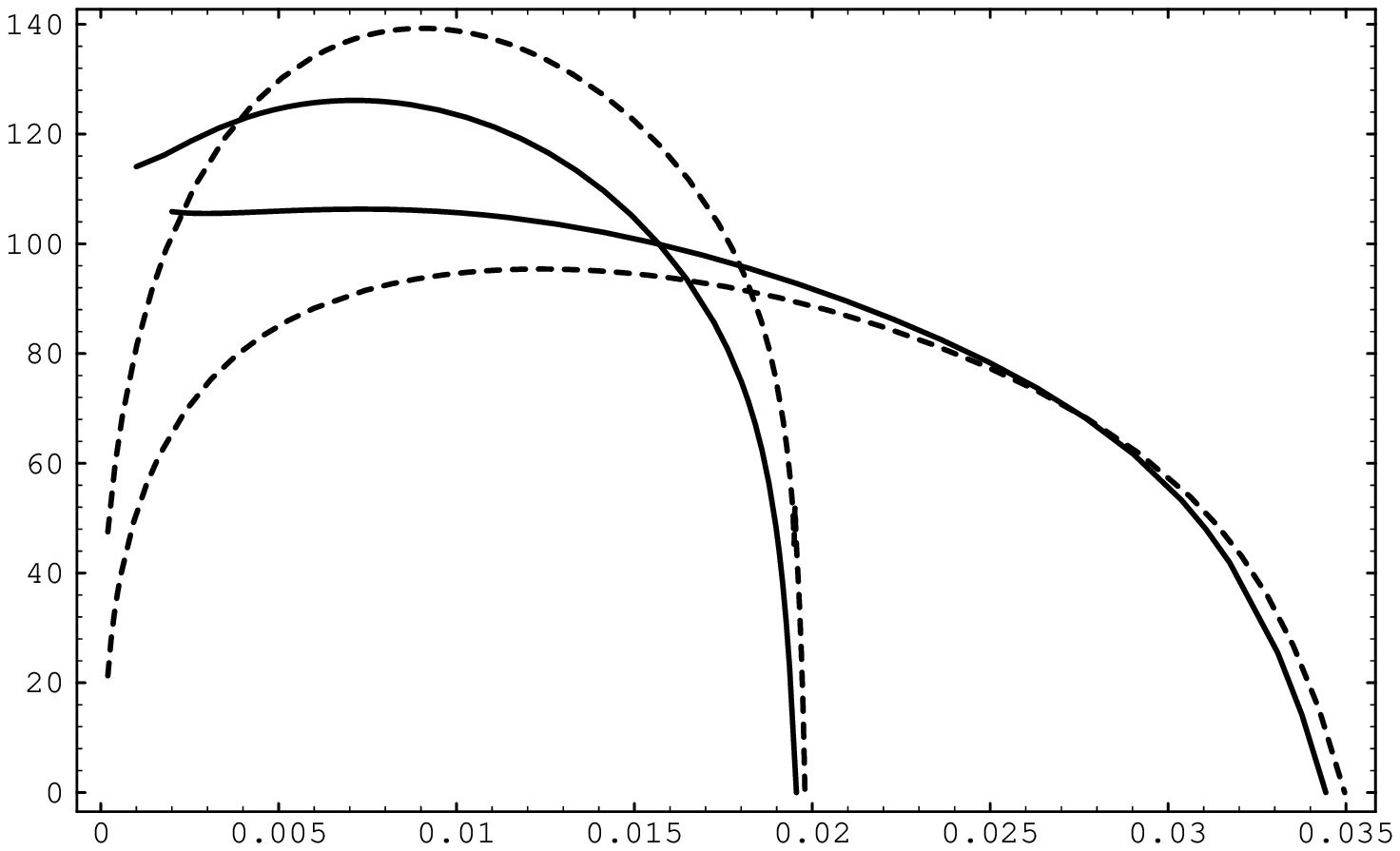}
        }
    \end{picture}
\caption{The pion decay constant $f_\pi$ as a function of 
  $\ol{m}_{k_\Phi}^2/k_\Phi^2$ for $k_\Phi=630\MeV$ and initial values
  $\la_{i,I}=0$ and $h^2_I=300$ (solid line) as well as $h^2_I=10^4$
  (dashed line).}
\label{Fig8}
\end{figure}
\end{center}

Most importantly, one finds that the system of flow equations exhibits
a partial IR fixed point in the symmetric phase. As already pointed
out one expects $Z_{\Phi,k_\Phi}$ to be rather small.  In turn, one
may assume that, at least for the initial range of running in the
symmetric regime the mass parameter $m^2\sim Z_{\Phi,k}^{-1}$ is
large. This means, in particular, that all threshold functions with
arguments $\sim m^2$ may be neglected in this regime.  As a
consequence, the flow equations simplify considerably.  We find, for
instance, for the $U_L(2)\times U_R(2)$ model
\begin{eqnarray}
  \label{AAA08}
  \ds{\frac{d}{d t}\tilde{\la}_1} &\equiv& \ds{
    \frac{d}{d t}\frac{\la_1}{h^2}\simeq
    \frac{N_c}{4\pi^2}h^2\left[\hal\tilde{\la}_1-\frac{1}{N}\right]}\nnn
  \ds{\frac{d}{d t}\tilde{\la}_2} &\equiv& \ds{
    \frac{d}{d t}\frac{\la_2}{h^2}\simeq
    \frac{N_c}{4\pi^2}h^2\left[\hal\tilde{\la}_2-\frac{2}{N}\right]}\\[2mm]
  \ds{\frac{d}{d t}h^2} &\simeq& \ds{
    \frac{N_c}{8\pi^2}h^4}\nnn
  \ds{\eta_\Phi} &\simeq& \ds{ 
    \frac{N_c}{8\pi^2}h^2\; ,\;\;\;
    \eta_\psi\simeq0}\nonumber\; .
\end{eqnarray} 
This system possesses an attractive IR fixed point for the
quartic scalar self interactions 
\begin{equation}
  \label{AAA10}
  \tilde{\la}_{1*}=\hal\tilde{\la}_{2*}=\frac{2}{N}\; .  
\end{equation}
Furthermore it is exactly soluble.$^{\citen{JW95-1}}$ Because of the strong
Yukawa coupling the quartic couplings $\tilde{\la}_1$ and $\tilde{\la}_2$
generally approach their fixed point values rapidly, long before the systems
enters the broken phase ($m\ra0$) and the approximation of large $m$ breaks
down.  In addition, for large initial values $h^2_I$ the Yukawa coupling at
the scale $k_{\chi SB}$ where $m$ vanishes (or becomes small) only depends on
the initial value $\ol{m}_{k_\Phi}$.  Hence, the system is approximately
independent in the IR upon the initial values of $\la_1$, $\la_2$ and $h^2$,
the only ``relevant'' parameter being $\ol{m}_{k_\Phi}$. (Once quark masses
and a proper treatment of the chiral anomaly are included for $N=3$ one
expects that $m_q$ and $\nu$ are additional relevant parameters. Their values
may be fixed by using the masses of $\pi$, $K$ and $\eta^\prime$ as
phenomenological input.) In other words, the effective action looses almost
all its ``memory'' in the far IR of where in the UV it came from. This feature
of the flow equations leads to a perhaps surprising degree of predictive
power.  In addition, also the dependence of $f_\pi=2\si_0$ on
$\ol{m}_{k_\Phi}$ is not very strong for a large range in $\ol{m}_{k_\Phi}$,
as shown in fig.~\ref{Fig8}.
The relevant parameter $\ol{m}_{k_\Phi}$ can be fixed by using the
constituent quark mass $M_q\equiv h\si_0\simeq350\MeV$ as a
phenomenological input.  One obtains for the $O(4)$--model
\begin{equation}
 \ds{\frac{\ol{m}_{k_\Phi}^2}{k_\Phi^2}} \simeq \ds{0.02}\; .
\end{equation}
The resulting value for the decay constant is
\begin{equation}
  \label{AAA52}
  f_\pi=(91-100)\MeV
\end{equation}
for $h^2_I=10^4-300$. It is striking that this comes close to the real
value $f_\pi=92.4\MeV$ but we expect that the uncertainty in the
determination of the compositeness scale $k_\Phi$ and the truncation
errors exceed the influence of the variation of $h^2_I$.  We have
furthermore used this result for an estimate of the chiral condensate:
\begin{equation}
 \VEV{\ol{\psi}\psi}\equiv
 -\ol{m}_{k_\Phi}^2f_\pi
 Z_{\Phi,k=0}^{-1/2}A\simeq-(195\MeV)^3 
\end{equation} 
where the factor $A\simeq1.7$ accounts for the change of the normalization
scale of $\VEV{\ol{\psi}\psi}$ from $k_\Phi$ to the commonly used value
$1\GeV$. Our value is in reasonable agreement with results from sum
rules.$^{\citen{JM95-1}}$ This is non--trivial since not only
$\ol{m}_{k_\Phi}$ and $f_\pi$ enter but also the IR value $Z_{\Phi,k=0}$.
Integrating~(\ref{AAA08}) for $\eta_\Phi$ one finds
\begin{equation}
  \label{AAA09}
  Z_{\Phi,k}=Z_{\Phi,k_\Phi}+
  \frac{N_c}{8\pi^2}\ln\frac{k_\Phi}{k}\; .
\end{equation}
Thus $Z_{\Phi,k}$ will indeed be practically independent of its
initial value $Z_{\Phi,k_\Phi}$ already after some running as long as
$Z_{\Phi,k_\Phi}$ is small compared to $0.01$.

The alert reader may have noticed that the beta functions~(\ref{AAA08})
correspond exactly to those obtained in the one--quark--loop approximation or,
in other words, to the leading order in the large--$N_c$ expansion for the
Nambu--Jona-Lasinio model.$^{\citen{BHJ94-1}}$ The fixed point~(\ref{AAA10})
is then nothing but the large--$N_c$ boundary condition on the evolution of
$\la_1$ and $\la_2$ in this model. Yet, we wish to stress that nowhere we have
made the assumption that $N_c$ is a large number.  On the contrary, the
physical value $N_c=3$ suggests that the large--$N_c$ expansion should a
priori only be trusted on a quantitatively rather crude level. The reason why
we expect~(\ref{AAA08}) to nevertheless give rather reliable results is based
on the fact that for small $Z_{\Phi,k_\Phi}$ all (renormalized) meson masses
are much larger than the scale $k$ for the initial part of the running.  This
implies that the mesons are effectively decoupled and their contribution to
the beta functions is negligible leading to the one--quark--loop
approximation. Yet, already after some relatively short period of running the
renormalized meson masses approach zero and our approximation of neglecting
mesonic threshold functions breaks down.  Hence, the one--quark--loop
approximation is reasonable only for scales close to $k_\Phi$ but is bound to
become inaccurate around $k_{\chi{\rm SB}}$ and in the SSB regime. What is
important in our context is not the numerical value of the partial fixed
points~(\ref{AAA10}) but rather their mere existence and the presence of a
large coupling $h^2$ driving the $\tilde{\la}_i$ fast towards them.  This is
enough for the IR values of all couplings to become almost independent of the
initial values $\la_{i,I}$.  Similar features of IR stability are expected if
the truncation is enlarged, for instance, to a more general form of the
effective potential $U_{k_\Phi}$ as will be discussed in the next section.

\section{Flow equation for the scalar potential}
\label{sec8}

In this and the remaining sections we consider the $O(4)$--symmetric quark
meson model without truncating its effective average potential to a polynomial
form.$^{\citen{BJW97-1}}$ A comparison with the results of the last sections
will give us a feeling for the size of the errors induced by the quartic
truncation used so far.  Furthermore, such an approach is well suited for a
study of the chiral phase transition close to which the form of the potential
deviates substantially from a polynomial.

It is convenient to work with dimensionless and renormalized variables
therefore eliminating all explicit $k$--dependence. With
($t=\ln(k/k_\Phi)$)
\begin{equation}
  \label{AAA190}
  u(t,\tilde{\rho})\equiv k^{-d}U_k(\rho)\; ,\;\;\;
  \tilde{\rho}\equiv Z_{\Phi,k} k^{2-d}\rho
\end{equation}
and using~(\ref{GammaEffective}) as a first truncation of the effective
average action $\Gamma_k$ one obtains the flow equation in arbitrary
dimensions $d$
\begin{equation}
  \label{AAA68}
  \begin{array}{rcl}
    \ds{\frac{\prl}{\prl t}u} &=& \ds{
      -d u+\left(d-2+\eta_\Phi\right)
      \tilde{\rho}u^\prime}\\[2mm]
    &+& \ds{
      2v_d\left\{
      3l_0^d(u^\prime;\eta_\Phi)+
      l_0^d(u^\prime+2\tilde{\rho}u^{\prime\prime};\eta_\Phi)-
      2^{\frac{d}{2}+1}N_c
      l_0^{(F)d}(\frac{1}{2}\tilde{\rho}h^2;\eta_\psi)
      \right\} }\; .
  \end{array}
\end{equation}
Here $v_d^{-1}\equiv2^{d+1}\pi^{d/2}\Gamma(d/2)$ and primes denote derivatives
with respect to $\tilde{\rho}$.  We will always use in the following for the
number of quark colors $N_c=3$. ~(\ref{AAA68}) is a partial differential
equation for the effective potential $u(t,\tilde{\rho})$ which has to be
supplemented by the flow equation for the Yukawa coupling and expressions for
the anomalous dimensions $\eta_\Phi$, $\eta_\psi$. The definition of the
threshold functions $l_n^d$, $l_n^{(F)d}$ can be found in
Ref.$^{\citen{JW95-1}}$.

The dimensionless renormalized expectation value
$\kappa\equiv2k^{2-d}Z_{\Phi,k}\ol{\si}_{0,k}^2$, with
$\ol{\si}_{0,k}$ the $k$--dependent VEV of $\Phi$, may be computed
for each $k$ directly from the condition
\begin{equation}
  \label{AAA90}
  u^\prime(t,\kappa)=\frac{\jmath}{\sqrt{2\kappa}}
  k^{-\frac{d+2}{2}}Z_{\Phi,k}^{-1/2}\equiv
  \epsilon_g
\end{equation}
where$^{\citen{BJW97-1}}$
\begin{equation}
  \label{LAL001}
  \jmath=2\ol{m}_{k_\Phi,k}^2\hat{m}
\end{equation}
and $\hat{m}=(m_u+m_d)/2$ denotes the average light current quark mass
normalized at $k_\Phi$.  Note that $\kappa\equiv0$ in the symmetric regime for
vanishing source term.~(\ref{AAA90}) allows us to follow the flow of $\kappa$
according to
\begin{eqnarray}
  \label{AAA91}
  \ds{\frac{d}{d t}\kappa} &=& \ds{
    \frac{\kappa}{\epsilon_g+2\kappa\la}
    \Bigg\{\left[\eta_\Phi-d-2\right]\epsilon_g-
    2\frac{\prl}{\prl t}u^\prime(t,\kappa)\Bigg\} }
\end{eqnarray}
with $\la\equiv u^{\prime\prime}(t,\kappa)$.  We define the Yukawa coupling
for $\tilde{\rho}=\kappa$ and its flow equation reads$^{\citen{JW95-1}}$
\begin{equation}
  \label{AAA70n}
  \begin{array}{rcl}
  \ds{\frac{d}{d t}h^2} &=& \ds{
  \left(d-4+2\eta_\psi+\eta_\Phi\right)h^2-
    2v_d h^4\Bigg\{
    3l_{1,1}^{(F B)d}(\frac{1}{2}h^2\kappa,
    \epsilon_g;\eta_\psi,\eta_\Phi) }\\[2mm]
  &-& \ds{
    l_{1,1}^{(F B)d}(\frac{1}{2}h^2\kappa,
    \epsilon_g+2\la\kappa;
    \eta_\psi,\eta_\Phi)
  \Bigg\} }\; .
  \end{array}
\end{equation}
Similarly, the the scalar and quark anomalous
dimensions are inferred from
\begin{equation}
  \label{AAA69n}
  \begin{array}{rcl}
    \ds{\eta_\Phi} &\equiv& \ds{
      -\frac{d}{d t}\ln Z_{\Phi,k}=
      4\frac{v_d}{d}\Bigg\{
      4\kappa\la^2
      m_{2,2}^d(\epsilon_g,\epsilon_g+2\la\kappa;
      \eta_\Phi) }\nnn
    &+& \ds{
      2^{\frac{d}{2}}N_c h^2
      m_4^{(F)d}(\frac{1}{2}h^2\kappa;
      \eta_\psi)
      \Bigg\}\; , }\nnn
    \ds{\eta_\psi} &\equiv& \ds{
      -\frac{d}{d t}\ln Z_{\psi,k}=
      2\frac{v_d}{d}h^2\Bigg\{
      3m_{1,2}^{(F B)d}(\frac{1}{2}h^2\kappa,
      \epsilon_g;\eta_\psi,\eta_\Phi) }\\[2mm]
    &+& \ds{
      m_{1,2}^{(F B)d}(\frac{1}{2}h^2\kappa,
      \epsilon_g+2\la\kappa;
      \eta_\psi,\eta_\Phi)
      \Bigg\}\; , }
  \end{array}
\end{equation}
which is a linear set of equations for the anomalous dimensions.  The
definitions of the threshold functions $l_{n_1,n_2}^{(FB)d}$, $m_{n_1,n_2}^d$,
$m_4^{(F)d}$ and $m_{n_1,n_2}^{(FB)d}$ are again specified in
Ref.$^{\citen{JW95-1}}$.

The flow equations (\ref{AAA68}), (\ref{AAA91})---(\ref{AAA69n}), constitute a
coupled system of ordinary and partial differential equations which can be
integrated numerically.$^{\citen{ABBFTW95-1,BW97-1}}$ Here we take the
effective current quark mass dependence of $h$, $Z_{\Phi,k}$ and $Z_{\psi,k}$
into account by stopping the evolution according to Eqs.(\ref{AAA70n}),
(\ref{AAA69n}), evaluated for the chiral limit, below the pion mass $m_\pi$.

Similarly to the case of the quartic truncation of the effective
average potential described in the preceding section one finds for
$d=4$ that chiral symmetry breaking indeed occurs for a wide range of
initial values of the parameters. These include the presumably
realistic case of large renormalized Yukawa coupling and a bare mass
$\ol{m}_{k_\Phi}$ of order $100\MeV$.

Most importantly, the approximate partial IR fixed point behavior encountered
for the quartic potential approximation before, carries over to the truncation
of $\Gamma_k$ which maintains the full effective average
potential.$^{\citen{BJW97-1}}$ To see this explicitly we study the flow
equations (\ref{AAA68}), (\ref{AAA91})---(\ref{AAA69n}) subject to the
condition $Z_{\Phi,k_\Phi}\ll1$. For the relevant range of $\tilde{\rho}$ both
$u^\prime(t,\tilde{\rho})$ and $u^\prime(t,\tilde{\rho})+
2\tilde{\rho}u^{\prime\prime}(t,\tilde{\rho})$ are then much larger than
$\tilde{\rho}h^2(t)$ and we may therefore neglect in the flow equations all
scalar contributions with threshold functions involving these large masses.
This yields the simplified equations ($d=4,v_4^{-1}=32\pi^2$)
\begin{equation}
  \label{AAA110}
  \begin{array}{rcl}
    \ds{\frac{\prl}{\prl t}u} &=& \ds{
      -4u+\left(2+\eta_\Phi\right)
      \tilde{\rho}u^\prime
      -\frac{N_c}{2\pi^2}
      l_0^{(F)4}(\frac{1}{2}\tilde{\rho}h^2)\; ,
      }\nnn
    \ds{\frac{d}{d t}h^2} &=& \ds{
      \frac{N_c}{8\pi^2}h^4\; ,
      }\nnn
      \ds{\eta_\Phi} &=& \ds{
        \frac{N_c}{8\pi^2}h^2 \; ,\;\;\;
        \eta_\psi=0}\; .
    \end{array}
\end{equation}
Again, this approximation is only valid for the initial range of running below
$k_\Phi$ before the (dimensionless) renormalized scalar mass squared
$u^\prime(t,\tilde{\rho}=0)$ approaches zero near the chiral symmetry breaking
scale.  The system~(\ref{AAA110}) is exactly soluble. We find
\begin{equation}
  \label{AAA113}
  \begin{array}{rcl}
    \ds{h^2(t)} &=& \ds{
      Z_\Phi^{-1}(t)=
      \frac{h_I^2}{1-\frac{N_c}{8\pi^2}h_I^2 t}\; ,\;\;\;
      Z_\psi(t)=1\; ,
      }\nnn
    \ds{u(t,\tilde{\rho})} &=& \ds{
      e^{-4t}u_I(e^{2t}\tilde{\rho}\frac{h^2(t)}{h_I^2})-
      \frac{N_c}{2\pi^2}\int_0^t d r e^{-4r}
      l_0^{(F)4}(\frac{1}{2}h^2(t)\tilde{\rho}e^{2r}) }\; .
  \end{array}
\end{equation}
(The integration over $r$ on the right hand side of the solution for
$u$ can be carried out by first exchanging it with the one over
momentum implicit in the definition of the threshold function
$l_0^{(F)4}$.) Here $u_I(\tilde{\rho})\equiv u(0,\tilde{\rho})$
denotes the effective average potential at the compositeness scale and
$h_I^2$ is the initial value of $h^2$ at $k_\Phi$, i.e. for $t=0$.
For simplicity we will use an expansion of the initial value effective
potential $u_I(\tilde{\rho})$ in powers of $\tilde{\rho}$ around
$\tilde{\rho}=0$
\begin{equation}
  \label{AAA140}
  u_I(\tilde{\rho})=
  \sum_{n=0}^\infty
  \frac{u_I^{(n)}(0)}{n!}\tilde{\rho}^n
\end{equation}
even though this is not essential for the forthcoming reasoning.  Expanding
also $l_0^{(F)4}$ in~(\ref{AAA113}) in powers of its argument one finds for
$n>2$
\begin{equation}
  \label{LLL00}
  \ds{\frac{u^{(n)}(t,0)}{h^{2n}(t)}} = \ds{
    e^{2(n-2)t}\frac{u_I^{(n)}(0)}{h_I^{2n}}+
    \frac{N_c}{\pi^2}
    \frac{(-1)^n (n-1)!}{2^{n+2}(n-2)}
    l_n^{(F)4}(0)
    \left[1-e^{2(n-2)t}\right]}\; .
\end{equation}
For decreasing $t\ra-\infty$ the initial values $u_I^{(n)}$ become
rapidly unimportant and $u^{(n)}/h^{2n}$ approaches a fixed point.
For $n=2$, i.e., for the quartic coupling, one finds
\begin{equation}
  \label{LLL01}
  \frac{u^{(2)}(t,0)}{h^2(t)}=
  1-\frac{1-\frac{u_I^{(2)}(0)}{h_I^2}}
  {1-\frac{N_c}{8\pi^2}h_I^2 t}
\end{equation}
leading to the fixed point value $(u^{(2)}/h^2)_*=1$ already encountered
in~(\ref{AAA10}). As a consequence of this fixed point behavior the system
looses all its ``memory'' on the initial values $u_I^{(n\ge2)}$ at the
compositeness scale $k_\Phi$! This typically happens before the approximation
$u^\prime(t,\tilde{\rho}),u^\prime(t,\tilde{\rho})+
2\tilde{\rho}u^{\prime\prime}(t,\tilde{\rho})\gg\tilde{\rho}h^2(t)$ breaks
down and the solution~(\ref{AAA113}) becomes invalid.  Furthermore, the
attraction to partial infrared fixed points continues also for the range of
$k$ where the scalar fluctuations cannot be neglected anymore.  As for the
quartic truncation the initial value for the bare dimensionless mass parameter
\begin{equation}
  \label{AAA142}
  \frac{u_I^\prime(0)}{h_I^2}=
  \frac{\ol{m}^2_{k_\Phi}}{k_\Phi^2}
\end{equation}
is never negligible. (In fact, using the values for $\ol{m}^2_{k_\Phi}$ and
$k_\Phi$ computed previously$^{\citen{EW94-1}}$ for a pure quark effective
action as described above one obtains
$\ol{m}^2_{k_\Phi}/k_\Phi^2\simeq0.036$.)  For large $h_I$ (and dropping the
constant piece $u_I(0)$) the solution~(\ref{AAA113}) therefore behaves with
growing $\abs{t}$ as
\begin{equation}
  \label{AAA150}
  \begin{array}{rcl}
    \ds{Z_\Phi(t)} &\simeq& \ds{
      -\frac{N_c}{8\pi^2}t\; ,\;\;\;
      Z_\psi(t)=1\; ,
      }\nnn
    \ds{h^2(t)} &\simeq& \ds{
      -\frac{8\pi^2}{N_c t}\; ,
      }\nnn
    \ds{u(t,\tilde{\rho})} &\simeq& \ds{
      \frac{u_I^\prime(0)}{h_I^2} e^{-2t} h^2(t)\tilde{\rho}
      -\frac{N_c}{2\pi^2}\int_0^t d r e^{-4r}
      l_0^{(F)4}(\frac{1}{2}h^2(t)\tilde{\rho}e^{2r}) }\; .  
  \end{array}
\end{equation}
In other words, for $h_I\ra\infty$ the IR behavior of the linear quark meson
model will depend (in addition to the value of the compositeness scale
$k_\Phi$ and the quark mass $\hat{m}$) only on one parameter,
$\ol{m}^2_{k_\Phi}$.  We have numerically verified this feature by starting
with different values for $u_I^{(2)}(0)$.  Indeed, the differences in the
physical observables were found to be small.  This IR stability of the flow
equations leads to a perhaps surprising degree of predictive power: Not only
the scalar wave function renormalization but even the full effective potential
$U(\rho)=\lim_{k\ra0}U_k(\rho)$ is (approximately) fixed for
$Z_{\Phi,k_\Phi}\ll1$ once $\ol{m}_{k_\Phi}$ is known!  For definiteness we
will perform our numerical analysis of the full system of flow equations
(\ref{AAA68}), (\ref{AAA91})---(\ref{AAA69n}) with the idealized initial value
$u_I(\tilde{\rho})=u_I^\prime(0)\tilde{\rho}$ in the limit $h_I^2\ra\infty$.
It should be stressed, though, that deviations from this idealization will
lead only to small numerical deviations in the IR behavior of the linear quark
meson model as long as the condition $Z_{\Phi,k_\Phi}\ll1$ holds, say
for$^{\citen{JW96-1}}$ $h_I\gta15$.

With this knowledge at hand we may now fix the remaining three parameters of
our model, $k_\Phi$, $\ol{m}^2_{k_\Phi}$ and $\hat{m}$ by using
$f_\pi=92.4\MeV$, the pion mass $M_\pi=135\MeV$ and the constituent quark mass
$M_q$ as phenomenological input.  This approach differs from that of the
preceding sections where we took an earlier determination of $k_\Phi$ as input
for the computation of $f_\pi$.  It will be better suited for precision
estimates of the high temperature behavior in the following sections.  Because
of the uncertainty regarding the precise value of $M_q$ we give in
tab.~\ref{tab1} the results for several values of $M_q$.
\begin{table}
  \caption{\footnotesize The table shows the dependence on the
    constituent quark mass $M_q$ of the input parameters $k_\Phi$,
    $\ol{m}^2_{k_\Phi}/k_\Phi^2$ and $\jmath$ as well as some of our
    ``predictions''. The phenomenological input used here besides $M_q$
    is $f_\pi=92.4\MeV$, $m_\pi=135\MeV$. The first line corresponds to
    the values for $M_q$ and $\la_I$ used in the remainder of this
    work. The other three lines demonstrate the insensitivity of our
    results with respect to the precise values of these
    parameters.}
  \label{tab1}
  \begin{center}
    \begin{tabular}{|c|c||c|c|c||c|c|c|} 
      \hline
      $\frac{M_q}{\MeV}$ &
      $\frac{\la_I}{h_I^2}$ &
      $\frac{k_\Phi}{\MeV}$ &
      $\frac{\ol{m}^2_{k_\Phi}}{k_\Phi^2}$ &
      $\frac{\hat{m}(k_\Phi)}{\MeV}$ &
      $\frac{\hat{m}(1\GeV)}{\MeV}$ &
      $\frac{\VEV{\ol{\psi}\psi}(1\GeV)}{\MeV^3}$ &
      $\frac{f_\pi^{(0)}}{\MeV}$
      \\[0.5mm] \hline\hline
      $303$ &
      $1$ &
      $618$ &
      $0.0265$ &
      $14.7$ &
      $11.4$ &
      $-(186)^3$ &
      $80.8$
      \\ \hline
      $300$ &
      $0$ &
      $602$ &
      $0.026$ &
      $15.8$ &
      $12.0$ &
      $-(183)^3$ &
      $80.2$
      \\ \hline
      $310$ &
      $0$ &
      $585$ &
      $0.025$ &
      $16.9$ &
      $12.5$ &
      $-(180)^3$ &
      $80.5$
      \\ \hline
      $339$ &
      $0$ &
      $552$ &
      $0.0225$ &
      $19.5$ &
      $13.7$ &
      $-(174)^3$ &
      $81.4$
      \\ \hline
    \end{tabular}
  \end{center}
\end{table}
The first line of tab.~\ref{tab1} corresponds to the choice of $M_q$ and
$\la_I\equiv u_I^{\prime\prime}(\kappa)$ which we will use for the forthcoming
analysis of the model at finite temperature.  As argued analytically above the
dependence on the value of $\la_I$ is weak for large enough $h_I$ as
demonstrated numerically by the second line.  Moreover, we notice that our
results, and in particular the value of $\jmath$, are rather insensitive with
respect to the precise value of $M_q$. It is remarkable that the values for
$k_\Phi$ and $\ol{m}_{k_\Phi}$ are not very different from those
computed$^{\citen{EW94-1}}$ from four--quark interactions as described above.
As compared to the analysis of the preceding sections the present truncation
of $\Gamma_k$ is of a higher level of accuracy: We now consider an arbitrary
form of the effective average potential instead of a polynomial approximation
and we have included the pieces in the threshold functions which are
proportional to the anomalous dimensions. It is encouraging that the results
are rather robust with respect to these improvements.

Once the parameters $k_\Phi$, $\ol{m}^2_{k_\Phi}$ and $\hat{m}$ are fixed
there are a number of ``predictions'' of the linear meson model which can be
compared with the results obtained by other methods or direct experimental
observation. First of all one may compute the value of $\hat{m}$ at a scale of
$1\GeV$ which is suitable for comparison with results obtained from chiral
perturbation theory$^{\citen{Leu96-1}}$ and sum rules.$^{\citen{JM95-1}}$ For
this purpose one has to account for the running of this quantity with the
normalization scale from $k_\Phi$ as given in tab.~\ref{tab1} to the commonly
used value of $1\GeV$: $\hat{m}(1\GeV)=A^{-1}\hat{m}(k_\Phi)$. A reasonable
estimate of the factor $A$ is obtained from the three loop running of
$\hat{m}$ in the $\ol{MS}$ scheme.$^{\citen{JM95-1}}$ For $M_q\simeq300\MeV$
corresponding to the first two lines in tab.~\ref{tab1} its value is
$A\simeq1.3$.  The results for $\hat{m}(1\GeV)$ are in acceptable agreement
with recent results from other methods$^{\citen{Leu96-1,JM95-1}}$ even though
they tend to be somewhat larger.  Closely related to this is the value of the
chiral condensate
\begin{equation}
  \label{LLL04}
  \VEV{\ol{\psi}\psi}(1\GeV)\equiv
  -A\ol{m}^2_{k_\Phi}
  \left[f_\pi Z_{\Phi,k=0}^{-1/2}-
    2\hat{m}\right]
    \; .
\end{equation}
These results are quite non--trivial since not only $f_\pi$ and
$\ol{m}^2_{k_\Phi}$ enter but also the computed IR value
$Z_{\Phi,k=0}$.  We emphasize in this context that there may be
substantial corrections both in the extrapolation from $k_\Phi$ to
$1\GeV$ and because of the neglected influence of the strange quark
which may be important near $k_\Phi$. These uncertainties have only
little effect on the physics at lower scales as relevant for our
analysis of the temperature effects. Only the value of $\jmath$ which
is fixed by $m_\pi$ enters here.

A further more qualitative test concerns the mass of the sigma
resonance or radial mode in the limit $k\ra0$ whose renormalized mass
squared is given by
\begin{equation}
  \label{LAL000}
  m_\si^2=Z_{\Phi,k_\Phi}^{-1/2}
  \frac{\ol{m}_{k_\Phi}^2\hat{m}}
  {\si_0}+4\la\si_0^2\; .
\end{equation}
From our numerical analysis we obtain $\la_{k=0}\simeq20$ which translates
into $m_\si\simeq 430\MeV$.  One should note, though, that this result is
presumably not very accurate as we have employed in this work the
approximation of using the Goldstone boson wave function renormalization
constant also for the radial mode. Furthermore, the explicit chiral symmetry
breaking contribution to $m_\si^2$ is certainly underestimated as long as the
strange quark is neglected.  In any case, we observe that the sigma meson is
significantly heavier than the pions. This is a crucial consistency check for
the linear quark meson model. A low sigma mass would be in conflict with the
numerous successes of chiral perturbation theory$^{\citen{Leu95-1}}$ which
requires the decoupling of all modes other than the Goldstone bosons in the
IR--limit of QCD. The decoupling of the sigma meson is, of course, equivalent
to the limit $\la\ra\infty$ which formally describes the transition from the
linear to the non--linear sigma model and which appears to be reasonably well
realized by the large IR--values of $\la$ obtained in our analysis. We also
note that the issue of the sigma mass is closely connected to the value of
$f_\pi^{(0)}$, the value of $f_\pi$ in the chiral limit $\hat{m}=0$ also given
in tab.~\ref{tab1}.  To lowest order in $(f_\pi-f_\pi^{(0)})/f_\pi$ or,
equivalently, in $\hat{m}$ one has
\begin{equation}
  \label{ABC00}
  f_\pi-f_\pi^{(0)}=\frac{\jmath}{Z_\Phi^{1/2}m_\si^2}=
  \frac{f_\pi m_\pi^2}{m_\si^2}\; .
\end{equation}
A larger value of $m_\si$ would therefore reduce the difference
between $f_\pi^{(0)}$ and $f_\pi$.

\begin{figure}
  \unitlength1.0cm
  \begin{center}
    \begin{picture}(10.,8.5)
      \put(0.0,0.5){
        \epsfysize=11.cm
        \rotate[r]{\epsffile{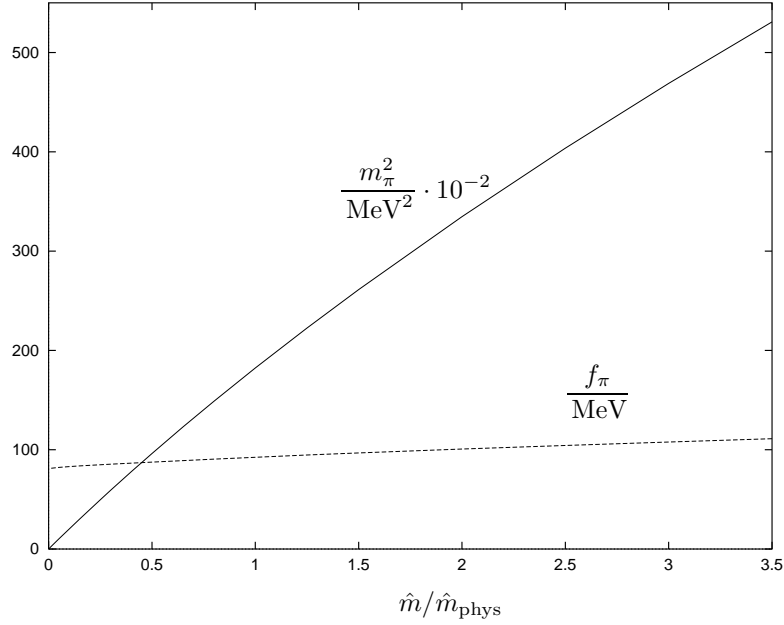}}
        }
      \put(8.0,2.8){\bf $\ds{\frac{f_\pi}{\MeV}}$}
      \put(5.0,5.5){\bf $\ds{\frac{m^2_\pi}{\MeV^2}\cdot10^{-2}}$}
      \put(5.8,-0.0){\bf $\ds{\hat{m}/\hat{m}_{\rm phys}}$}
    \end{picture}
  \end{center}
  \caption{\footnotesize The plot shows $m_\pi^2$ (solid line) and
    $f_\pi$ (dashed line) as functions of the current quark mass
    $\hat{m}$ in units of the physical value $\hat{m}_{\rm phys}$.}
  \label{mm}
\end{figure}

In fig.~\ref{mm} we show the dependence of the pion mass and decay constant on
the average current quark mass $\hat{m}$.
These curves depend very little on the values of the initial parameters as
demonstrated in tab.~\ref{tab1} by $f_\pi^{(0)}$. We observe a relatively
large difference of $12\MeV$ between the pion decay constants at
$\hat{m}=\hat{m}_{\rm phys}$ and $\hat{m}=0$.  According to~(\ref{ABC00}) this
difference is related to the mass of the sigma particle and will be modified
in the three flavor case. We will later find that the critical temperature
$T_c$ for the second order phase transition in the chiral limit is almost
independent of the initial conditions. The values of $f_\pi^{(0)}$ and $T_c$
essentially determine the non--universal amplitudes in the critical scaling
region (see below). In summary, we find that the behavior of our model for
small $k$ is quite robust as far as uncertainties in the initial conditions at
the scale $k_\Phi$ are concerned. We will see that the difference of
observables between non--vanishing and vanishing temperature is entirely
determined by the flow of couplings in the range $0<k\lta3T$.

\section{Chiral phase transition of two flavor QCD}
\label{ChrialPhaseTransitionInTwoFlavorQCD}

Strong interactions in thermal equilibrium at high temperature $T$ --- as
realized in early stages of the evolution of the Universe --- differ in
important aspects from the well tested vacuum or zero temperature properties.
A phase transition at some critical temperature $T_c$ or a relatively sharp
crossover may separate the high and low temperature
physics.$^{\citen{MO96-1}}$ Many experimental activities at heavy ion
colliders$^{\citen{QM96}}$ search for signs of such a transition.  It was
realized early that the transition should be closely related to a qualitative
change in the chiral condensate according to the general observation that
spontaneous symmetry breaking tends to be absent in a high temperature
situation. A series of stimulating
contributions$^{\citen{PW84-1,RaWi93-1,Raj95-1}}$ pointed out that for
sufficiently small up and down quark masses, $m_u$ and $m_d$, and for a
sufficiently large mass of the strange quark, $m_s$, the chiral transition is
expected to belong to the universality class of the $O(4)$ Heisenberg model.
This means that near the critical temperature only the pions and the sigma
particle play a role for the behavior of condensates and long distance
correlation functions. It was suggested$^{\citen{RaWi93-1,Raj95-1}}$ that a
large correlation length may be responsible for important fluctuations or lead
to a disoriented chiral condensate.$^{\citen{Ans88-1}}$ This was even
related$^{\citen{RaWi93-1,Raj95-1}}$ to the spectacular ``Centauro
events''$^{\citen{LFH80-1}}$ observed in cosmic rays. The question how small
$m_u$ and $m_d$ would have to be in order to see a large correlation length
near $T_c$ and if this scenario could be realized for realistic values of the
current quark masses remained, however, unanswered.  The reason was the
missing link between the universal behavior near $T_c$ and zero current quark
mass on one hand and the known physical properties at $T=0$ for realistic
quark masses on the other hand.

It is the purpose of the remaining sections of these lectures to provide this
link.$^{\citen{BJW97-1}}$ We present here the equation of state for two flavor
QCD within an effective quark meson model. The equation of state expresses the
chiral condensate $\VEV{\ol{\psi}\psi}$ as a function of temperature and the
average current quark mass $\hat{m}=(m_u+m_d)/2$.  This connects explicitly
the universal critical behavior for $T\ra T_c$ and $\hat{m}\ra0$ with the
temperature dependence for a realistic value $\hat{m}_{\rm phys}$.  Since our
discussion covers the whole temperature range $0\le T \,\ltap\, 1.7\, T_c$ we
can fix $\hat{m}_{\rm phys}$ such that the (zero temperature) pion mass is
$m_\pi=135\MeV$. The condensate $\VEV{\ol{\psi}\psi}$ plays here the role of
an order parameter.  Fig.~\ref{ccc_T} shows our results for
$\VEV{\ol{\psi}\psi}(T,\hat{m})$:
\begin{figure}
\unitlength1.0cm

\begin{center}
  \begin{picture}(10.,9.0)
    \put(0.0,1.0){
      \epsfysize=11.cm
      \rotate[r]{\epsffile{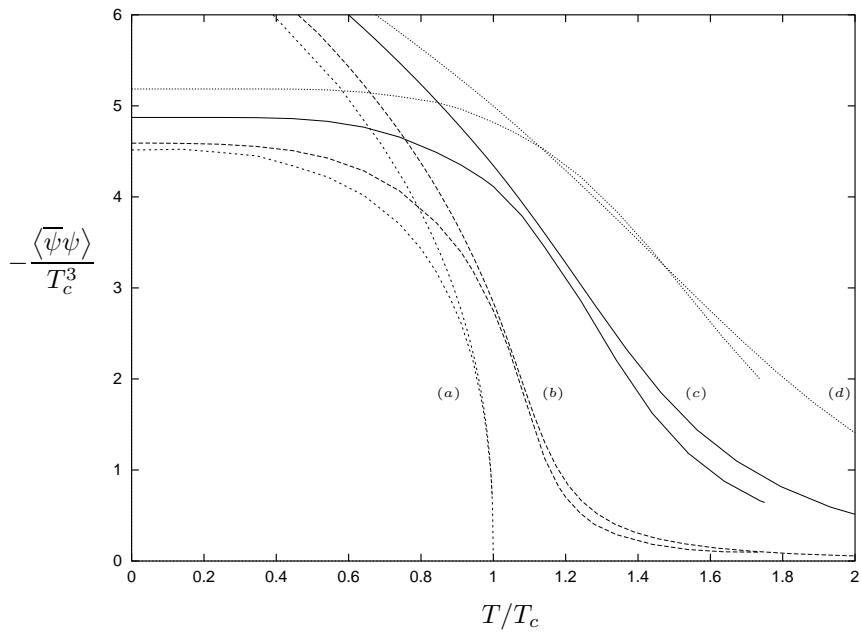}}
      }
    \put(-0.5,5.2){\bf $\ds{-\frac{\VEV{\ol{\psi}\psi}}{T_{c}^3}}$}
    \put(5.8,0.5){\bf $\ds{T/T_{c}}$}
    \put(5.2,3.5){\tiny $(a)$}
    \put(6.6,3.5){\tiny $(b)$}
    \put(8.5,3.5){\tiny $(c)$}
    \put(10.4,3.5){\tiny $(d)$}
  \end{picture}
\end{center}
\caption{\footnotesize The plot shows the chiral condensate
  $\VEV{\ol{\psi}\psi}$ as a function of temperature $T$.  Lines
  $(a)$, $(b)$, $(c)$, $(d)$ correspond at zero temperature to
  $m_\pi=0,45\MeV,135\MeV,230\MeV$, respectively. For each pair of
  curves the lower one represents the full $T$--dependence of
  $\VEV{\ol{\psi}\psi}$ whereas the upper one shows for comparison the
  universal scaling form of the equation of state for the $O(4)$
  Heisenberg model. The critical temperature for zero quark mass is
  $T_c=100.7\MeV$. The chiral condensate is normalized at a scale
  $k_{\Phi}\simeq 620\MeV$.}
\label{ccc_T}
\end{figure}
Curve $(a)$ gives the temperature dependence of $\VEV{\ol{\psi}\psi}$ in the
chiral limit $\hat{m}=0$. Here the lower curve is the full result for
arbitrary $T$ whereas the upper curve corresponds to the universal scaling
form of the equation of state for the $O(4)$ Heisenberg model.  We see perfect
agreement of both curves for $T$ sufficiently close to $T_c=100.7 \MeV$. This
demonstrates the capability of our method to cover the critical behavior and,
in particular, to reproduce the critical exponents of the $O(4)$--model.  We
have determined (see below) the universal critical equation of state as well
as the non--universal amplitudes.  This provides the full functional
dependence of $\VEV{\ol{\psi}\psi} (T,\hat{m})$ for small $T-T_c$ and
$\hat{m}$.  The curves $(b)$, $(c)$ and $(d)$ are for non--vanishing values of
the average current quark mass $\hat{m}$.  Curve $(c)$ corresponds to
$\hat{m}_{\rm phys}$ or, equivalently, $m_\pi(T=0)=135\MeV$. One observes a
crossover in the range $T=(1.2-1.5)T_c$. The $O(4)$ universal equation of
state (upper curve) gives a reasonable approximation in this temperature
range. The transition turns out to be much less dramatic than for $\hat{m}=0$.
We have also plotted in curve $(b)$ the results for comparably small quark
masses $\simeq1\MeV$, i.e.~$\hat{m}=\hat{m}_{\rm phys}/10$, for which the
$T=0$ value of $m_\pi$ equals $45\MeV$. The crossover is considerably sharper
but a substantial deviation from the chiral limit remains even for such small
values of $\hat{m}$. In order to facilitate comparison with lattice
simulations which are typically performed for larger values of $m_\pi$ we also
present results for $m_\pi(T=0)=230\MeV$ in curve $(d)$. One may define a
``pseudocritical temperature'' $T_{pc}$ associated to the smooth crossover as
the inflection point of $\VEV{\ol{\psi}\psi}(T)$ as often done in lattice
simulations. Our results for this definition of $T_{pc}$ are denoted by
$T_{pc}^{(1)}$ and are presented in tab.~\ref{tab11} for the four different
values of $\hat{m}$ or, equivalently, $m_\pi(T=0)$.
\begin{table}
  \caption{\footnotesize The table shows the critical and
    ``pseudocritical'' temperatures for various values of the zero
    temperature pion mass. Here $T_{pc}^{(1)}$ is defined as the
    inflection point of $\VEV{\ol{\psi}\psi}(T)$ whereas $T_{pc}^{(2)}$
    is the location of the maximum of the sigma correlation length.}
  \label{tab11}
  \begin{center}
    \begin{tabular}{|c||c|c|c|c|} \hline
      $\stackrel{ }{\frac{m_\pi}{\MeV}}$ &
      $0$ &
      $45$ &
      $135$ &
      $230$
      \\[1.0mm] \hline
      $\stackrel{ }{\frac{T_{pc}^{(1)}}{\MeV}}$ &
      $100.7$ &
      $\simeq110$ &
      $\simeq130$ &
      $\simeq150$
      \\[1mm] \hline
      $\stackrel{ }{\frac{T_{pc}^{(2)}}{\MeV}}$ &
      $100.7$ &
      $$113 &
      $$128 &
      $$---
      \\[1mm] \hline
    \end{tabular}
  \end{center}
\end{table}
The value for the pseudocritical temperature for $m_{\pi}=230\MeV$ compares
well with the lattice results for two flavor QCD (cf.~the discussion below).
One should mention, though, that a determination of $T_{p c}$ according to
this definition is subject to sizeable numerical uncertainties for large pion
masses as the curve in fig.~\ref{ccc_T} is almost linear around the inflection
point for quite a large temperature range.  A difficult point in lattice
simulations with large quark masses is the extrapolation to realistic values
of $m_\pi$ or even to the chiral limit. Our results may serve here as an
analytic guide. The overall picture shows the approximate validity of the
$O(4)$ scaling behavior over a large temperature interval in the vicinity of
and above $T_c$ once the (non--universal) amplitudes are properly computed.

\begin{figure}
  \unitlength1.0cm
  \begin{center}
    \begin{picture}(10.,9.0)
      \put(-1.0,1.0){
        \epsfysize=11.cm
        \rotate[r]{\epsffile{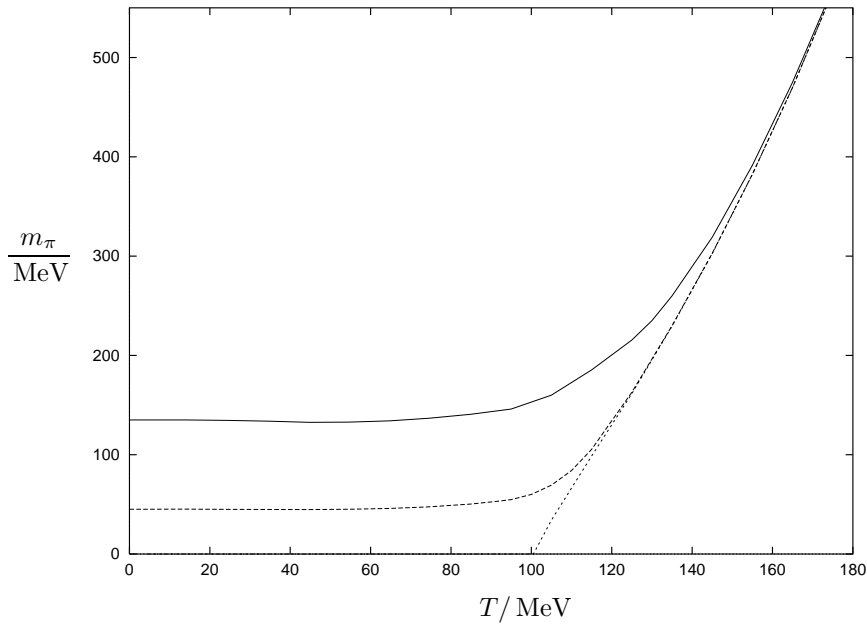}}
        }
      \put(-1.5,5.2){\bf $\ds{\frac{m_\pi}{\MeV}}$}
      \put(4.8,0.5){\bf $\ds{T/\MeV}$}
    \end{picture}
  \end{center}
  \caption{\footnotesize The plot shows $m_\pi$ as a function of
    temperature $T$ for three different values of the average light
    current quark mass $\hat{m}$. The solid line corresponds to the
    realistic value $\hat{m}=\hat{m}_{\rm phys}$ whereas the dotted line
    represents the situation without explicit chiral symmetry breaking,
    i.e., $\hat{m}=0$. The intermediate, dashed line assumes 
    $\hat{m}=\hat{m}_{\rm phys}/10$.}
  \label{mpi_T}
\end{figure}

A second important result of our investigations is the temperature dependence
of the space--like pion correlation length $m_\pi^{-1}(T)$. (We will often
call $m_\pi(T)$ the temperature dependent pion mass since it coincides with
the physical pion mass for $T=0$.) The plot for $m_\pi(T)$ in fig.~\ref{mpi_T}
again shows the second order phase transition in the chiral limit $\hat{m}=0$.
For $T<T_c$ the pions are massless Goldstone bosons whereas for $T>T_c$ they
form together with the sigma particle a degenerate vector of $O(4)$ whose mass
increases as a function of temperature.  For $\hat{m}=0$ the behavior for
small positive $T-T_c$ is characterized by the critical exponent $\nu$, i.e.
$m_\pi(T)=\left(\xi^+\right)^{-1}T_c \left( (T-T_c)/T_c\right)^\nu$ and we
obtain $\nu=0.787$, $\xi^+=0.270$. For $\hat{m}>0$ we find that $m_\pi(T)$
remains almost constant for $T\lta T_c$ with only a very slight dip for $T$
near $T_c/2$. For $T>T_c$ the correlation length decreases rapidly and for
$T\gg T_c$ the precise value of $\hat{m}$ becomes irrelevant. We see that the
universal critical behavior near $T_c$ is quite smoothly connected to $T=0$.
The full functional dependence of $m_\pi(T,\hat{m})$ allows us to compute the
overall size of the pion correlation length near the critical temperature and
we find $ m_\pi(T_{pc})\simeq 1.7 m_\pi(0)$ for the realistic value
$\hat{m}_{\rm phys}$. This correlation length is even smaller than the vacuum
($T=0$) one and shows no indication for strong fluctuations of pions with long
wavelength. It would be interesting to see if a decrease of the pion
correlation length at and above $T_c$ is experimentally observable.  It should
be emphasized, however, that a tricritical behavior with a massless excitation
remains possible for three flavors. This would not be characterized by the
universal behavior of the $O(4)$--model. We also point out that the present
investigation for the two flavor case does not take into account a speculative
``effective restoration'' of the axial $U_A(1)$ symmetry at high
temperature.$^{\citen{PW84-1,Shu94-1}}$ We will comment on these issues in the
last section. In the next sections we will describe the formalism which leads
to these results.

\section{Thermal equilibrium and dimensional reduction}
\label{FiniteTemperatureFormalism}

The extension of flow equations to thermal equilibrium situations at
non--vanishing temperature $T$ is straightforward.$^{\citen{TW93-1}}$ In the
Euclidean formalism non--zero temperature results in (anti--)periodic boundary
conditions for (fermionic) bosonic fields in the Euclidean time direction with
periodicity$^{\citen{Kap89-1}}$ $1/T$.  This leads to the replacement
\begin{equation}
  \label{AAA120}
  \int\frac{d^d q}{(2\pi)^d}f(q^2)\ra
  T\sum_{l\in\ZZZ}\int\frac{d^{d-1}\vec{q}}{(2\pi)^{d-1}}
  f(q_0^2(l)+\vec{q}^{\,2})
\end{equation}
in the trace in~(\ref{ERGE}) when represented as a momentum integration, with
a discrete spectrum for the zero component
\begin{equation}
  \label{AAA121}
  q_0(l)=\left\{
  \begin{array}{lll}
    2l\pi T &{\rm for}& {\rm bosons}\\
    (2l+1)\pi T &{\rm for}& {\rm fermions}\; .
  \end{array}\right.
\end{equation}
Hence, for $T>0$ a four--dimensional QFT can be interpreted as a
three--dimensional model with each bosonic or fermionic degree of freedom now
coming in an infinite number of copies labeled by $l\in\ZZ$ (Matsubara modes).
Each mode acquires an additional temperature dependent effective mass term
$q_0^2(l)$. In a high temperature situation where all massive Matsubara modes
decouple from the dynamics of the system one therefore expects to observe an
effective three--dimensional theory with the bosonic zero modes as the only
relevant degrees of freedom. In other words, if the characteristic length
scale associated with the physical system is much larger than the inverse
temperature the compactified Euclidean ``time'' dimension cannot be resolved
anymore. This phenomenon is known as ``dimensional
reduction''.$^{\citen{Gin80-1}}$

\begin{center}
\begin{figure}
  \unitlength1.0cm
  \begin{center}
    \begin{picture}(10.,12.5)
      \put(0.0,6.5){
        \epsfysize=9.cm
        \rotate[r]{\epsffile{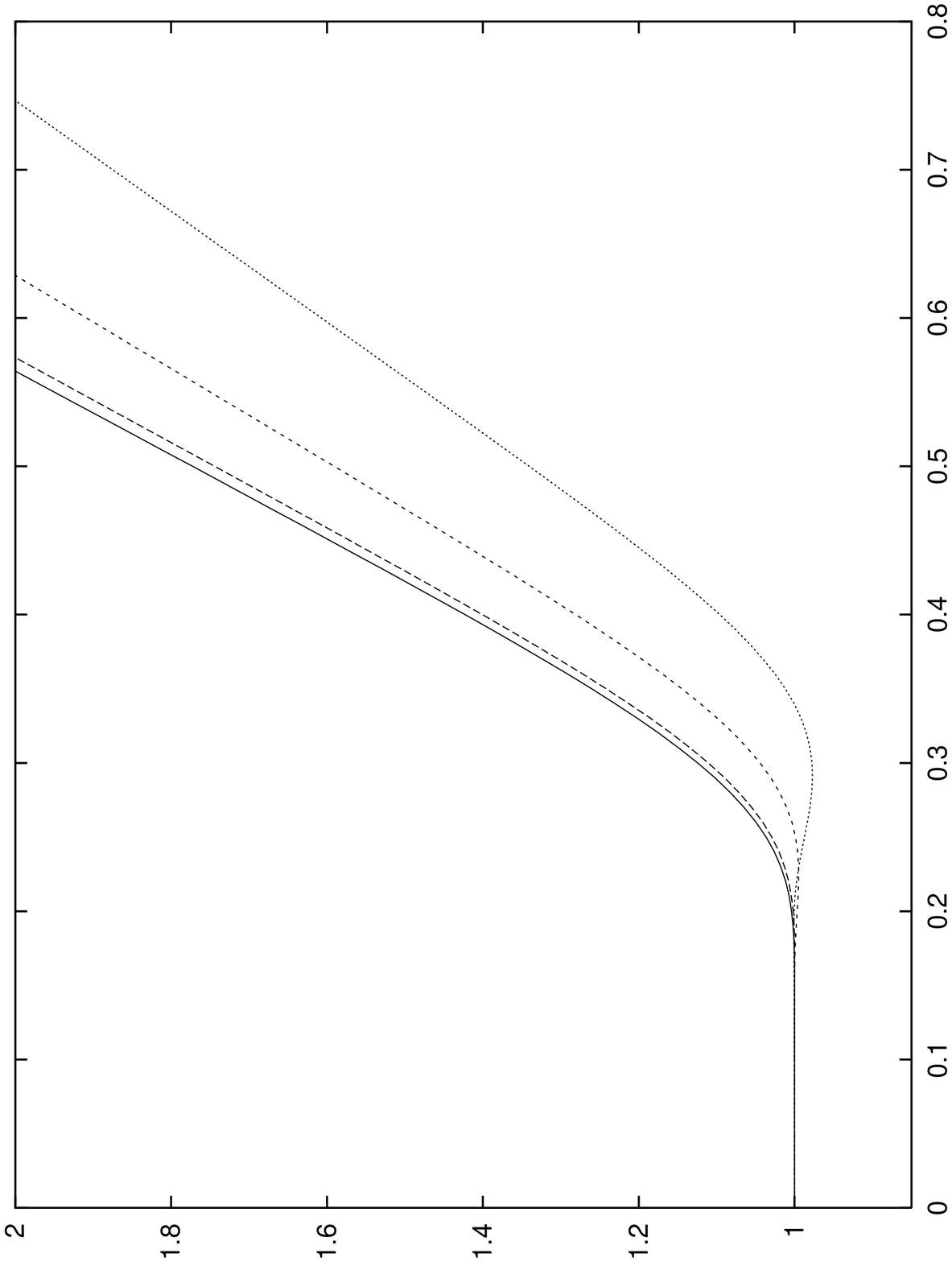}}
        }
      \put(-1.2,10.9){\bf $\ds{\frac{l_1^4\left(w,\ds{\frac{T}{k}}\right)}
          {l_1^4(w)}}$}
      \put(1.5,12.0){\bf $\ds{(a)}$}
      \put(0.0,-0.0){
        \epsfysize=9.cm
        \rotate[r]{\epsffile{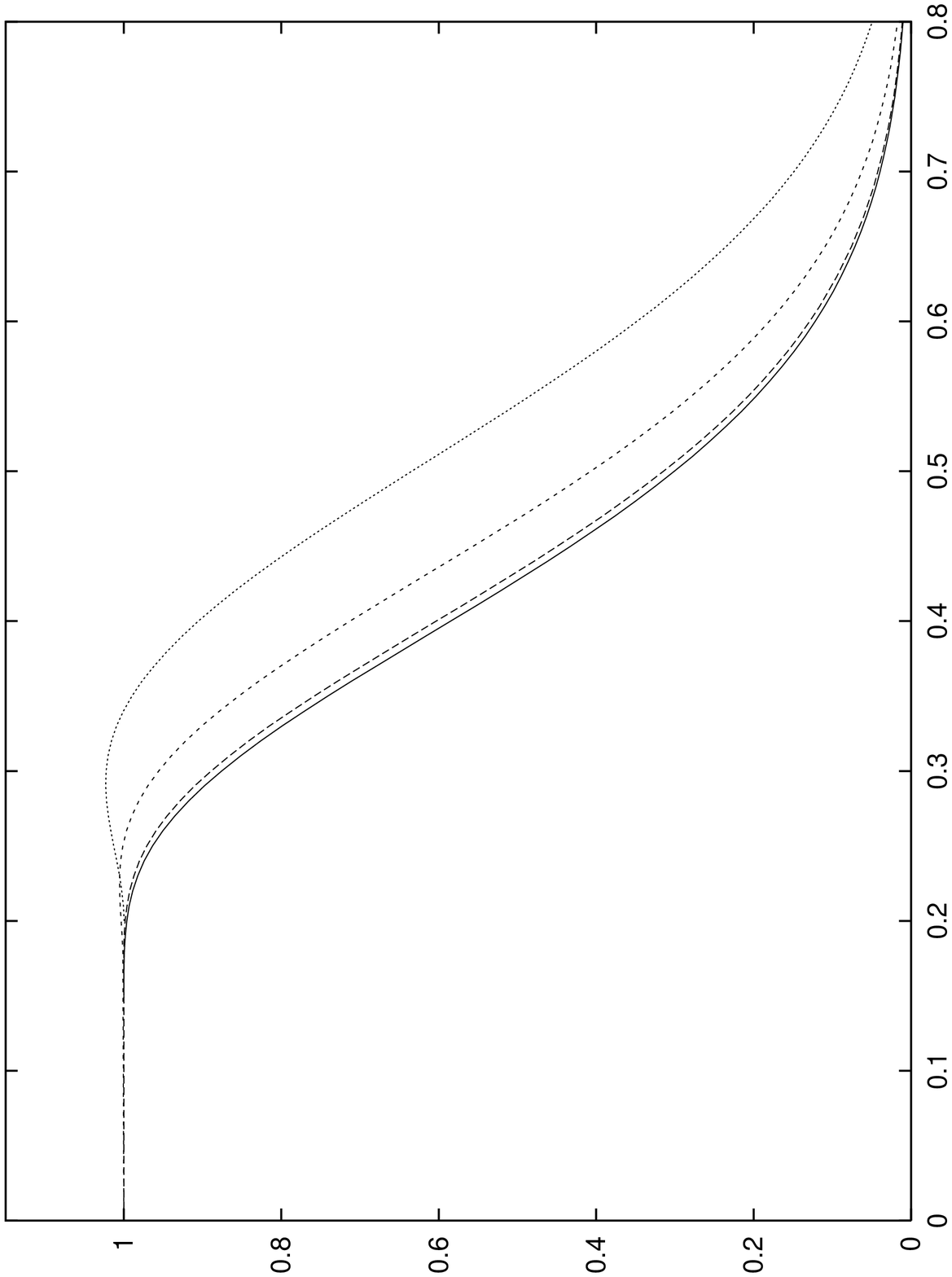}}
        }
      \put(-1.4,4.5){\bf
        $\ds{\frac{l_1^{(F)4}\left(w,\ds{\frac{T}{k}}\right)}
          {l_1^{(F)4}(w)}}$}
      \put(4.7,-0.3){\bf $\ds{T/k}$}
      \put(1.5,1.0){\bf $\ds{(b)}$}
    \end{picture}
  \end{center}
  \caption{\footnotesize The plot shows the temperature 
    dependence of the bosonic (a) and the fermionic (b) threshold
    functions $l_1^4(w,T/k)$ and $l_1^{(F)4}(w,T/k)$, respectively, for
    different values of the dimensionless mass term $w$.  The solid line
    corresponds to $w=0$ whereas the dotted ones correspond to $w=0.1$,
    $w=1$ and $w=10$ with decreasing size of the dots.  For $T \gg k$
    the bosonic threshold function becomes proportional to $T/k$ whereas
    the fermionic one tends to zero.  In this range the theory with
    properly rescaled variables behaves as a classical
    three--dimensional theory.}
  \label{Thresh}
\end{figure}
\end{center}

The formalism of the effective average action automatically provides the tools
for a smooth decoupling of the massive Matsubara modes as the scale $k$ is
lowered from $k\gg T$ to $k\ll T$.  It therefore allows us to directly link
the low--$T$, four--dimensional QFT to the effective three--dimensional
high--$T$ theory. The replacement~(\ref{AAA120}) in~(\ref{ERGE}) manifests
itself in the flow equations~(\ref{AAA68}), (\ref{AAA91})---(\ref{AAA69n})
only through a change to $T$--dependent threshold functions.  For instance,
the dimensionless functions $l_n^d(w;\eta_\Phi)$ defined in~(\ref{AAA85n}) are
replaced by
\begin{eqnarray}
  \label{AAA200}
  \ds{l_n^d(w,\frac{T}{k};\eta_\Phi)} &\equiv& \ds{
    \frac{n+\delta_{n,0}}{4}v_d^{-1}k^{2n-d}}\nnn
  &\times& \ds{
    T\sum_{l\in\ZZZ}\int
    \frac{d^{d-1}\vec{q}}{(2\pi)^{d-1}}
    \left(\frac{1}{Z_{\Phi,k}}\frac{\prl R_k(q^2)}{\prl t}\right)
      \frac{1}{\left[P(q^2)+k^2 w\right]^{n+1}} }
\end{eqnarray}
where $q^2=q_0^2+\vec{q}^{\,2}$ and $q_0=2\pi l T$. A list of the various
temperature dependent threshold functions appearing in the flow equations can
be found.$^{\citen{BJW97-1}}$ There we also discuss some subtleties regarding
the definition of the Yukawa coupling and the anomalous dimensions for
$T\neq0$.  In the limit $k\gg T$ the sum over Matsubara modes approaches the
integration over a continuous range of $q_0$ and we recover the zero
temperature threshold function $l_n^d(w;\eta_\Phi)$.  In the opposite limit
$k\ll T$ the massive Matsubara modes ($l\neq0$) are suppressed and we expect
to find a $d-1$ dimensional behavior of $l_n^d$. In fact, one obtains
from~(\ref{AAA200})
\begin{equation}
  \label{AAA201}
  \begin{array}{rclcrcl}
    \ds{l_n^d(w,T/k;\eta_\Phi)} &\simeq& \ds{
      l_n^{d}(w;\eta_\Phi)}
    &{\rm for}& \ds{T\ll k}\; ,\nnn
    \ds{l_n^d(w,T/k;\eta_\Phi)} &\simeq& \ds{
      \frac{T}{k}\frac{v_{d-1}}{v_d}
      l_n^{d-1}(w;\eta_\Phi)}
    &{\rm for}& \ds{T\gg k}\; .
  \end{array}
\end{equation}
For our choice of the infrared cutoff function $R_k$, (\ref{AAA61}), the
temperature dependent Matsubara modes in $l_n^d(w,T/k;\eta_\Phi)$ are
exponentially suppressed for $T\ll k$ whereas the behavior is more complicated
for other threshold functions appearing in the flow equations (\ref{AAA68}),
(\ref{AAA91})---(\ref{AAA69n}).  Nevertheless, all bosonic threshold functions
are proportional to $T/k$ for $T\gg k$ whereas those with fermionic
contributions vanish in this limit\footnote{For the present choice of $R_k$
  the temperature dependence of the threshold functions is considerably
  smoother than in that of previous investigations.$^{\citen{TW93-1}}$.} This
behavior is demonstrated in fig.~\ref{Thresh} where we have plotted the
quotients $l_1^4(w,T/k)/l_1^4(w)$ and $l_1^{(F)4}(w,T/k)/l_1^{(F)4}(w)$ of
bosonic and fermionic threshold functions, respectively.
One observes that for $k\gg T$ both threshold functions essentially behave as
for zero temperature. For growing $T$ or decreasing $k$ this changes as more
and more Matsubara modes decouple until finally all massive modes are
suppressed. The bosonic threshold function $l^4_1$ shows for $k \ll T$ the
linear dependence on $T/k$ derived in~(\ref{AAA201}).  In particular, for the
bosonic excitations the threshold function for $w\ll1$ can be approximated
with reasonable accuracy by $l_n^4(w;\eta_\Phi)$ for $T/k<0.25$ and by
$(4T/k)l_n^3(w;\eta_\Phi)$ for $T/k>0.25$. The fermionic threshold function
$l_1^{(F)4}$ tends to zero for $k\ll T$ since there is no massless fermionic
zero mode, i.e.~in this limit all fermionic contributions to the flow
equations are suppressed.  On the other hand, the fermions remain
quantitatively relevant up to $T/k\simeq0.6$ because of the relatively long
tail in fig.~\ref{Thresh}b.  The transition from four to three--dimensional
threshold functions leads to a {\em smooth dimensional reduction} as $k$ is
lowered from $k\gg T$ to $k\ll T$!  Whereas for $k\gg T$ the model is most
efficiently described in terms of standard four--dimensional fields $\Phi$ a
choice of rescaled three--dimensional variables $\Phi_{3}=\Phi/\sqrt{T}$
becomes better adapted for $k\ll T$.  Accordingly, for high temperatures one
will use a potential
\begin{equation}
  \label{CCC01}
  u_{3}(t,\tilde{\rho}_{3})=\frac{k}{T}
  u(t,\tilde{\rho})\; ;\;\;\;
  \tilde{\rho}_{3}=\frac{k}{T}\tilde{\rho}\; .
\end{equation}
In this regime $\Gamma_{k\ra0}$ corresponds to the free energy of
classical statistics and $\Gamma_{k>0}$ is a classical coarse grained
free energy.

For our numerical calculations at non--vanishing temperature we exploit the
discussed behavior of the threshold functions by using the zero temperature
flow equations in the range $k\ge10T$. For smaller values of $k$ we
approximate the infinite Matsubara sums (cf.~(\ref{AAA200})) by a finite
series such that the numerical uncertainty at $k=10T$ is better than
$10^{-4}$. This approximation becomes exact in the limit $k\ll10T$.

\section{The quark meson model at $T\neq0$}
\label{TheQuarkMesonModelAtTNeq0}

So far we have considered the relevant fluctuations that contribute to the
flow of $\Gamma_k$ in dependence on the scale $k$. In a thermal equilibrium
situation $\Gamma_k$ also depends on the temperature $T$ and one may ask for
the relevance of thermal fluctuations at a given scale $k$.  In particular,
for not too high values of $T$ the ``initial condition'' $\Gamma_{k_\Phi}$ for
the solution of the flow equations should essentially be independent of
temperature.  This will allow us to fix $\Gamma_{k_\Phi}$ from
phenomenological input at $T=0$ and to compute the temperature dependent
quantities in the infrared ($k \to 0$).  We note that the thermal fluctuations
which contribute to the r.h.s.~of the flow equation for the meson
potential~(\ref{AAA68}) are effectively suppressed for $T \lta k/4$ as
discussed in detail in the last section.  Clearly for $T \gta k_{\Phi}/3$
temperature effects become important at the compositeness scale. We expect the
linear quark meson model with a compositeness scale $k_{\Phi} \simeq 600 \MeV$
to be a valid description for two flavor QCD below a temperature of about $170
\MeV$. We note that there will be an effective temperature dependence of
$\Gamma_{k_{\Phi}}$ induced by the fluctuations of other degrees of freedom
besides the quarks, the pions and the sigma which are taken into account here.
We will comment on this issue in the last section.  For realistic three flavor
QCD the thermal kaon fluctuations will become important for $T\gta170\MeV$.

We compute the quantities of interest for temperatures $T\lta170\MeV$ by
numerically solving the $T$--dependent version of the flow
equations$^{\citen{BJW97-1}}$ (\ref{AAA68}), (\ref{AAA91})---(\ref{AAA69n}) by
lowering $k$ from $k_\Phi$ to zero.  For this range of temperatures we use the
initial values as given in the first line of tab.~\ref{tab1}.  This
corresponds to choosing the zero temperature pion mass and the pion decay
constant ($f_{\pi}=92.4 \MeV$ for $m_{\pi}=135 \MeV$) as phenomenological
input. The only further input is the constituent quark mass $M_q$ which we
vary in the range $M_q \simeq 300 - 350 \MeV$. We observe only a minor
dependence of our results on $M_q$ for the considered range of values. In
particular, the value for the critical temperature $T_c$ of the model remains
almost unaffected by this variation.

\begin{figure}
  \unitlength1.0cm
  \begin{center}
    \begin{picture}(10.,9.0)
      \put(-0.5,1.0){
        \epsfysize=11.cm
        \rotate[r]{\epsffile{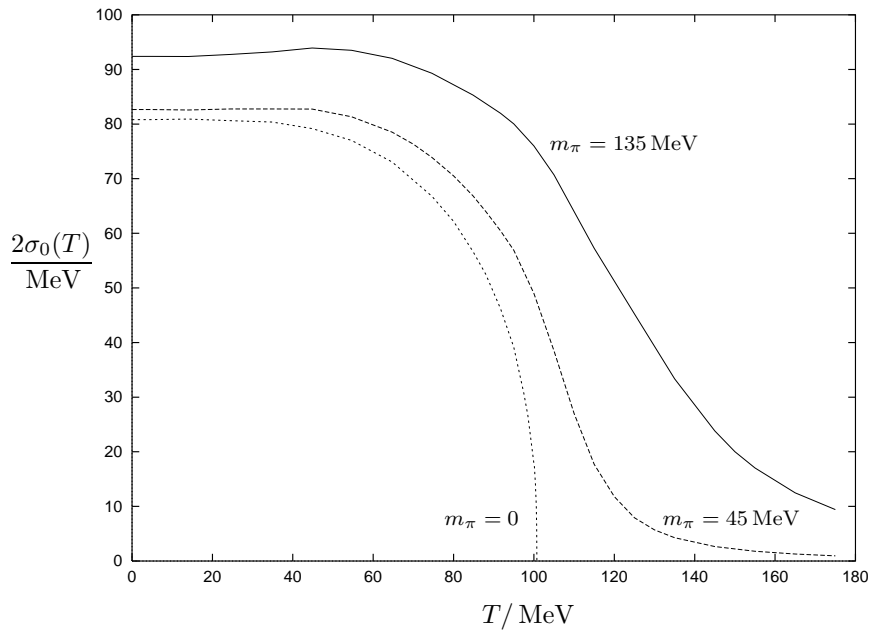}}
        }
      \put(-1.0,5.2){\bf $\ds{\frac{2\si_0(T)}{\MeV}}$}
      \put(5.3,0.5){\bf $\ds{T/\MeV}$}
      \put(4.8,1.8){\footnotesize\bf $m_\pi=0$}
      \put(7.7,1.8){\footnotesize\bf $m_\pi=45\MeV$}
      \put(6.2,6.8){\footnotesize\bf $m_\pi=135\MeV$}
    \end{picture}
  \end{center}
  \caption{\footnotesize The expectation value $2\si_0$ is shown as a
    function of temperature $T$ for three different values of the
    zero temperature pion mass.}
  \label{fpi_T}
\end{figure}

We have plotted in fig.~\ref{fpi_T} the renormalized expectation value
$2\si_0$ of the scalar field as a function of temperature for three different
values of the average light current quark mass $\hat{m}$. (We remind that
$2\si_0(T=0)=f_{\pi}$.)
For $\hat{m}=0$ the order parameter $\si_0$ of chiral symmetry breaking
continuously goes to zero for $T\ra T_c = 100.7\MeV$ characterizing the phase
transition to be of second order.  The universal behavior of the model for
small $T-T_c$ and small $\hat{m}$ is discussed in more detail in the following
section.  We point out that the value of $T_c$ corresponds to
$f_\pi^{(0)}=80.8\MeV$, i.e.~the value of the pion decay constant for
$\hat{m}=0$, which is significantly lower than $f_\pi=92.4\MeV$ obtained for
the realistic value $\hat{m}_{\rm phys}$.  If we would fix the value of the
pion decay constant to be $92.4\MeV$ also in the chiral limit ($\hat{m}=0$),
the value for the critical temperature would raise to $115\MeV$.  The nature
of the transition changes qualitatively for $\hat{m}\neq0$ where the second
order transition is replaced by a smooth crossover.  The crossover for a
realistic $\hat{m}_{\rm phys}$ or $m_{\pi}(T=0)=135 \MeV$ takes place in a
temperature range $T \simeq(120-150)\MeV$.  The middle curve in
fig.~\ref{fpi_T} corresponds to a value of $\hat{m}$ which is only a tenth of
the physical value, leading to a zero temperature pion mass $m_\pi=45\MeV$.
Here the crossover becomes considerably sharper but there remain substantial
deviations from the chiral limit even for such small quark masses
$\hat{m}\simeq 1 \MeV$. The temperature dependence of $m_\pi$ has already been
mentioned (see fig.~\ref{mpi_T}) for the same three values of $\hat{m}$.  As
expected, the pions behave like true Goldstone bosons for $\hat{m}=0$,
i.e.~their mass vanishes for $T\le T_c$. Interestingly, $m_\pi$ remains almost
constant as a function of $T$ for $T<T_c$ before it starts to increase
monotonically. We therefore find for two flavors no indication for a
substantial decrease of $m_\pi$ around the critical temperature.

\begin{center}
\begin{figure}
  \unitlength1.0cm
  \begin{center}
    \begin{picture}(10.,9.0)
      \put(-0.5,1.0){
        \epsfysize=11.cm
        \rotate[r]{\epsffile{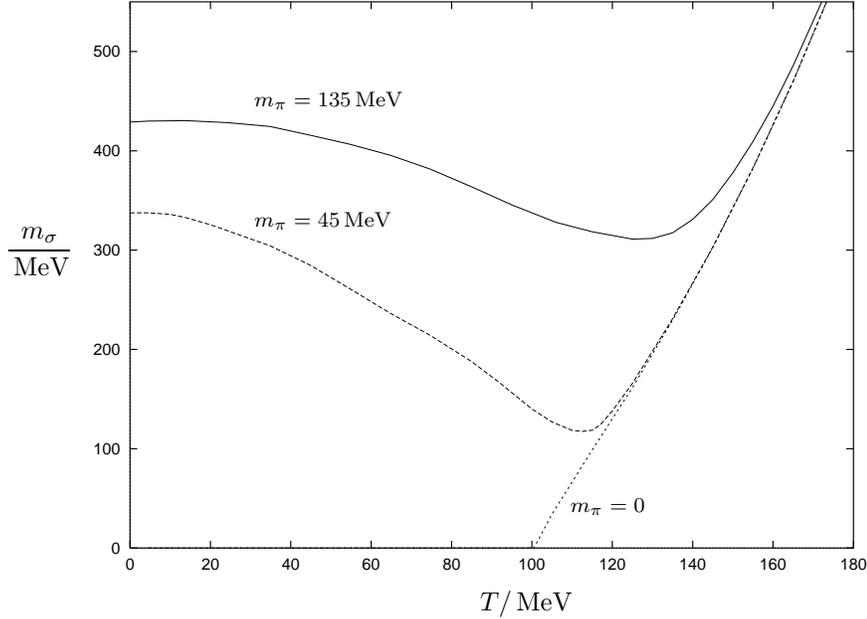}}
        }
      \put(-1.0,5.2){\bf $\ds{\frac{m_\si}{\MeV}}$}
      \put(5.3,0.5){\bf $\ds{T/\MeV}$}
      \put(6.5,1.8){\footnotesize\bf $m_\pi=0$}
      \put(2.3,5.6){\footnotesize\bf $m_\pi=45\MeV$}
      \put(2.3,7.2){\footnotesize\bf $m_\pi=135\MeV$}
    \end{picture}
  \end{center}
  \caption{\footnotesize The plot shows the $m_\si$ as a function of
    temperature $T$ for three different values of the 
    zero temperature pion mass.}
  \label{ms_T}
\end{figure}
\end{center}

The dependence of the mass of the sigma resonance $m_\si$ on the temperature
is displayed in fig.~\ref{ms_T} for the above three values of $\hat{m}$
In the absence of explicit chiral symmetry breaking, $\hat{m}=0$, the sigma
mass vanishes for $T\le T_c$. For $T<T_c$ this is a consequence of the
presence of massless Goldstone bosons in the Higgs phase which drive the
renormalized quartic coupling $\la$ to zero.  In fact, $\la$ runs linearly
with $k$ for $T \gta k/4$ and only evolves logarithmically for $T \lta k/4$.
Once $\hat{m}\neq0$ the pions acquire a mass even in the spontaneously broken
phase and the evolution of $\la$ with $k$ is effectively stopped at $k\sim
m_\pi$. Because of the temperature dependence of $\si_{0,k=0}$
(cf.~fig.~\ref{fpi_T}) this leads to a monotonically decreasing behavior of
$m_\si$ with $T$ for $T\lta T_c$. This changes into the expected monotonic
growth once the system reaches the symmetric phase for $T>T_c$. For low enough
$\hat{m}$ one may use the minimum of $m_{\si}(T)$ for an alternative
definition of the (pseudo-)critical temperature denoted as $T_{pc}^{(2)}$.
Tab.~\ref{tab11} in the introduction shows our results for the pseudocritical
temperature for different values of $\hat{m}$ or, equivalently,
$m_{\pi}(T=0)$. For a zero temperature pion mass $m_{\pi}=135 \MeV$ we find
$T_{pc}^{(2)}=128 \MeV$. At larger pion masses of about $230 \MeV$ we observe
no longer a characteristic minimum for $m_{\si}$ apart from a very broad,
slight dip at $T \simeq 90 \MeV$.  A comparison of our results with lattice
data is given below in the next section.

Our results for the chiral condensate $\VEV{\ol{\psi}\psi}$ as a function of
temperature for different values of the average current quark mass are
presented in fig.~\ref{ccc_T}. We will compare
$\VEV{\ol{\psi}\psi}(T,\hat{m})$ with its universal scaling form for the
$O(4)$ Heisenberg model in the following section.

\begin{figure}
  \unitlength1.0cm
  \begin{center}
    \begin{picture}(11.,9.0)
      \put(0.0,1.0){
        \epsfysize=11.cm
        \rotate[r]{\epsffile{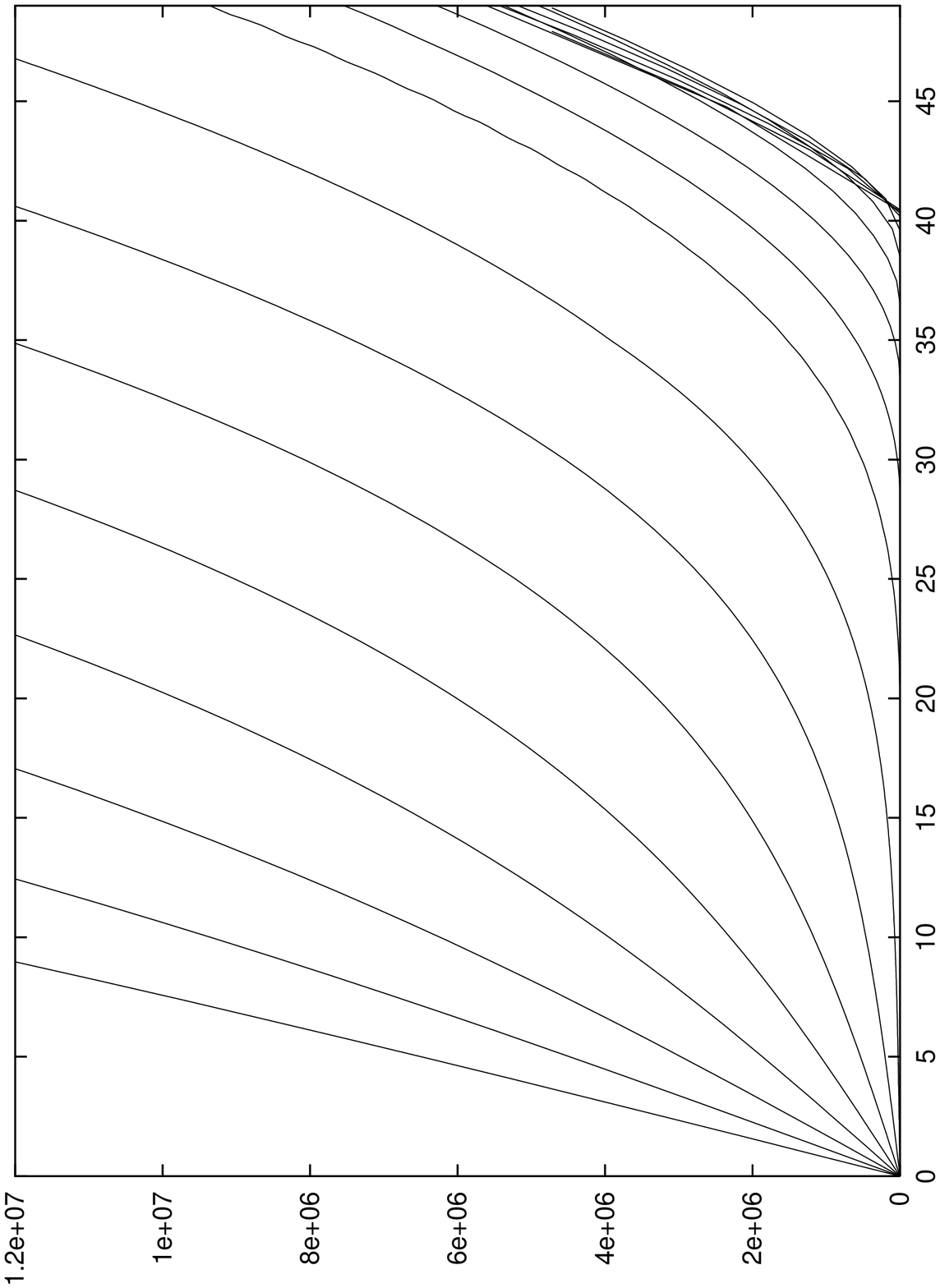}}
        }
      \put(-1.3,5.2){\bf $\ds{\frac{\partial U(T)/\partial \phi_R}
          {\MeV^{3}}}$}
      \put(5.8,0.5){\bf $\ds{\phi_R/\MeV}$}
    \end{picture}
  \end{center}
  \caption{\footnotesize The plot shows the derivative of the
    meson potential $U(T)$ with respect to the renormalized field
    $\phi_R=(Z_\Phi\rho/2)^{1/2}$ for different values of $T$.  The
    first curve on the left corresponds to $T=175 \MeV$. The successive
    curves to the right differ in temperature by $\Delta T=10 \MeV$ down
    to $T=5 \MeV$. }
  \label{Usig}
\end{figure}

Our ability to compute the complete temperature dependent effective meson
potential $U$ is demonstrated in fig.~\ref{Usig} where we display the
derivative of the potential with respect to the renormalized field
$\phi_R=(Z_\Phi\rho/2)^{1/2}$, for different values of $T$.
The curves cover a temperature range $T = (5 - 175) \MeV$.  The first
one to the left corresponds to $T=175 \MeV$ and neighboring curves
differ in temperature by $\Delta T = 10 \MeV$. One observes only a
weak dependence of $\partial U(T)/\partial\phi_R$ on the temperature
for $T\lta60\MeV$.  Evaluated for $\phi_R=\si_{0}$ this
function connects the renormalized field expectation value with
$m_{\pi}(T)$, the source $\jmath$ and the mesonic wave function
renormalization $Z_{\Phi}(T)$ according to
\begin{equation}
  \label{Usigeq}
  \ds{\frac{\partial U(T)}{\partial\phi_R}}
  (\phi_R=\si_{0})=
  \ds{\frac{2\jmath}{Z_{\Phi}^{1/2}(T)}}=4 \si_{0}(T) 
  m_{\pi}^2(T) \; .
\end{equation} 

We close this section with a short assessment of the validity of our effective
quark meson model as an effective description of two flavor QCD at
non--vanishing temperature.  The identification of qualitatively different
scale intervals which appear in the context of chiral symmetry breaking, as
presented in the preceding sections for the zero temperature case, can be
generalized to $T \neq 0$: For scales below $k_{\Phi}$ there exists a hybrid
description in terms of quarks and mesons. For $k_{\chi SB} \leq k \lta 600
\MeV$ chiral symmetry remains unbroken where the symmetry breaking scale
$k_{\chi SB}(T)$ decreases with increasing temperature. Also the constituent
quark mass decreases with $T$. The running Yukawa coupling depends only mildly
on temperature for $T\lta120\MeV$.  (Only near the critical temperature and
for $\hat{m}=0$ the running is extended because of massless pion
fluctuations.) On the other hand, for $k\lta4T$ the effective
three--dimensional gauge coupling increases faster than at $T=0$
leading$^{\citen{RW93-1}}$ to an increase of $\Lambda_{\rm QCD}(T)$ with $T$.
As $k$ gets closer to the scale $\Lambda_{\rm QCD}(T)$ it is no longer
justified to neglect in the quark sector confinement effects which go beyond
the dynamics of our present quark meson model.  Here it is important to note
that the quarks remain quantitatively relevant for the evolution of the meson
degrees of freedom only for scales $k \gta T/0.6$ (cf.~fig.~\ref{Thresh}).  In
the limit $k \ll T/0.6$ all fermionic Matsubara modes decouple from the
evolution of the meson potential according to the temperature dependent
version of~(\ref{AAA68}).  Possible sizeable confinement corrections to the
meson physics may occur if $\Lambda_{\rm QCD}(T)$ becomes larger than the
maximum of $M_q(T)$ and $T/0.6$.  This is particularly dangerous for small
$\hat{m}$ in a temperature interval around $T_c$. Nevertheless, the situation
is not dramatically different from the zero temperature case since only a
relatively small range of $k$ is concerned. We do not expect that the
neglected QCD non--localities lead to qualitative changes.  Quantitative
modifications, especially for small $\hat{m}$ and $\abs{T-T_c}$ remain
possible. This would only effect the non--universal amplitudes which will be
discussed in the next section.  The size of these corrections depends on the
strength of (non--local) deviations of the quark propagator and the Yukawa
coupling from the values computed in the quark meson model.

\section{Critical behavior near the chiral phase transition}
\label{CriticalBehavior}

In this section we study the linear quark meson model in the vicinity
of the critical temperature $T_c$ close to the chiral limit
$\hat{m}=0$. In this region we find that the sigma mass
$m_\si^{-1}$ is much larger than the inverse temperature $T^{-1}$,
and one observes an effectively three--dimensional behavior of the
high temperature quantum field theory.  We also note that the fermions
are no longer present in the dimensionally reduced system as has been
discussed above. We therefore have to deal with a purely bosonic
$O(4)$--symmetric linear sigma model.  At the phase transition the
correlation length becomes infinite and the effective
three--dimensional theory is dominated by classical statistical
fluctuations. In particular, the critical exponents which describe the
singular behavior of various quantities near the second order phase
transition are those of the corresponding classical system.

Many properties of this system are universal, i.e.~they only depend
on its symmetry ($O(4)$), the dimensionality of space (three) and its
degrees of freedom (four real scalar components). Universality means
that the long--range properties of the system do not depend on the
details of the specific model like its short distance
interactions. Nevertheless, important properties as the value of the
critical temperature are non--universal. We emphasize that although we
have to deal with an effectively three--dimensional bosonic theory,
the non--universal properties of the system crucially depend on the
details of the four--dimensional theory and, in particular, on the
fermions. 

Our aim is a computation of the critical equation of state which
relates for arbitrary $T$ near $T_c$ 
the derivative of the free energy or effective potential $U$
to the average current quark mass $\hat{m}$. The equation of state
then permits to study the temperature and quark mass dependence of
properties of the chiral phase transition.

At the critical temperature and in the chiral limit there is no scale present
in the theory. In the vicinity of $T_c$ and for small enough $\hat{m}$ one
therefore expects a scaling behavior of the effective average potential $u_k$
and accordingly a universal scaling form of the equation of
state.$^{\citen{TW94-1}}$ There are only two independent scales close to the
transition point which can be related to the deviation from the critical
temperature, $T-T_c$, and to the explicit symmetry breaking through the quark
mass $\hat{m}$.  As a consequence, the properly rescaled potential can only
depend on one scaling variable.  A possible choice for the parameterization of
the rescaled ``unrenormalized'' potential is the use of the Widom scaling
variable$^{\citen{Wid65-1}}$
\begin{equation}
  \label{XXX20}
  x=\frac{\left( T-T_c\right)/T_c}
  {\left(2\ol{\si}_0/T_c\right)^{1/\beta}}\; .
\end{equation}
Here $\beta$ is the critical exponent of the order parameter $\ol{\si}_0$ in
the chiral limit $\hat{m}=0$ (see~(\ref{NNN21})).  With
$U^\prime(\rho=2\ol{\si}_0^2)=\jmath/(2\ol{\si}_0)$ the Widom scaling form of
the equation of state reads$^{\citen{Wid65-1}}$
\begin{equation}
  \label{XXX21}
  \frac{\jmath}{T_c^3}=
  \left(\frac{2\ol{\si}_0}{T_c}\right)^\delta f(x)
\end{equation}
where the exponent $\delta$ is related to the behavior of the order parameter
according to~(\ref{NNN21b}).  The equation of state~(\ref{XXX21}) is written
for convenience directly in terms of four--dimensional quantities.  They are
related to the corresponding effective variables of the three--dimensional
theory by appropriate powers of $T_c$.  The source $\jmath$ is determined by
the average current quark mass $\hat{m}$ as
$\jmath=2\ol{m}^2_{k_\Phi}\hat{m}$.  The mass term at the compositeness scale,
$\ol{m}^2_{k_\Phi}$, also relates the chiral condensate to the order parameter
according to $\VEV{\ol{\psi}\psi}=-2\ol{m}^2_{k_\Phi}(\ol{\si}_0-\hat{m})$.
The critical temperature of the linear quark meson model was found above to be
$T_c=100.7\MeV$.

\begin{figure}[htb]
  \unitlength1.0cm
  \begin{center}
    \begin{picture}(11.5,10.0)
      \put(-0.5,6.0){$\ds{\frac{2\ol{\si}_0/T_c}
          {(\jmath/T_c^3 D)^{1/\dt}}}$}
      \put(5.5,0.7){$\ds{\frac{(T-T_c)/T_c}
          {(\jmath/T_c^3 B^{\dt} D)^{1/\bt \dt}}}$}
      \put(6.5,3.5){\footnotesize $\mbox{ERGE}$}
      \put(3.5,9.45){\footnotesize $\mbox{ERGE}$}
      \put(8.5,3.45){\footnotesize $\eps$}
      \put(2.9,9.55){\footnotesize $\eps$}
      \put(9.2,2.5){\footnotesize $\rm{M C}$}
      \put(3.6,8.2){\footnotesize $\rm{M C}$}
      \put(10.4,2.4){\footnotesize $\rm{M F}$}
      \put(5.1,8.1){\footnotesize $\rm{M F}$}
      \put(-0.47,-3.3){
        \epsfysize=15.75cm
        \epsfxsize=13.5cm
        \epsffile{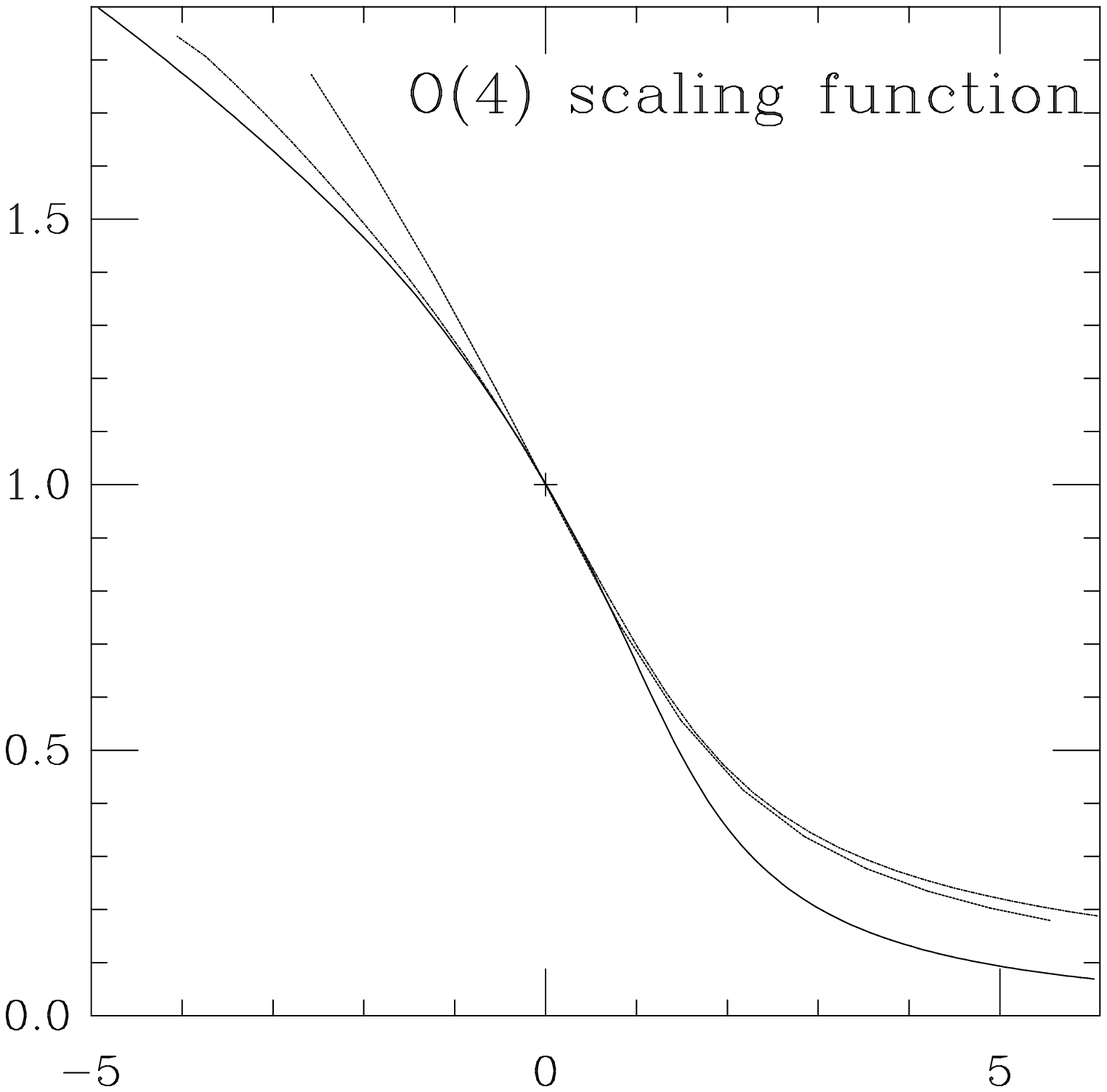}}
      \put(1.25,1.483){
        \epsfysize=10.0875cm
        \epsfxsize=8.415cm
        \rotate[r]{\epsffile{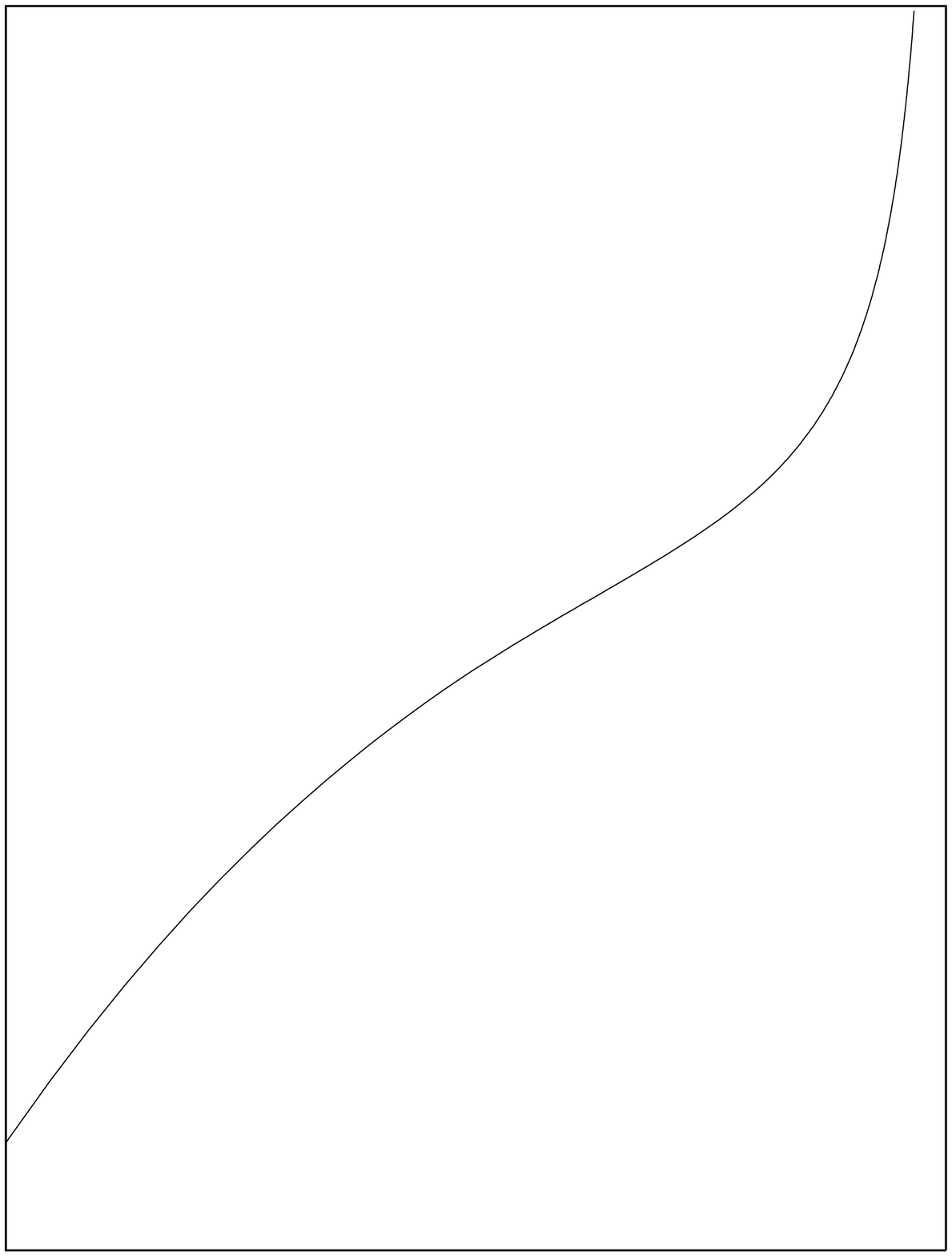}}}
    \end{picture}
  \end{center}
  \caption[]{\footnotesize
    The figure shows a comparison of our results, denoted by ``average
    action'', with results of other methods for the scaling function of the
    three--dimensional $O(4)$ Heisenberg model. We have labeled the axes for
    convenience in terms of the expectation value $\ol{\si}_0$ and the source
    $\jmath$ of the corresponding four--dimensional theory.  The constants $B$
    and $D$ specify the non--universal amplitudes of the model
    (cf.~(\ref{norm}).  The curve labeled by ``MC'' represents a fit to
    lattice Monte Carlo data. The second order epsilon
    expansion$^{\citen{BWW73-1}}$ and mean field results are denoted by
    ``$\epsilon$'' and ``mf'', respectively.  Apart from our results the
    curves are taken from Ref.$^{\citen{Tou97-1}}$.
    \label{scalfunc}
    }
\end{figure}

The scaling function $f$ is universal up to the model specific
normalization of $x$ and itself. Accordingly, all models in the same
universality class can be related by a rescaling of $\ol{\si}_0$
and $T-T_c$. The non--universal normalizations for the quark meson
model discussed here are defined according to
\begin{equation}
  \label{norm}
  f(0)=D\quad, \qquad f(-B^{-1/\beta})=0\; .
\end{equation}
We find $D=1.82\cdot10^{-4}$, $B=7.41$ and our result for $\beta$ is given in
tab.~\ref{tab2}. Apart from the immediate vicinity of the zero of $f(x)$ we
find the following two parameter fit for the scaling
function,$^{\citen{BTW96-1}}$
\begin{equation}
  \label{ffit}
  \begin{array}{rcl}
    \ds{f_{\rm fit}(x)}&=&\ds{1.816 \cdot 10^{-4} (1+136.1\, x)^2 \,
      (1+160.9\, \theta\,
      x)^{\Delta}}\nnn
    &&
    \ds{(1+160.9\, (0.9446\, \theta^{\Delta})^{-1/(\gamma-2-\Delta)} 
      \, x)^{\gamma-2-\Delta}}
  \end{array}
\end{equation}
to reproduce the numerical results for $f$ and $df/dx$ at the $1-2\%$ level
with $\theta=0.625$ $(0.656)$, $\Delta=-0.490$ $(-0.550)$ for $x > 0$ $(x <
0)$ and $\gamma$ as given in tab.~\ref{tab2}.  The universal properties of the
scaling function can be compared with results obtained by other methods for
the three--dimensional $O(4)$ Heisenberg model.  In fig.~\ref{scalfunc} we
display our results along with those obtained from lattice Monte Carlo
simulation,$^{\citen{Tou97-1}}$ second order epsilon
expansion$^{\citen{BWW73-1}}$ and mean field theory.
We observe a good agreement of average action, lattice and epsilon expansion
results within a few per cent for $T < T_c$. Above $T_c$ the average action
and the lattice curve go quite close to each other with a substantial
deviation from the epsilon expansion and mean field scaling function. (We note
that the question of a better agreement of the curves for $T<T_c$ or $T>T_c$
depends on the chosen non--universal normalization conditions for $x$ and $f$
(cf.~(\ref{norm})).)

Before we use the scaling function $f(x)$ to discuss the general
temperature and quark mass dependent case, we consider the limits
$T=T_c$ and $\hat{m}=0$, respectively.  In these limits the behavior
of the various quantities is determined solely by critical amplitudes
and exponents. In the spontaneously broken phase ($T<T_c$) and in the
chiral limit we observe that the renormalized and unrenormalized order
parameters scale according to
\begin{equation}
  \label{NNN21}
  \begin{array}{rcl}
    \ds{\frac{2\si_0(T)}{T_c}} &=& \ds{
      \left(2E\right)^{1/2}
        \left(\frac{T_c-T}{T_c}\right)^{\nu/2}
      }\; ,\nnn
    \ds{\frac{2\ol{\si}_0(T)}{T_c}} &=& \ds{
      B \left(\frac{T_c-T}{T_c}\right)^{\beta}
      }\; ,
  \end{array}
\end{equation}
respectively, with $E=0.814$ and the value of $B$ given above.  In the
symmetric phase the renormalized mass $m=m_\pi=m_\si$ and the
unrenormalized mass $\ol{m}=Z_\Phi^{1/2}m$ behave as
\begin{equation}
  \label{NNN21a}
  \begin{array}{rcl}
    \ds{\frac{m(T)}{T_c}} &=& \ds{
      \left(\xi^+\right)^{-1}
      \left(\frac{T-T_c}{T_c}\right)^\nu
      }\; ,\nnn
    \ds{\frac{\ol{m}(T)}{T_c}} &=& \ds{
      \left( C^+\right)^{-1/2}
      \left(\frac{T-T_c}{T_c}\right)^{\gamma/2}
       \; , }
  \end{array}
\end{equation}
where $\xi^+=0.270$, $C^+=2.79$. For $T=T_c$ and
non--vanishing current quark mass we have
\begin{equation}
  \label{NNN21b}
  \begin{array}{rcl}
    \ds{\frac{2\ol{\si}_0}{T_c}} &=& \ds{
      D^{-1/\delta}
        \left(\frac{\jmath}{T_c^3}\right)^{1/\delta}
      }
  \end{array}
\end{equation}
with the value of $D$ given above. 

Though the five amplitudes $E$, $B$, $\xi^+$, $C^+$ and $D$ are not
universal there are ratios of amplitudes which are invariant under a
rescaling of $\ol{\si}_0$ and $T-T_c$. Our results for the
universal amplitude ratios are
\begin{equation}
  \label{ABC01}
  \begin{array}{rcl}
    \ds{R_\chi} &=& \ds{C^+ D B^{\delta-1}=1.02}\; ,\nnn
    \ds{\tilde{R}_\xi} &=& \ds{
      (\xi^+)^{\beta/\nu}D^{1/(\delta+1)}B=0.852}\; ,\nnn
    \ds{\xi^+ E} &=& \ds{0.220}\; .
  \end{array}
\end{equation}
Those for the critical exponents are given in tab.~\ref{tab2}.
\begin{table}
  \caption[]{\footnotesize The table shows the critical exponents 
    corresponding to the three--dimensional $O(4)$--Heisenberg model.
    Our results are denoted by ``AA'' whereas ``$3d$ PT''
    labels the exponents obtained from a $3d$ perturbative expansion
    using Pad\'e techniques.$^{\citen{BMN78-1,Zin93-1}}$ The bottom line
    contains lattice Monte Carlo results.$^{\citen{KK95-1}}$
    \label{tab2}}
  \begin{center}
    \begin{tabular}{|c||l|l|l|l|l|} \hline
      &
      $\nu$ &
      $\gamma$ &
      $\delta$ &
      $\beta$ &
      $\eta$
      \\[0.5mm] \hline\hline
      AA &
      $0.787$ &
      $1.548$ &
      $4.80$ &
      $0.407$ &
      $0.0344$
      \\ \hline
      $3d$ PT &
      $0.73(2)$ &
      $1.44(4)$ &
      $4.82(5)$ &
      $0.38(1)$ &
      $0.03(1)$
      \\ \hline
      MC &
      $0.7479(90)$ &
      $1.477(18)$ &
      $4.851(22)$ &
      $0.3836(46)$ &
      $0.0254(38)$
      \\ \hline
    \end{tabular}
  \end{center}
\end{table}
Here the value of $\eta$ is obtained from the temperature dependent version
of~(\ref{AAA69n}) at the critical temperature.$^{\citen{BJW97-1}}$ For
comparison tab.~\ref{tab2} also gives the results$^{\citen{BMN78-1,Zin93-1}}$
of a $3d$ perturbative computation as well as lattice Monte Carlo
results$^{\citen{KK95-1}}$ which have been used for the lattice form of the
scaling function in fig.~\ref{scalfunc}.~\footnote{See also
  Ref.$^{\citen{MT97-1}}$ and references therein for a recent calculation of
  critical exponents using similar methods as in this work. For high precision
  estimates of the critical exponents see also
  Refs$^{\citen{BC95-1,Rei95-1}}$.}  There are only two independent amplitudes
and critical exponents, respectively.  They are related by the usual scaling
relations of the three--dimensional scalar $O(N)$--model$^{\citen{Zin93-1}}$
which we have explicitly verified by the independent calculation of our
exponents.

We turn to the discussion of the scaling behavior of the chiral condensate
$\VEV{\ol{\psi}\psi}$ for the general case of a temperature and quark mass
dependence.  In fig.~\ref{ccc_T} we have displayed our results for the scaling
equation of state in terms of the chiral condensate
\begin{equation}
  \label{XXX30}
  \VEV{\ol{\psi}\psi}=
  -\ol{m}^2_{k_\Phi}T_c
  \left(\frac{\jmath/T_c^3}{f(x)}\right)^{1/\delta}+
    \jmath
\end{equation}
as a function of $T/T_c=1+x(\jmath/T_c^3 f(x))^{1/\beta\delta}$ for different
quark masses or, equivalently, different values of $\jmath$.  The curves shown
in fig.~\ref{ccc_T} correspond to quark masses $\hat{m}=0$,
$\hat{m}=\hat{m}_{\rm phys}/10$, $\hat{m}=\hat{m}_{\rm phys}$ and
$\hat{m}=3.5\hat{m}_{\rm phys}$ or, equivalently, to zero temperature pion
masses $m_\pi=0$, $m_\pi=45\MeV$, $m_\pi=135\MeV$ and $m_\pi=230\MeV$,
respectively (cf.~fig.~\ref{mm}). One observes that the second order phase
transition with a vanishing order parameter at $T_c$ for $\hat{m}=0$ is turned
into a smooth crossover in the presence of non--zero quark masses.

The scaling form~(\ref{XXX30}) for the chiral condensate is exact only in the
limit $T\to T_c$, $\jmath\ra0$.  It is interesting to find the range of
temperatures and quark masses for which $\VEV{\ol{\psi}\psi}$ approximately
shows the scaling behavior~(\ref{XXX30}).  This can be inferred from a
comparison (see fig.~\ref{ccc_T}) with our full non--universal solution for
the $T$ and $\jmath$ dependence of $\VEV{\ol{\psi}\psi}$. For $m_\pi=0$ one
observes approximate scaling behavior for temperatures $T\gta90\MeV$. This
situation persists up to a pion mass of $m_\pi=45\MeV$. Even for the realistic
case, $m_\pi=135\MeV$, and to a somewhat lesser extent for $m_\pi=230\MeV$ the
scaling curve reasonably reflects the physical behavior for $T\gta T_c$. For
temperatures below $T_c$, however, the zero temperature mass scales become
important and the scaling arguments leading to universality break down.

The above comparison may help to shed some light on the use of universality
arguments away from the critical temperature and the chiral limit. One
observes that for temperatures above $T_c$ the scaling assumption leads to
quantitatively reasonable results even for a pion mass almost twice as large
as the physical value. This in turn has been used for two flavor lattice QCD
as theoretical input to guide extrapolation of results to light current quark
masses.  From simulations based on a range of pion masses $0.3\lta
m_\pi/m_\rho\lta0.7$ and temperatures $0<T\lta250\MeV$ a ``pseudocritical
temperature'' of approximately $140\MeV$ with a weak quark mass dependence is
reported.$^{\citen{MILC97-1}}$ Here the ``pseudocritical temperature''
$T_{pc}$ is defined as the inflection point of $\VEV{\ol{\psi}\psi}$ as a
function of temperature.  The values of the lattice action parameters used
in$^{\citen{MILC97-1}}$ with $N_t=6$ were $a\hat{m}=0.0125$, $6/g^2=5.415$ and
$a\hat{m}=0.025$, $6/g^2=5.445$. For comparison with lattice data we have
displayed in fig.~\ref{ccc_T} the temperature dependence of the chiral
condensate for a pion mass $m_\pi=230\MeV$.  From the free energy of the
linear quark meson model we obtain in this case a pseudocritical temperature
of about $150\MeV$ in reasonable agreement with the results of
Ref.$^{\citen{MILC97-1}}$.  In contrast, for the critical temperature in the
chiral limit we obtain $T_c=100.7\MeV$.  This value is considerably smaller
than the lattice results of about $(140 - 150) \MeV$ obtained by extrapolating
to zero quark mass in Ref.$^{\citen{MILC97-1}}$.  We point out that for pion
masses as large as $230\MeV$ the condensate $\VEV{\ol{\psi}\psi}(T)$ is almost
linear around the inflection point for quite a large range of temperature.
This makes a precise determination of $T_c$ somewhat difficult.  Furthermore,
fig.~\ref{ccc_T} shows that the scaling form of $\VEV{\ol{\psi}\psi}(T)$
underestimates the slope of the physical curve. Used as a fit with $T_c$ as a
parameter this can lead to an overestimate of the pseudocritical temperature
in the chiral limit.  We also mention here the results of
Ref.$^{\citen{Got97-1}}$.  There two values of the pseudocritical temperature,
$T_{pc}=150(9)\MeV$ and $T_{pc}=140(8)$, corresponding to $a\hat{m}=0.0125$,
$6/g^2=5.54(2)$ and $a\hat{m}=0.00625$, $6/g^2=5.49(2)$, respectively, (both
for $N_t=8$) were computed.  These values show a somewhat stronger quark mass
dependence of $T_{pc}$ and were used for a linear extrapolation to the chiral
limit yielding $T_c=128(9)\MeV$.

The linear quark meson model exhibits a second order phase transition for two
quark flavors in the chiral limit. As a consequence the model predicts a
scaling behavior near the critical temperature and the chiral limit which can,
in principle, be tested in lattice simulations. For the quark masses used in
the present lattice studies the order and universality class of the transition
in two flavor QCD remain a partially open question. Though there are results
from the lattice giving support for critical
scaling$^{\citen{Kar94-1,IKKY97-1}}$ there are also recent simulations with
two flavors that reveal significant finite size effects and problems with
$O(4)$ scaling.$^{\citen{BKLO96-1,Uka97-1}}$

\section{Additional degrees of freedom}
\label{AdditionalDegreesOfFreedom}

So far we have investigated the chiral phase transition of QCD as described by
the linear $O(4)$--model containing the three pions and the sigma resonance as
well as the up and down quarks as degrees of freedom. Of course, it is clear
that the spectrum of QCD is much richer than the states incorporated in our
model. It is therefore important to ask to what extent the neglected degrees
of freedom like the strange quark, strange (pseudo)scalar mesons,
(axial)vector mesons, baryons, etc., might be important for the chiral
dynamics of QCD.  Before doing so it is perhaps instructive to first look into
the opposite direction and investigate the difference between the linear quark
meson model described here and chiral perturbation theory based on the
non--linear sigma model.$^{\citen{Leu95-1}}$ In some sense, chiral
perturbation theory is the minimal model of chiral symmetry breaking
containing only the Goldstone degrees of freedom. By construction it is
therefore only valid in the spontaneously broken phase and can not be expected
to yield realistic results for temperatures close to $T_c$ or for the
symmetric phase.  However, for small temperatures (and momentum scales) the
non--linear model is expected to describe the low--energy and low--temperature
limit of QCD reliably as an expansion in powers of the light quark masses. For
vanishing temperature it has been demonstrated
recently$^{\citen{JW96-1,JW96-3,JW97-1}}$ that the results of chiral
perturbation theory can be reproduced within the linear meson model once
certain higher dimensional operators in its effective action are taken into
account for the three flavor case.  Moreover, some of the parameters of chiral
perturbation theory ($L_4,\ldots,L_8$) can be expressed and therefore also
numerically computed in terms of those of the linear model. For non--vanishing
temperature one expects agreement only for low $T$ whereas deviations from
chiral perturbation theory should become large close to $T_c$.  Yet, even for
$T\ll T_c$ small quantitative deviations should exist because of the
contributions of (constituent) quark and sigma meson fluctuations in the
linear model which are not taken into account in chiral perturbation theory.

\begin{figure}
  \unitlength1.0cm
  \begin{center}
    \begin{picture}(11.,9.0)
      \put(0.0,1.0){
        \epsfysize=11.cm
        \rotate[r]{\epsffile{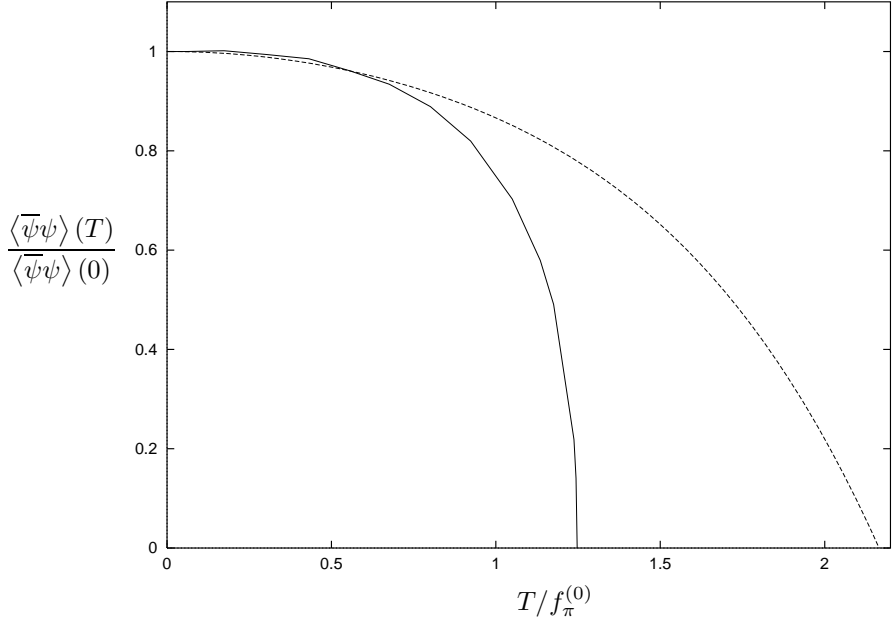}}
        }
      \put(-1.0,5.2){\bf 
        $\ds{\frac{\VEV{\ol{\psi}\psi}(T)}{\VEV{\ol{\psi}\psi}(0)} }$}
      \put(5.8,0.5){\bf $\ds{T/f_\pi^{(0)}}$}
    \end{picture}
  \end{center}
  \caption{\footnotesize The plot displays the chiral condensate
    $\VEV{\ol{\psi}\psi}$ as a function of $T/f_\pi^{(0)}$. The solid
    line corresponds to our results for vanishing average current quark
    mass $\hat{m}=0$ whereas the dashed line shows the corresponding
    three--loop chiral perturbation theory result for
    $\Gamma_1=470\MeV$.}
  \label{cc_T}
\end{figure}

From$^{\citen{GL87-1}}$ we infer the three--loop result for the temperature
dependence of the chiral condensate in the chiral limit for $N$ light flavors
\begin{equation}
  \label{BBB100}
  \begin{array}{rcl}
    \ds{\VEV{\ol{\psi}\psi}(T)_{\chi PT}} &=& \ds{
      \VEV{\ol{\psi}\psi}_{\chi PT}(0)
      \Bigg\{1-\frac{N^2-1}{N}\frac{T^2}{12F_0^2}-
        \frac{N^2-1}{2N^2}
        \left(\frac{T^2}{12F_0^2}\right)^2}\nnn
    &+& \ds{
      N(N^2-1)\left(\frac{T^2}{12F_0^2}\right)^3
      \ln\frac{T}{\Gamma_1}
      \Bigg\} +\Oc(T^8)}\; .
  \end{array}
\end{equation}
The scale $\Gamma_1$ can be determined from the $D$--wave isospin zero
$\pi\pi$ scattering length and is given by $\Gamma_1=(470\pm100)\MeV$.  The
constant $F_0$ is (in the chiral limit) identical to the pion decay constant
$F_0=f_\pi^{(0)}=80.8\MeV$. In fig.~\ref{cc_T} we have plotted the chiral
condensate as a function of $T/F_0$ for both, chiral perturbation theory
according to~(\ref{BBB100}) and for the linear quark meson model.
As expected the agreement for small $T$ is very good. Nevertheless,
the anticipated small numerical deviations present even for $T\ll T_c$
due to quark and sigma meson loop contributions are manifest.  For
larger values of $T$, say for $T\gta0.8f_\pi^{(0)}$ the deviations
become significant because of the intrinsic inability of chiral
perturbation theory to correctly reproduce the critical behavior of
the system near its second order phase transition.

Within the language of chiral perturbation theory the neglected effects of
thermal quark fluctuations may be described by an effective temperature
dependence of the parameter $F_0(T)$. We notice that the temperature at which
these corrections become important equals approximately one third of the
constituent quark mass $M_q(T)$ or the sigma mass $m_\si(T)$, respectively, in
perfect agreement with fig.~\ref{Thresh}. As suggested by this figure the
onset of the effects from thermal fluctuations of heavy particles with a
$T$--dependent mass $m_H(T)$ is rather sudden for $T\gta m_H(T)/3$.  These
considerations also apply to our two flavor quark meson model.  Within full
QCD we expect temperature dependent initial values at $k_\Phi$.

The dominant contribution to the temperature dependence of the initial
values presumably arises from the influence of the mesons containing
strange quarks as well as the strange quark itself.  Here the quantity
$\ol{m}^2_{k_\Phi}$ seems to be the most important one.  (The
temperature dependence of higher couplings like $\la(T)$ is not
very relevant if the IR attractive behavior remains valid, i.e.~if
$Z_{\Phi,k_\Phi}$ remains small for the range of temperatures
considered. We neglect a possible $T$--dependence of the current quark
mass $\hat{m}$.) In particular, for three flavors the potential
$U_{k_\Phi}$ contains a term
\begin{equation}
  \label{LLL12}
  -\frac{1}{2}\ol{\nu}_{k_\Phi}
  \left(\det\Phi+\det\Phi^\dagger\right)=
  -\ol{\nu}_{k_\Phi}\vph_s\Phi_{uu}\Phi_{dd}+\ldots
\end{equation}
which reflects the axial $U_A(1)$ anomaly. It yields a contribution to
the effective mass term proportional to the expectation value
$\VEV{\Phi_{ss}}\equiv\vph_s$, i.e.
\begin{equation}
  \label{LLL13}
  \Delta\ol{m}^2_{k_\Phi}=
  -\frac{1}{2}\ol{\nu}_{k_\Phi}\vph_s\; .
\end{equation}
Both, $\ol{\nu}_{k_\Phi}$ and $\vph_s$, depend on $T$.  We expect
these corrections to become relevant only for temperatures exceeding
$m_K(T)/3$ or $M_s(T)/3$. We note that the temperature dependent kaon
and strange quark masses, $m_K(T)$ and $M_s(T)$, respectively, may be
somewhat different from their zero temperature values but we do not
expect them to be much smaller. A typical value for these scales is
around $500\MeV$. Correspondingly, the thermal fluctuations neglected
in our model should become important for $T\gta170\MeV$. It is even
conceivable that a discontinuity appears in $\vph_s(T)$ for
sufficiently high $T$ (say $T\simeq170\MeV$). This would be reflected
by a discontinuity in the initial values of the $O(4)$--model leading
to a first order transition within this model.  Obviously, these
questions should be addressed in the framework of the three flavor
$SU_L(3)\times SU_R(3)$ quark meson model. Work in this direction is
in progress.

We note that the temperature dependence of $\ol{\nu}(T)\vph_s(T)$ is closely
related to the question of an effective high temperature restoration of the
axial $U_A(1)$ symmetry.$^{\citen{PW84-1,Shu94-1}}$ The $\eta^\prime$ mass
term is directly proportional to this combination,$^{\citen{JW96-1}}$
$m_{\eta^\prime}^2(T)-m_\pi^2(T)\simeq\frac{3}{2}\ol{\nu}(T) \vph_s(T)$.
Approximate $U_A(1)$ restoration would occur if $\vph_s(T)$ or $\ol{\nu}(T)$
would decrease sizeable for large $T$.  For realistic QCD this question should
be addressed by a three flavor study. Within two flavor QCD the combination
$\ol{\nu}_k\vph_s$ is replaced by an effective anomalous mass term
$\ol{\nu}_k^{(2)}$. The temperature dependence of $\ol{\nu}^{(2)}(T)$ could be
studied by introducing quarks and the axial anomaly in the two flavor matrix
model of Ref.$^{\citen{BW97-1}}$.  We add that this question has also been
studied within full two flavor QCD in lattice
simulations.$^{\citen{BKLO96-1,MILC97-2,KLS97-1}}$ So far there does not seem
to be much evidence for a restoration of the $U_A(1)$ symmetry near $T_c$ but
no final conclusion can be drawn yet.

To summarize, we have found that the effective two flavor quark meson
model presumably gives a good description of the temperature effects
in two flavor QCD for a temperature range $T\lta170\MeV$. Its
reliability should be best for low temperature where our results agree
with chiral perturbation theory.  However, the range of validity is
considerably extended as compared to chiral perturbation theory and
includes, in particular, the critical temperature of the second order
phase transition in the chiral limit.  We have explicitly connected
the universal critical behavior for small $\abs{T-T_c}$ and small
current quark masses with the renormalized couplings at $T=0$ and
realistic quark masses. The main quantitative uncertainties from
neglected fluctuations presumably concern the values of $f_\pi^{(0)}$
and $T_c$ which, in turn, influence the non--universal amplitudes $B$
and $D$ in the critical region. We believe that our overall picture is
rather solid. Where applicable our results compare well with numerical
simulations of full two flavor QCD.

\section{Conclusions}

Our conclusions may be summarized in the following points:
\begin{enumerate}
\item The connection between short distance perturbative QCD and long
  distance meson physics by analytical methods seems to emerge step by
  step. The essential ingredients are nonperturbative flow equations
  as approximations of exact renormalization group equations and a
  formalism which allows a change of variables by introducing
  meson--like composite fields.
\item The relevance of meson--like $\ol{\psi}\psi$ composite objects
  is established in this framework. A typical scale where the mesonic
  bound states appear is the compositeness scale
  $k_\Phi\simeq(600-700)\MeV$. The occurrence of chiral symmetry
  breaking depends on the effective meson mass $\ol{m}_{k_\Phi}$ at
  the compositeness scale. For a certain range of values of the ratio
  $\ol{m}_{k_\Phi}^2/k_\Phi^2$ spontaneous chiral symmetry breaking is
  induced by quark fluctuations. A definite analytical establishment
  of spontaneous chiral symmetry breaking from ``first principles''
  (i.e., short distance QCD) still awaits a reliable calculation of
  this ratio.
\item Phenomenologically, the ratio $\ol{m}_{k_\Phi}^2/k_\Phi^2$ may
  be determined from the value of the constituent quark mass in units
  of the pion decay constant $f_\pi$. For this value the ratios
  $f_\pi/k_\Phi$ and $\VEV{\ol{\psi}\psi}/k_\Phi^3$ can be computed.
  Together with an earlier estimate of $k_\Phi$ this yields rather
  encouraging results: $f_\pi\simeq100\MeV$ and
  $\VEV{\ol{\psi}\psi}\simeq-(190\MeV)^3$ for two flavor QCD. It will
  be very interesting to generalize these results to the realistic
  three flavor case.
\item The formalism presented here naturally leads to an effective
  linear quark meson model for the description of mesons below the
  compositeness scale $k_\Phi$. For this model the standard
  non--linear sigma model of chiral perturbation theory emerges as a
  low energy approximation due to the large sigma mass.
\item The old puzzle about the precise connection between the current
  and the constituent quark mass reveals new interesting aspects in
  this formalism. In the context of the effective average action one
  may define these masses by the quark propagator at zero or very
  small momentum, either in the quark gluon picture (current quark
  mass) or the effective quark meson model (constituent quark mass).
  There is no conceptual difference between the two situations. As a
  function of $k$ the running quark mass smoothly interpolates between
  the standard current quark mass $m_q$ for high $k$ and the standard
  constituent quark mass $M_q$ for low $k$. (This holds at least as
  long as the minimum of the effective scalar potential is unique.) A
  rapid quantitative change occurs for $k\simeq(400-500)\MeV$ because
  of the onset of chiral symmetry breaking. For $k=0$ one expects this
  behavior to carry over to the momentum dependence of the quark
  propagator: For small $q^2$ the inverse quark propagator is
  dominated by $M_q$. In contrast, for high $q^2$ the constant term in
  the inverse propagator is reduced to $m_q$ since $q^2$ replaces
  $k^2$ as an effective infrared cutoff.  One expects a smooth
  interpolation between the two limits and it would be interesting to
  know the form of the propagator for momenta in the transition
  region. Furthermore, the symmetry breaking source term which
  determines the pion mass in the linear or non--linear meson model
  can be related to the current quark mass at the compositeness scale.
  This constitutes a bridge between the low energy meson properties
  and the running quark mass at a scale which is not too far from the
  validity of perturbation theory.
\item A particular version of the Nambu--Jona-Lasinio model appears in
  our formalism as a limiting case (infinitely strong renormalized
  Yukawa coupling at the compositeness scale $k_\Phi$). Here the
  ultraviolet cutoff which is implicit in the NJL model is dictated by
  the momentum dependence of the infrared cutoff function $R_k$ in the
  effective average action. The characteristic cutoff scale is
  $k_\Phi$. In this context our results can be interpreted as an
  approximative solution of the NJL model. Our method includes many
  contributions beyond the leading order contribution of the $1/N_c$
  expansion.
\item Based on a satisfactory understanding of the meson properties in
  the vacuum we have described their behavior in a thermal equilibrium
  situation. Our method should remain valid for temperature below
  $\sim170\MeV$. This extends well beyond the validity of chiral
  perturbation theory. In particular, in the chiral limit of vanishing
  quark masses we can describe the universal critical behavior near a
  second order phase transition of two flavor QCD. The universal
  behavior is quantitatively connected to observed quantities at zero
  temperature and realistic quark masses.
\end{enumerate}

\section*{Acknowledgments}
We thank J.~Berges and B.~Bergerhoff for collaboration on many
subjects covered by these lectures. This work was supported in part by the
{\em Deutsche Forschungsgemeinschaft}.

\end{document}